\documentclass[american,twocolumn,longbibliography,aps]{revtex4-2}

\usepackage[T1]{fontenc}
\usepackage{textcomp}
\usepackage[utf8]{inputenc}
\usepackage{color}
\usepackage{babel}
\usepackage{verbatim}
\usepackage{refstyle}
\usepackage{amsmath}
\usepackage{amssymb}
\usepackage{graphicx}
\usepackage{wasysym}
\usepackage{esint}
\usepackage{microtype}
\usepackage[bookmarks=true,bookmarksnumbered=false,bookmarksopen=false,
 breaklinks=false,pdfborder={0 0 0},pdfborderstyle={},backref=false,colorlinks=true]
 {hyperref}
\hypersetup{pdftitle={Influence of Trotterization Error on Single-particle Tunneling},
 pdfauthor={Anton Khvalyuk},
 pdfsubject={Quantum simulations}}

\makeatletter

\AtBeginDocument{\providecommand\appref[1]{\ref{app:#1}}}
\AtBeginDocument{\providecommand\secappref[1]{\ref{secapp:#1}}}
\AtBeginDocument{\providecommand\subsecappref[1]{\ref{subsecapp:#1}}}

\RS@ifundefined{subsecref}{
  \newref{subsec}{
        name      = \RSsectxt,
        names     = \RSsecstxt,
        Name      = \RSSectxt,
        Names     = \RSSecstxt,
        refcmd    = {\S\ref{#1}},
        rngtxt    = \RSrngtxt,
        lsttwotxt = \RSlsttwotxt,
        lsttxt    = \RSlsttxt
  }
}{}

\usepackage{chngcntr} 

\newref{fig}{ refcmd={Fig.~\ref{#1}} }
\newref{tab}{ refcmd={Tab.~\ref{#1}} }
\newref{eq}{ refcmd={(\ref{#1})} }
\newref{subsec}{ refcmd={Sec.~\ref{#1}} }
\newref{sec}{ refcmd={Sec.~\ref{#1}} }
\newref{app}{ refcmd={Appendix~\ref{#1}} }

\let\secappref=\relax 
\newref{secapp}{ refcmd={Appendix~\ref{#1}} } 

\let\subsecappref=\relax 
\newref{subsecapp}{ refcmd={Appendix Sec.~\ref{#1}} } 

\makeatother

\begin{document}
\title{Influence of Trotterization Error on Single-particle Tunneling}
\author{Anton V. Khvalyuk}
\email{anton.khvalyuk@ens-lyon.fr}

\affiliation{LPMMC, University Grenoble-Alpes, France}
\author{Kostyantyn Kechedzhi}
\affiliation{Google Research, Mountain View, CA, USA}
\author{Vadim S. Smelyansky}
\affiliation{Google Research, Mountain View, CA, USA}
\author{Lev B. Ioffe}
\affiliation{Google Research, Mountain View, CA, USA}
\date{\today{}}
\begin{abstract}
Existing superconducting quantum devices struggle to simulate spatial
quantum tunneling as it is exponentially slow and inherently nonlocal.
The Suzuki-Trotter algorithm (STA) offers a practical solution for
existing quantum hardware by approximating nonlocal evolution with
a local two-qubit gate sequence that requires no ancilla qubits and
preserves physical interpretability. However, estimating the required
gate count from naive fidelity-based error bounds yields values exceeding
the constraints imposed by noise and decoherence in real devices.
Here, we fully characterize STA for a semiclassical tunneling model
by adapting the Wentzel-Kramers-Brillouin approximation. We find that,
in the general case, tunneling is suppressed by resonance detuning.
In contrast, if STA preserves the target resonance, tunneling rates
are exponentially enhanced via non-perturbative renormalization, dramatically
reducing the gate count required to observe tunneling and thus benefiting
from large Trotter steps---equally natural for existing quantum hardware.
However, for the latter regime, we further find that Floquet mixing
of quasi-energy levels reconstructs the low-energy spectrum and strongly
deforms tunneling dynamics by introducing additional allowed regions.
Based on the semiclassical picture, we conjecture interplay of tunneling
and large-step STA in the presence of topology-altering modifications,
such as staggered potential modulation---a subject of future work.
Our findings enable direct tunneling simulations on existing superconducting
platforms, providing a controlled platform for studying the nontrivial
interplay of tunneling and STA, with the possibility of addressing
the effects of noise, disorder, and interaction.
\end{abstract}
\maketitle

\section{Introduction}

The notoriously hard problem of simulating many-body quantum systems,
often intractable even for the most powerful classical supercomputers,
can be addressed by employing a well-controlled quantum system as
a simulator of other quantum systems~\citep{quantum-simulation_review_2014}.
Superconducting devices in the form of a 2D grid of two-level systems
(qubits) exemplify this approach~\citep{arute-2019_quantum-supremacy-paper,nature-2021_long-coherence-times},
with the grid structure making them a natural platform for simulating
one- and two-dimensional spin-$1/2$ models~\citep{neill-2021_fermionic-ring-simulation,satzinger-2021_realizing-toric-code-on-quantum-proccessor}.
However, the grid architecture of superconducting platforms limits
simultaneously applicable operations to one- and two-qubit unitary
operations (gates) on non-overlapping pairs of adjacent qubits. This
limitation gives rise to the challenge of approximating nonlocal unitary
evolution of the simulated system with a sequence of gates, possibly
with the use of ancilla qubits. Crucially, conventional imperfections
of physical hardware, such as relaxation, decoherence, accumulation
of gate errors, and leakage to non-computational subspace, strongly
penalize approximations that require too many gates or exploit too
many ancilla qubits~\citep{neill-2021_fermionic-ring-simulation,miInformationScramblingQuantum2021,arute-2019_quantum-supremacy-paper,satzinger-2021_realizing-toric-code-on-quantum-proccessor}.

This restriction is particularly acute for simulation of tunneling.
Tunneling is exponentially slow in the potential barrier strength,
thus requiring a long simulation to observe a tunneling event or,
equivalently, an exponentially precise eigenvalue estimation (for
signal processing-based algorithms~\citep{lowOptimalHamiltonianSimulation2017a,lowHamiltonianSimulationQubitization2019}).
In addition, tunneling is inherently a coherent nonlocal process,
which is in conflict with the local-only gate capabilities of the
superconducting platforms. At the same time, tunneling plays an important
role both as a fundamental part of quantum dynamics and phase transitions~\citep{kaganQuantumTunnellingCondensed1992,colemanFateFalseVacuum1977},
and as the main practical tool behind the advantage of quantum algorithms
for combinatorial optimization and other computationally intensive
tasks~\citep{arute-2019_quantum-supremacy-paper,smelyanskiy_2020_nonergodic,AQCReview,NISQReview}.
It is therefore of practical importance to characterize the approximation
algorithm specifically in application to tunneling.

In this work, we conduct a detailed study of the second-order Suzuki-Trotter
algorithm (STA)~\citep{suzuki-1991}, also known as the product formula~\citep{childs_quantum-simulations-with-speedup}
or Trotterization. STA splits the full evolution into $M=t/\delta t$
steps of length $\delta t$, and for each step, approximates the simulated
evolution operator $U(\delta t)$ as a product of exponentials implementable
via local gates. For example, for $U=\exp\left\{ -i\delta t(A_{1}+A_{2}...)\right\} $
containing two components $A_{1,2}$ that can be exponentiated exactly
but only one at a time, the second-order approximation reads $U_{\text{appr}}(\delta t)=e^{-i\delta tA_{1}/2}e^{-i\delta tA_{2}}e^{-i\delta tA_{1}/2}$;
details follow in \secref{Solvable-model}. In contrast to other methods~\citep{lowOptimalHamiltonianSimulation2017a,lowHamiltonianSimulationQubitization2019,childs_quantum-simulations-with-speedup},
STA has several practical advantages that make it the only feasible
algorithm for existing superconducting platforms: \emph{i)}~no qubits
are required in excess of those directly storing the configuration
of the target spin system, and \emph{ii)}~STA can be viewed as a
distorted evolution of the simulated system's degrees of freedom,
implying similar locality and physical interpretability~\citep{neill-2021_fermionic-ring-simulation}
and, consequently, more direct understanding of the influence of device
imperfections~\citep{miInformationScramblingQuantum2021}.

The performance of STA is conventionally described by its fidelity
error $\epsilon=1-\left|\left\langle U_{\text{appr}}(t)\psi|U(t)\psi\right\rangle \right|^{2}$,
which is bounded from above for sufficiently large $M$ as $\epsilon\sim\Lambda t\,\left(\Lambda t/M\right)^{1/2}$~\citep{childs_quantum-simulations-with-speedup,poulin-2014_Trotter-step_for-quantum-chemistry,childs-2019_Nearly-optimal-simulation-by-product-formulas,kivlichan-2020_Trotterizaiton-Fermionic-Hamiltonians},
where $\Lambda$ is the typical local energy scale of the simulated
system. Importantly, using this bound to estimate $M$ required to
simulate tunneling would render a prohibitively large number due to
the exponentially long simulation time~$t$. However, upper bounds
are often not saturated by the actual evolution~\citep{childs_quantum-simulations-with-speedup,childs-2019_Nearly-optimal-simulation-by-product-formulas,childs-2019_faster-simulations-by-randomization},
leaving one wondering about the true value of~$\epsilon$. Moreover,
even large~$\epsilon$ does not necessarily translate to a comparable
error in physical observables, as those might be insensitive to the
sectors of the Hilbert space that host the largest contribution to~$\epsilon$.

In this work, we fully characterize the STA for a simple model of
semiclassical tunneling, readily implementable in existing hardware.
By adapting the Wentzel-Kramers-Brillouin approximation (WKBA) to
the product formula structure of STA, we show that, for a general
tunneling process, the error bound correctly conveys the strongly
detrimental effect of STA. However, our analysis uncovers the regime
in which STA exponentially enhances tunneling rates through non-perturbative
renormalization of the tunneling dynamics. We thus clearly demonstrate
that controlling the error only in the physically relevant part of
the Hilbert space dramatically reduces the required gate count and
gives access to larger system sizes on existing hardware. These results
further prompt the study of the non-perturbative regime of large Trotter
steps $\delta t\,\Lambda\sim1$, to which end we construct a non-perturbative
extension of WKBA to respect the Floquet theorem. We then analyze
nontrivial STA dynamics in the regime $\delta t\,\Lambda\sim1$, demonstrating
a drastic deformation of tunneling as a crucial manifestation of finite
quasi-energy period~$\Omega=2\pi/\delta t$. These findings are particularly
relevant because the regime $\delta t\,\Lambda\sim1$ is equally natural
for existing superconducting platforms~\citep{neill-2021_fermionic-ring-simulation,satzinger-2021_realizing-toric-code-on-quantum-proccessor,miInformationScramblingQuantum2021}.

The paper is organized as follows: \secref{Solvable-model} introduces
the model Hamiltonian, the associated STA, and reviews the semiclassical
tunneling dynamics. \secref{Semiclassical-Approach-to-Trotterization}
describes the WKBA and its extension to treat non-perturbative in~$\delta t$
effects. \secref{Perturbative-regime} presents the error analysis
in the regime of small Trotter step, $\Lambda\delta t\ll1$, highlighting
the dominant effects of resonance detuning due to local energy shifts
and tunneling rate enhancement. \secref{Large-Trotter-steps} then
analyzes the nontrivial tunneling dynamics under STA with large Trotter
step, including non-perturbative spectrum reconstruction via Floquet
mixing. Finally, \secref{Discussion} summarizes the key qualitative
findings, explains the relative merits and limitations of rigorous
error bounds, describes the experimental realization, and outlines
future research directions. The paper is supplemented by Appendices
containing technical details and experiment design.

\section{Solvable Model of Trotterization\label{sec:Solvable-model}}

\subsection{Hamiltonian\label{subsec:model-Hamiltonian}}

We consider the Hamiltonian of a chain of $L$ spins $1/2$ with $XY$
nearest-neighbor interaction:
\begin{equation}
H=\sum^{L}_{i=1}h_{i}S^{z}_{i}-\sum^{L-1}_{i=1}J\left(S^{x}_{i}S^{x}_{i+1}+S^{y}_{i}S^{y}_{i+1}\right),
\label{eq:Hamiltonian}
\end{equation}
where $h_{i}$ is position-dependent and $J>0$ is a constant. The
spatial profile of $h_{i}$ is assumed to be smooth, with precise
criteria provided shortly. %
The Hamiltonian~(\ref{eq:Hamiltonian}) conserves the $z$-projection
of the total spin $S^{z}_{\text{total}}=\sum_{i}S^{z}_{i}$. 

Crucially, Hamiltonian~(\ref{eq:Hamiltonian}) can be mapped to a
system of non-interacting fermions via the Jordan-Wigner transformation~\citep{lieb-shultz-mattis_XY-model}:
\begin{equation}
S^{z}_{i}=a^{\dagger}_{i}a_{i}-\frac{1}{2},\,\,\,S^{x}_{i}S^{x}_{i+1}+S^{y}_{i}S^{y}_{i+1}=\frac{a^{\dagger}_{i}a_{i+1}+a^{\dagger}_{i+1}a_{i}}{2},
\label{eq:Jordan-Wigner_transform}
\end{equation}
where $a^{\dagger}_{i}$ and $a_{i}$ are fermionic creation and annihilation
operators for site $i$. The vacuum state with no fermions corresponds
to the state with all spins down $\left|\Omega\right\rangle =\otimes_{i}\left|\downarrow_{i}\right\rangle $,
and, more generally, the subspace with $N$ fermions is identified
with the subspace of~$S^{z}_{\text{total}}=N-L/2$.%

\subsection{Approximate Evolution by the Suzuki-Trotter Algorithm\label{subsec:Model_Suzuki-Trotter-alg}}

The problem of simulating the unitary evolution under Hamiltonian~(\ref{eq:Hamiltonian})
for time $t$ amounts to approximating the unitary $U=\exp\left\{ -itH\right\} $
by a product of local one- and two-qubit gates available on the quantum
hardware. However, the superconducting quantum hardware only supports
local one- and two-qubit operations~\citep{neill-2021_fermionic-ring-simulation}
corresponding to the evolution under the following local Hamiltonians:
\begin{equation}
H_{k,k+1}=-J\left(S^{x}_{k}S^{x}_{k+1}+S^{y}_{k}S^{y}_{k+1}\right),\,\,\,H_{k}=h_{k}S^{z}_{k}.
\label{eq:local-hamiltonians_def}
\end{equation}
The target Hamiltonian, Eq.~(\ref{eq:Hamiltonian}), is then decomposed
as
\begin{equation}
H=K_{\text{even}}+K_{\text{odd}}+P,
\label{eq:Ham-decomposition}
\end{equation}
\begin{equation}
K_{\text{even}(\text{odd})}=\sum_{k:\text{even}(\text{odd})}H_{k,k+1},\,\,\,P=\sum^{L}_{k=1}H_{k}.
\label{eq:Ham-non-commuting terms}
\end{equation}
Each term in Eq.~(\ref{eq:Ham-decomposition}) consists of mutually
commuting operations and can thus be implemented exactly. However,
because these three terms do not commute with each other, $U=\exp\left\{ -itH\right\} $
cannot be reproduced exactly in hardware.

The second-order Suzuki-Trotter approximation (STA)~\citep{suzuki-1991,berry-2007_trotterization-and-no-fast-forward}
addresses this problem by providing an approximation of $U=e^{-itH}$
with Hamiltonian $H=A_{1}+..+A_{n}$, where each term $A_{k},\,k=1,...,n$
can be exponentiated individually, but only one at a time. The whole
evolution is partitioned into identical steps of length $\delta t$
(assuming that $t/\delta t$ is an integer): 
\begin{equation}
U_{\text{appr}}(t)=\left\{ U_{\text{appr}}(\delta t)\right\} ^{t/\delta t},
\label{eq:total-evolution_operators}
\end{equation}
with each step performed according to the approximate evolution operator
\begin{equation}
U_{\text{appr}}(\delta t)=e^{-iA_{n}\delta t/2}...e^{-iA_{1}\delta t/2}\times e^{-iA_{1}\delta t/2}...e^{-iA_{n}\delta t/2},
\label{eq:Suzuki-Trotter_2nd-order}
\end{equation}
where the second half of the product applies the operators in reverse
order compared to the first half. For a given decomposition of $H$
into $n$ admissible terms $A_{k}$, there are $n!$ possible orderings
of $A_{k}$ for Eq.~(\ref{eq:Suzuki-Trotter_2nd-order}), although
some $U_{\text{appr}}(\delta t)$ differ only by a gauge transform.
The physically distinct orderings for the case of Eq.~(\ref{eq:Ham-decomposition})
are discussed in \appref{Gate-ordering}.

In the limit $\delta t\rightarrow0$, $U_{\text{appr}}(\delta t)$
approaches $e^{-i\delta tH}$, while finite~$\delta t$ causes a
distortion that can be characterized by the effective Hamiltonian:
\begin{equation}
H_{\text{eff}}:=-\frac{1}{i\delta t}\ln\,U_{\text{appr}}(\delta t),
\label{eq:effective-Hamiltonian-def}
\end{equation}
where $U_{\text{appr}}(t)=\exp\left\{ -iH_{\text{eff}}t\right\} $
describes the actual evolution after $t/\delta t\in\mathbb{N}$ steps,
and $H_{\text{eff}}$ is close to $H$~\citep{suzuki-1991}:
\begin{equation}
H_{\text{eff}}-H=\delta t^{2}D+O\left(\delta t^{4}\right).
\label{eq:Suzuki-Trotter_2nd-order_error-term}
\end{equation}
The error operator $D$, for the particular case of Eqs.~(\ref{eq:Ham-decomposition})--(\ref{eq:Ham-non-commuting terms}),
can be estimated as $D\sim\max\left\{ J,P\right\} $, where $J$ and
$P=\max_{i}h_{i}-\min_{i}h_{i}$ are the characteristic scales of
the kinetic and potential terms, respectively (see \appref{trotter_perturb-corrections}
for the exact expression for~$D$). According to Eq.~(\ref{eq:Suzuki-Trotter_2nd-order_error-term}),
admissible values of $\delta t$ are then determined from the naive
comparison of characteristic energy scales: 
\begin{equation}
J\delta t,\,P\delta t\ll1.
\label{eq:Suzuki-Trotter_naive-applicability-criteria}
\end{equation}
However, if the problem contains other relevant energy scales, one
should also require these scales to be small compared to $\delta t^{-1}$,
which might lead to a much stronger requirement on the step size $\delta t$,
as shown below.

\subsection{Reduction to a Single-particle Problem\label{subsec:Reduction-to-a-single-particle-problem}}

For non-interacting fermions, the calculation of $H_{\text{eff}}$
reduces to a single-fermion problem. Indeed, each term in the product
formula, Eq.~(\ref{eq:Suzuki-Trotter_2nd-order}), corresponds to
a Hamiltonian that is quadratic in fermionic operators:
\begin{equation}
A_{k}=\sum_{ij}a^{\dagger}_{i}A^{ij}_{k}a_{j},
\end{equation}
where the sum goes over all sites $i,j$, and the matrix elements
$A^{ij}_{k}$ are obtained from Eqs.~(\ref{eq:local-hamiltonians_def})--(\ref{eq:Ham-non-commuting terms})
by substitution~(\ref{eq:Jordan-Wigner_transform}). By Wick's theorem,
the full product in Eq.~(\ref{eq:Suzuki-Trotter_2nd-order}) is also
generated by a quadratic Hamiltonian:
\begin{equation}
U_{\text{appr}}(\delta t)=\exp\left\{ -i\delta t\sum_{ij}a^{\dagger}_{i}\mathcal{H}^{ij}_{\text{eff}}a_{j}\right\} .
\label{eq:fermionic-evolution-operator}
\end{equation}
The effective single-particle Hamiltonian $\mathcal{H}_{\text{eff}}$
is defined via its evolution operator $\mathcal{U}=e^{-i\delta t\mathcal{H}_{\text{eff}}}$
by the same STA~(\ref{eq:Suzuki-Trotter_2nd-order}), but with each
$A_{k}$ replaced by a single-particle matrix:
\begin{equation}
H_{k,k+1}\mapsto\mathcal{H}^{ij}_{k,k+1}=-\frac{J}{2}\left(\delta_{ik}\delta_{j,k+1}+\delta_{i+1,k}\delta_{jk}\right),
\end{equation}
\begin{equation}
H_{k}\mapsto\mathcal{H}^{ij}_{k}=h_{k}\delta_{ik}\delta_{jk},
\end{equation}
where we have dropped the additional $1/2$ in the expression for
$S^{z}_{i}$ in Eq.~(\ref{eq:Jordan-Wigner_transform}) as it produces
a constant energy shift. The behavior of a given spin state under
$U_{\text{appr}}$ is thus entirely described by the single-particle
effects and the Fermi statistics of the Jordan-Wigner fermions.

\subsection{Single-Particle Tunneling Problem\label{subsec:Single-particle-tunneling-problem}}

In this work, we are interested in the single spin flip subspace corresponding
to just a single Jordan-Wigner fermion described by the following
single-particle Hamiltonian:
\begin{equation}
\mathcal{H}=\sum^{L}_{k=1}h_{k}\,\left|k\left\rangle \right\langle k\right|-\sum^{L-1}_{k=1}\frac{J}{2}\left(\left|k\left\rangle \right\langle k+1\right|+\left|k+1\left\rangle \right\langle k\right|\right),
\label{eq:single-particle_Hamiltonian}
\end{equation}
Unless stated otherwise, we consider smooth potential profiles of
the form $h_{n}=h(an)$, where $h(x)$ is a smooth function, and $a\ll1$
plays the role of a small lattice constant. We are then interested
in describing the problem of tunneling from a local minimum of the
potential. In terms of spins, this corresponds to creating a single-spin
excitation with spin density localized in a local minimum of $h(x)$
and observing the spin density tunneling through the potential barrier
due to coherent virtual multi-spin processes.

\begin{figure}[t]
\begin{centering}
\includegraphics[width=0.95\columnwidth]{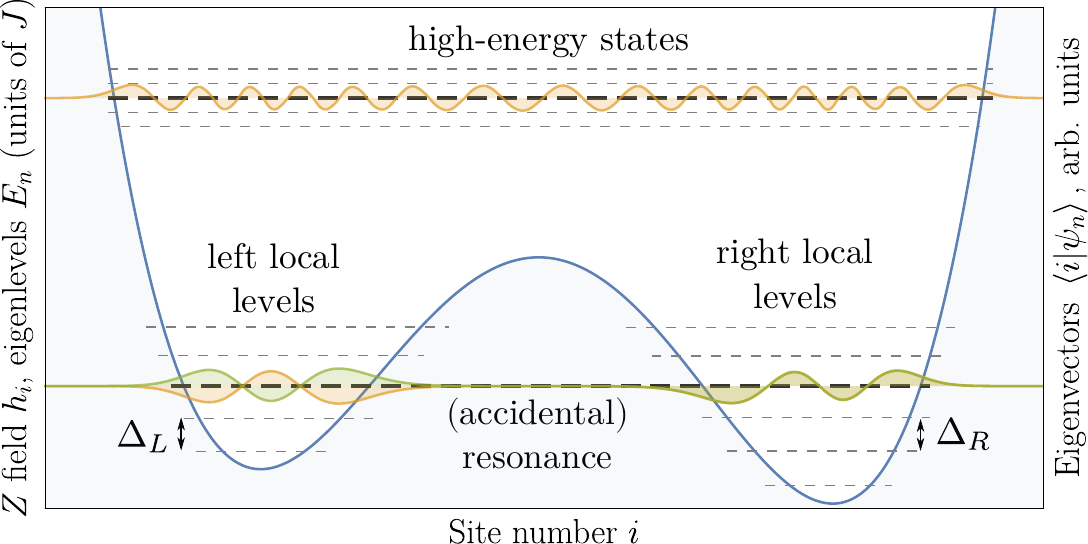}
\par\end{centering}
\caption{Sketch of the double-well potential profile (filled blue curve), the
resulting eigenlevels (horizontal dashed lines) and wave functions
(filled curves) for the tunneling problem. For a given energy, the
potential defines a set of classically allowed and forbidden regions.
If the allowed region is bounded, it hosts a series of discrete levels
with energy-dependent level spacing $\Delta$ (determined by the particular
geometry of the allowed region). Due to the finite height of the potential
barrier, the resonant states in the two wells are hybridized, with
a small energy splitting $\eta$. The resonance between left and right
metastable states can be either accidental (and thus characterized
by finite detuning $\varepsilon\sim\eta$) or protected by reflection
symmetry about the center. \label{fig:tunneling-problem_potential-sketch}}
\end{figure}

As a concrete example, consider the double-well potential in \figref{tunneling-problem_potential-sketch}.
If the barrier is infinite, each well hosts a series of discrete energy
levels characterized by level spacing~$\Delta$ (energy-dependent
for an anharmonic well). Suppose that a pair of those levels from
different wells has a small energy difference $\varepsilon=E_{\text{right}}-E_{\text{left}}\ll\Delta$,
either due to reflection symmetry of the potential (in which case
$\varepsilon=0$) or due to an accidental degeneracy. Finite transparency
of the potential barrier then leads to hybridization of those levels
according to the following reduced Hamiltonian of the resonant subspace:
\begin{equation}
\mathcal{H}_{\text{res.}}=\begin{pmatrix}-\varepsilon/2 & \eta\\
\eta & \varepsilon/2
\end{pmatrix},
\label{eq:resonant-subspace_Hamiltonian}
\end{equation}
where $\eta\ll\Delta$ is the tunneling amplitude between the two
resonant states, exponentially small for a large barrier. If the initial
state is close to a localized metastable state in one of the wells,
the dynamics of Hamiltonian~(\ref{eq:single-particle_Hamiltonian})
leads to tunneling to the other well, with physical observables exhibiting
Rabi oscillations.

This problem is readily implementable on existing quantum devices~\citep{neill-2021_fermionic-ring-simulation}.
However, the approximate evolution operator~(\ref{eq:Suzuki-Trotter_2nd-order})
distorts the evolution, and the aim of this work is to quantify this
distortion analytically. To verify our calculations, we use numerical
diagonalization of the single-particle Hamiltonian~(\ref{eq:single-particle_Hamiltonian})
and the STA evolution operator~(\ref{eq:Suzuki-Trotter_2nd-order}).
Importantly, this requires high-precision arithmetic (60 decimal digits
for this work) to accurately capture exponentially small quantities,
such as~$\eta$. Unless stated otherwise, numerical simulations use
the ordering $A_{1}=K_{\text{even}}$, $A_{2}=P$, $A_{3}=K_{\text{odd}}$
in Eq.~(\ref{eq:Suzuki-Trotter_2nd-order}). This reduces the two-qubit
gate count to that of the Trotter steps by merging of $K_{\text{even}}$
and $K_{\text{odd}}$ gates from adjacent Trotter steps---beneficial
for existing hardware (see \appref{Gate-ordering} for details).

\section{Semiclassical Approach to Trotterization\label{sec:Semiclassical-Approach-to-Trotterization}}

\subsection{Semiclassical Approximation for Single-Particle Tunneling\label{subsec:Semiclassical-approximation}}

The Wentzel-Kramers-Brillouin approximation (WKBA), also known as
the semiclassical approach, is a conventional tool for describing
tunneling through a large barrier. It consists of two parts~\citep{berry-1972_semiclassical-approximation}:
\emph{i)}~constructing semiclassical wave functions in the region
where the semiclassical approximation is applicable, and \emph{ii)}~asymptotic
matching of two semiclassical regions with a transfer matrix obtained
from an exact solution of the region where the semiclassical approach
is inapplicable. In our problem, there is a complication in the form
of discretized spatial dimension, but this does not reduce the power
of WKBA~\citep{braun-1993_discrete-WKB}. A detailed overview of
the procedure is given in \appref{Discrete-Semiclassical-Approximation},
while here we briefly summarize the technique.

Within WKBA, an eigenfunction of the Hamiltonian is approximated by
the following asymptotic series in powers of the small lattice constant
$a\ll1$ (being a close analog of Planck's constant):
\begin{equation}
\psi_{n}=\frac{e^{iS(an)/a}}{\sqrt{v(an)}}\left[1+O(a)\right],
\label{eq:semiclassical-wave-function}
\end{equation}
where $v,\,S$ are assumed to be smooth functions of $x$. The classical
action $S(x)$ obeys the Hamilton-Jacobi equation of motion:
\begin{equation}
\mathcal{H}(x,p)=E,\,\,\,p=S'(x),
\label{eq:classical-eq-of-motion}
\end{equation}
which also defines the classical momentum $p(x)$. The classical counterpart
of the Hamiltonian, $\mathcal{H}(x,p)$, is defined through the Maslov
symbol:
\begin{equation}
\mathcal{H}(x=an,p)=\sum_{m}\mathcal{H}^{nm}e^{ip(m-n)},
\label{eq:Maslov-symbol}
\end{equation}
where the sum in the right-hand side runs over all sites. The particular
case of Hamiltonian~(\ref{eq:single-particle_Hamiltonian}) evaluates
to:
\begin{equation}
\mathcal{H}(x,p)=-J\cos p+h(x).
\label{eq:classical-hamiltonian}
\end{equation}
 Finally, $v(x)$ in Eq.~(\ref{eq:semiclassical-wave-function})
is the classical group velocity:
\begin{equation}
v(x)=\left.\frac{\partial\mathcal{H}}{\partial p}\right|_{p=p(x)}=J\sin p(x).
\label{eq:group-velocity}
\end{equation}

Approximation~(\ref{eq:semiclassical-wave-function}) is applicable
in the region where de~Broglie wavelength $\lambda$ is locally well
defined, i.e., its relative variation at spatial scale $\lambda$
is small:
\begin{equation}
\lambda'(x)=\frac{a\,h'(x)}{p^{2}(x)\,v(x)}\ll1.
\label{eq:semiclassics_applicability-criteria}
\end{equation}
This condition is violated if the potential is not smooth (i.e., $h'(x)\sim1/a\gg1$),
or when either $p$ or~$v$ vanishes, which defines the classical
turning points. In addition to the standard turning points ($v=0,\,p=0$),
there are also anomalous turning points at $v=0,\,p=\pm\pi$---a
consequence of the problem discreteness~\citep{braun-1993_discrete-WKB}.
In particular, the existence of anomalous turning points explains
the Bloch oscillations~\citep{bloch-1929_Bloch-oscillations}.

In the regions where the semiclassical approximation breaks down,
the corresponding stationary Schrödinger equation has to be solved
exactly. The resulting asymptotic behavior is then matched to Eq.~(\ref{eq:semiclassical-wave-function}),
and the latter is required to be applicable in a sufficiently large
region, $\Delta x\gg\lambda(x)$, implying $S/a\gg1$. This procedure,
detailed in \appref{Discrete-Semiclassical-Approximation}, yields
a connection between the coefficients of the two linearly independent
semiclassical solutions on either side of the nonclassical region.
Requiring the wave function to decay into the forbidden region results
in the celebrated Bohr quantization rule~\citep{braun-1993_discrete-WKB,berry-1972_semiclassical-approximation,landau-lifshitz_quantum-mechanics-2013}
for the discrete energy levels in a potential well:
\begin{equation}
\frac{2S_{12}(E)}{a}=\frac{2}{a}\intop^{x_{2}(E)}_{x_{1}(E)}p(E;x)\,dx=2\pi N-\theta_{1}-\theta_{2},
\label{eq:Bohr-quantization-condition}
\end{equation}
where $N\in\mathbb{Z}$ is the level quantum number, $p(E,x)$ is
the classical momentum found as a positive solution of Eq.~(\ref{eq:classical-eq-of-motion}),
$x_{1,2}$ are the positions of the classical turning points defined
by vanishing group velocity $v\left(x_{i}\right)=0$ or by system
boundaries, and $\theta_{i}$ are the additional phases accumulated
at each turning point according to its type~\citep{braun-1993_discrete-WKB}:
\begin{equation}
\theta_{i}=\begin{cases}
-\pi/2, & p\left(x_{i}\right)=0,\\
\pi, & x_{i}=0\,\text{or}\,a(L+1),\\
\pi/2\pm2\pi x_{i}/a, & \left|p\left(x_{i}\right)\right|=\pi.
\end{cases}
\label{eq:anomalous-phases-def}
\end{equation}
In the last line, the sign is negative if the allowed region is to
the left of $x_{i}$, and positive otherwise. Notably, $\theta_{i}$
for $p\left(x_{i}\right)=\pm\pi$ depends on the exact position $n_{i}=x_{i}/a\in\mathbb{R}$
of the anomalous turning point \emph{between} the two sites, and thus
on energy. This is essential for the quantization conditions, as pointed
out in Ref~\citep{braun-1993_discrete-WKB}. In addition, the middle
line of Eq.~\eqref{anomalous-phases-def} places the turning points
due to system edges one site after the terminating physical site.
This placement can be verified in the exactly solvable case of a homogeneous
system (\emph{cf.} open boundary conditions, $\psi_{0,L+1}=0$).
Eq.~(\ref{eq:Bohr-quantization-condition}) produces the well-known
expression for the single-particle level spacing $\Delta=E_{N+1}-E_{N}$:
\begin{equation}
T_{12}(E)\,\Delta(E)=2\pi a,
\label{eq:single-particle_level-spacing}
\end{equation}
where $T_{12}=\left|2\,S_{12}'(E)+a\theta_{1}'(E)+a\theta_{2}'(E)\right|$
is the period of classical motion at energy~$E$ that includes the
scattering delay $a\theta_{i}'(E)$. 

Tunneling corresponds to motion with imaginary momenta in the classical
forbidden region. The tunnel splitting of two semiclassical levels
with close energies is described by~\citep[§50]{landau-lifshitz_quantum-mechanics-2013}
\begin{equation}
\eta=\frac{\sqrt{\Delta_{L}\Delta_{R}}}{2\pi}\exp\left\{ -\frac{1}{a}S_{B}\left(\frac{E_{L}+E_{R}}{2}\right)\right\} ,
\label{eq:semiclassical_tunneling-rate}
\end{equation}
where $\Delta_{L,R}$ are the local level spacings at energies $E_{L,R}$
in the corresponding wells, and 
\begin{equation}
S_{B}(E)=\intop^{x_{R}(E)}_{x_{L}(E)}\left|\text{Im}p(E;x)\right|\,dx,
\label{eq:underbarrier-action}
\end{equation}
is the under-barrier action integrating the complex momentum in the
forbidden region between the classical turning points~$x_{L},\,x_{R}$.

Eqs.~(\ref{eq:Bohr-quantization-condition})--(\ref{eq:underbarrier-action})
are applicable if the region of applicability of the semiclassical
approximation is sufficiently large, viz., $S_{12}/a,\,S_{B}/a\gg1$.
This condition is typically violated for the lowest energy levels
of the well or weak barriers, requiring more advanced methods. However,
even for such cases, the parametric dependence given by Eq.~(\ref{eq:semiclassical_tunneling-rate})
is correct up to a numeric coefficient of the order of unity~\citep{landau-lifshitz_quantum-mechanics-2013}
that can further be evaluated exactly by a more accurate matching
procedure~\citep{gargTunnelSplittingsOne2000}. Moreover, the quantization
condition~(\ref{eq:Bohr-quantization-condition}) correctly describes
the leading parametric dependence even for the ground state of a near-quadratic
potential minimum~\citep{gargTunnelSplittingsOne2000}\citep[§48]{landau-lifshitz_quantum-mechanics-2013}.

\subsection{Semiclassical Description for the Effective Hamiltonian\label{subsec:Matrix-WKBA}}

WKBA cannot be directly applied to STA, as the alternate evolution
according to even and odd parts of the kinetic energy, Eq.~(\ref{eq:Suzuki-Trotter_2nd-order}),
doubles the translation period of the system, introducing staggering
to matrix elements (see also \appref{trotter_perturb-corrections}).
However, the fundamental idea behind WKBA is that it is exact in a
translationally invariant system: the WKBA wave function, Eq.~(\ref{eq:semiclassical-wave-function}),
is just an eigenfunction of the evolution operator corresponding to
momentum $p$, the quantum number of the translationally invariant
eigenbasis. The effect of smooth spatial variations is to induce smooth
changes of~$p$, while the instantaneous eigenvalue $\Lambda(x,p)$
of the evolution operator produces the classical Hamiltonian as $H(x,p)=\frac{i}{\delta t}\ln\Lambda(x,p)$.
In this way, WKBA is naturally extended to describe STA. A rigorous
implementation of this logic is presented in \appref{Matrix-WKB},
whereas here we guide the reader through the results.

The first step is to consider the translationally invariant case,
$h_{i}=\text{const}$. The eigenfunction of both the evolution operator
$\mathcal{U}_{\text{appr}}$ and distance-two translation operator
$\mathcal{T}^{2}$ ($\left\langle n\left|\mathcal{T}^{2}\right|m\right\rangle =\delta_{n,m+2}$)
is described by two components, $\psi_{n}=\varphi_{+}\,e^{ipn}+\varphi_{-}\,(-1)^{n}e^{ipn}$
with constant $\varphi_{\pm}$. In these terms, the eigenproblem $\sum_{m}\mathcal{U}^{nm}_{\text{appr}}\psi_{m}(p)=\Lambda(p)\,\psi_{n}$
is solved by
\begin{equation}
\Lambda_{\pm}(p)=\exp\left[-i\delta th\pm2i\arcsin\left\{ \sin\frac{J\delta t}{2}\,\cos p\right\} \right],
\label{eq:trotterized-evolution_eigenvalue}
\end{equation}
\begin{equation}
\begin{pmatrix}\varphi_{+}\\
\varphi_{-}
\end{pmatrix}\propto\begin{pmatrix}a\mp\sqrt{a^{2}+b^{2}}\\
-ib
\end{pmatrix},
\label{eq:trotterized-evolution_eigenfunction}
\end{equation}
where
\begin{equation}
a=1-2\sin^{2}\frac{J\delta t}{4}\,\cos^{2}p,\,\,\,b=\sin^{2}\frac{J\delta t}{4}\,\sin2p.
\label{eq:trotterized-evolution_oscillating-component}
\end{equation}
To preserve the total number of states in the system, doubling the
number of solutions due to the possible choice of sign~$\pm$ in
Eqs.~(\ref{eq:trotterized-evolution_eigenvalue})--(\ref{eq:trotterized-evolution_eigenfunction})
has to be compensated by restricting $p$ to the halved Brillouin
zone, $p\in[-\pi/2,\pi/2]$. However, the complete set of states is
also reproduced by choosing only the ``plus'' sign and allowing
$p$ to sweep the entire Brillouin zone,~$p\in[-\pi,\pi]$.

For smoothly varying potential $h_{i}$, WKBA then amounts to promoting
$\varphi_{+},\,\varphi_{-},\,p$ to smooth functions of $x=an$ (see
also Ref.~\citep[Ch. 11]{maslov-2001-semiclassical-approx}):
\begin{equation}
\psi_{n}=\left[\varphi_{+}(an)+(-1)^{n}\,\varphi_{-}(an)\right]\,\exp\left\{ \frac{i}{a}S(an)\right\} ,
\label{eq:trotterized-evolution_WKB-ansatz}
\end{equation}
where $S(x),\,p(x)$ are determined by Eq.~(\ref{eq:classical-eq-of-motion})
as before, whereas the corresponding semiclassical Hamiltonian is
given by $-\frac{1}{i\delta t}\ln\Lambda_{+}$ and reads:
\begin{equation}
\mathcal{H}_{\text{eff}}(x,p)=-\frac{2}{\delta t}\arcsin\left\{ \sin\frac{J\delta t}{2}\,\cos p\right\} +h(x).
\label{eq:trotterized-classical-hamiltonian}
\end{equation}
Eqs.~(\ref{eq:trotterized-evolution_WKB-ansatz})--(\ref{eq:trotterized-classical-hamiltonian})
describe both classically allowed and classically forbidden regions,
with the latter characterized by purely imaginary momenta, as before. 

Equations~(\ref{eq:Bohr-quantization-condition})--(\ref{eq:underbarrier-action})
describing the spectrum of the Hamiltonian remain applicable with
minor modifications. The scattering phases in Eq.~\eqref{anomalous-phases-def}
represent a topological property~\citep{maslov-2001-semiclassical-approx}
and thus remain intact (as we have verified perturbatively in \appref{trotter_perturb-corrections}).
In contrast, the edge scattering phase $\theta=\pi$ becomes a function
of $J\delta t$ and $p$ that is computed in \appref{Matrix-WKB}.
Notably, even for finite $\delta t$, the corresponding turning point
in Eq.~(\ref{eq:anomalous-phases-def}) still lies at one site past
the terminating physical one.

Remarkably, the WKBA description does not depend on the operator ordering
in Eq.~(\ref{eq:Suzuki-Trotter_2nd-order})---a consequence of a
property of the problem with homogeneous potential~\footnote{The operator $\exp\left\{ -i\delta th_{i}\right\} $ with $h_{i}=\text{const}$
represents a pure phase and commutes with both kinetic terms in Eq.~(\ref{eq:Ham-decomposition}),
so these phases from all Trotter steps can be merged together and
applied at any moment. The product formulae~(\ref{eq:Suzuki-Trotter_2nd-order})
for two possible orderings of the two remaining non-commuting operators
$K_{\text{even}},\,K_{\text{odd}}$ differ from each other by the
time translation gauge transform discussed in \appref{Gate-ordering}.}. However, Eqs.~(\ref{eq:trotterized-evolution_eigenvalue})--(\ref{eq:trotterized-evolution_WKB-ansatz})
do not fix the wave function normalization. It is instead determined
by the subleading terms in the formal expansion in powers of~$a$
in the form of the transport equation~\citep{maslov-2001-semiclassical-approx}
and depend non-trivially on the operator ordering (see \secappref{matrix_eq-of-motion_transport-eq}).
Nevertheless, the normalization is manifestly regular away from the
turning points and thus does not affect the validity of WKBA quantization
conditions.

Crucially, Eqs.~(\ref{eq:trotterized-evolution_eigenvalue})--(\ref{eq:trotterized-evolution_oscillating-component})
are non-perturbative in $\delta t$: as explained in more detail in
\secappref{Correction-to-semiclassical-Hamiltonian}, it corresponds
to collecting largest contributions in all orders of $\delta t$---a
clear advantage over analysis of finite-order terms in $\delta t$
power series~\citep{childs-2019_Nearly-optimal-simulation-by-product-formulas}.
At the same time, the approach of \subsecref{Semiclassical-approximation}
is promptly reproduced in the limit $J\delta t\to0$: $\varphi_{-}$
in Eq.~(\ref{eq:trotterized-evolution_eigenfunction}) vanishes,
reducing Eq.~(\ref{eq:trotterized-evolution_WKB-ansatz}) to its
scalar counterpart, Eq.~(\ref{eq:semiclassical-wave-function}),
and $\frac{i}{\delta t}\ln\Lambda_{+}$ approaches $\mathcal{H}(x,p)$
from Eq.~(\ref{eq:classical-hamiltonian}). We also note that the
kinetic term of Eq.~(\ref{eq:trotterized-classical-hamiltonian})
features in Refs.~\footnote{$\delta t\mathcal{H}$ from Eq.~(\ref{eq:trotterized-classical-hamiltonian})
with $J\delta t=\pi/2$ and $h=0$ reproduces $\omega_{k}$ from
Eq.~(1) of Ref~\citep{neill-2021_fermionic-ring-simulation} with
$\Phi=0$ and $k=2p$ ($k$ is the true quasi-momentum of the unit
cell). $\delta t\mathcal{H}/2$ from Eq.~(\ref{eq:trotterized-classical-hamiltonian})
with $h=J$ coincides with $\omega$ in Eq.~(4) of Ref.~\citep{vakulchykUniformAndersonLocalization2023}
with $\theta=J\delta t/2$. The system in both Refs. corresponds
to first-order STA which, for a translationally invariant system,
is equivalent to second-order STA with twice the Trotter step, up
to the time translation gauge transform discussed in \appref{Gate-ordering}.}.

\section{Resonance Detuning and Acceleration of Tunneling at $\Lambda\delta t\ll1$\label{sec:Perturbative-regime}}

\subsection{Leading Correction to WKBA\label{subsec:Correction-to-semiclassics}}

Expanding Eq.~(\ref{eq:trotterized-classical-hamiltonian}) in powers
of $\delta t$ yields the leading correction to the classical Hamiltonian:%
\begin{equation}
\delta\mathcal{H}_{\text{eff}}(x,p)=J\frac{(J\delta t)^{2}}{24}\cos p\,\sin^{2}p.
\label{eq:effective-Hamiltonian_classical-correction}
\end{equation}
Crucially, $\delta\mathcal{H}_{\text{eff}}$ vanishes at $p=0,\pm\pi$,
so the positions of all classical turning points according to Eq.~(\ref{eq:classical-eq-of-motion})
remain intact. Correction~(\ref{eq:effective-Hamiltonian_classical-correction})
induces a change of the action in both allowed and forbidden regions.
By expanding the perturbed equation of motion~(\ref{eq:classical-eq-of-motion})
to find the correction to the classical momentum $\delta p$, one
obtains the following correction to classical action between two classical
turning points $x_{1,2}$:
\begin{equation}
\delta S_{12}(E)=-\frac{(J\delta t)^{2}}{48}\intop^{x_{2}(E)}_{x_{1}(E)}dx\,\sin2p(E;x),
\label{eq:correction-to-action}
\end{equation}
where $p(E;x')$ is computed from the unperturbed equation of motion~(\ref{eq:classical-eq-of-motion}).

This correction causes a shift in the energy levels according to Eq.~(\ref{eq:Bohr-quantization-condition}),
with the result given by%
\begin{equation}
\delta E=-\frac{2\,\delta S_{12}(E)}{T_{12}(E)},
\label{eq:semiclassical-energy-shift}
\end{equation}
where $E$ is the unperturbed value, and $T_{12}(E)=2\,\partial S_{12}/\partial E$
is the classical period of motion in the allowed region. Remarkably,
the answer remains correct up to a numerical prefactor of order unity
even for states not described by WKBA~\citep[par. 48]{landau-lifshitz_quantum-mechanics-2013}.%
{} By estimating $\left|\sin2p\right|\le1$ in Eq.~(\ref{eq:correction-to-action}),
this energy shift can be estimated as
\begin{equation}
\left|\delta E\right|\apprle\frac{\left(J\delta t\right)^{2}}{48\pi}\,n_{\text{cl}}(E)\,\Delta(E),
\label{eq:energy-shift-estimation}
\end{equation}
where $\Delta(E)$ is the semiclassical level spacing, Eq.~(\ref{eq:single-particle_level-spacing}),
and $n_{\text{cl}}(E)=\left|x_{2}(E)-x_{1}(E)\right|/a$ is the total
number of sites in the classically allowed region.

\subsection{Rabi Oscillations Are Suppressed by Detuning in General Case\label{subsec:Rabi-oscillations_detuning}}

The key distortion of the Rabi oscillations comes from the shift of
the quasi-energy levels in each well. Within WKBA, the detuning parameter~$\varepsilon$
in the resonant subspace Hamiltonian~(\ref{eq:resonant-subspace_Hamiltonian})
is given by
\begin{equation}
\varepsilon_{\text{eff}}=\varepsilon+\frac{\delta S_{R}}{T_{R}}-\frac{\delta S_{L}}{T_{L}},
\label{eq:effective-level-detuning}
\end{equation}
where the index $L,R$ describes the characteristics of left and right
wells, respectively, according to Eq.(\ref{eq:semiclassical-energy-shift}).
In general, the two energy shifts are different, and the additional
detuning in $\varepsilon_{\text{eff}}$ can be estimated as $(J\delta t)^{2}n_{\text{cl}}\Delta$
according to Eq.~(\ref{eq:energy-shift-estimation}), with $n_{\text{cl}}=n_{\text{cl},L}+n_{\text{cl},R}$
being the total number of sites in the allowed region. To observe
tunneling, $\varepsilon_{\text{eff}}$ should be smaller than the
tunneling amplitude $\eta$, Eq.~(\ref{eq:resonant-subspace_Hamiltonian}),
rendering another applicability condition of the STA:
\begin{equation}
(J\delta t)^{2}\apprle\eta\,n_{\text{cl}}/\Delta
\label{eq:criteria_absence-of-detuning}
\end{equation}
Crucially, $\eta$ is exponentially small in barrier size, making
Eq.~(\ref{eq:criteria_absence-of-detuning}) much more restrictive
for $\delta t$ than the naive criteria~(\ref{eq:Suzuki-Trotter_naive-applicability-criteria}).
In particular, this allows one to neglect all other perturbative effects
as long as one is tasked with observing tunneling in the general case.

\subsection{Enhancement of Tunneling in Absence of Detuning\label{subsec:Rabi-oscillations_no-detuning}}

The situation changes drastically if the effects of detuning are irrelevant,
as is the case for two \emph{symmetric} wells in \figref{tunneling-problem_potential-sketch}.
If STA respects the potential symmetry, it causes symmetric shifts
in the wells, and all pairs of quasi-energy levels remain perfectly
hybridized ($\varepsilon\equiv0$). At the same time, the tunneling
amplitude~$\eta_{N}$ of each resonant subspace, Eq.~(\ref{eq:semiclassical_tunneling-rate}),
is distorted:
\begin{equation}
\frac{\eta_{N,\text{eff}}}{\eta_{N}}\approx\exp\left\{ -\frac{1}{a}\left(\delta S_{B}\left(E_{N}\right)+\frac{\partial S_{B}\left(E_{N}\right)}{\partial E}\delta E_{N}\right)\right\} .
\label{eq:effective-tunneling-amplitude}
\end{equation}
The first term in the exponent reflects the change in the under-barrier
action due to perturbation, Eqs.~(\ref{eq:effective-Hamiltonian_classical-correction})--(\ref{eq:correction-to-action}):
\begin{equation}
\delta S_{B}(E)=-\frac{(J\delta t)^{2}}{48}\intop^{x_{R}(E)}_{x_{L}(E)}dx\,\sinh2\left|p(E;x)\right|,
\label{eq:correction-to-tunneling-action}
\end{equation}
with purely imaginary momentum $p$ found from the equation of motion~(\ref{eq:classical-eq-of-motion})
in the forbidden region defined by the classical turning points $x_{L,R}(E)$.
The second term in the exponent of Eq.~(\ref{eq:effective-tunneling-amplitude})
stands for the change of the under-barrier action~(\ref{eq:underbarrier-action})
due to the induced shift in the resonance subspace energy:
\begin{equation}
\delta E_{N}=-\frac{\delta S_{L}}{T_{L}}-\frac{\delta S_{R}}{T_{R}},
\label{eq:energy-shift-of-resonance-subspace}
\end{equation}
where the notation of Eq.~(\ref{eq:effective-level-detuning}) is
used, and the result is similarly estimated by Eq.~(\ref{eq:energy-shift-estimation}).
In Eq.~(\ref{eq:effective-tunneling-amplitude}), the change of the
sub-exponential prefactor of Eq.~(\ref{eq:semiclassical_tunneling-rate})
was neglected in comparison to the change of the exponent due to the
smoothness of the potential, as explained in~\secappref{Change-in-classical-period}. 

Importantly, the corrected value $\eta_{N,\text{eff}}$ is larger
than $\eta_{N}$ for the majority of states: The first term in the
exponent of Eq.~(\ref{eq:effective-tunneling-amplitude}) is always
negative, as evident from Eq.~(\ref{eq:correction-to-tunneling-action}),
while the second term is negative for states with $\left|p\right|$
either always above or always below $\pi/2$, see Eqs.~(\ref{eq:semiclassical-energy-shift})
and~(\ref{eq:underbarrier-action}). Furthermore, if $\delta t$
is not limited by the detuning condition, Eq.~(\ref{eq:criteria_absence-of-detuning}),
enhancement of $\eta_{N,\text{eff}}/\eta_{N}$ by several orders of
magnitude becomes possible, as the effect of next-to-nearest hopping
accumulates along the barrier. Indeed, the first term in the exponent
of Eq.~(\ref{eq:effective-tunneling-amplitude}) is estimated as
\begin{equation}
\frac{\delta S_{B}}{a}\sim-(J\delta t)^{2}n_{\text{barr}}\times\begin{cases}
\sqrt{P/J}, & P/J\ll1,\\
\left(P/J\right)^{2}, & P/J\gg1,
\end{cases}
\label{eq:underbarrier-action-change_direct_rough-estimation}
\end{equation}
where $n_{\text{barr}}$ is the number of sites in the forbidden region.
While it is difficult to provide a similar estimate for the second
term in the exponent of Eq.~(\ref{eq:effective-tunneling-amplitude}),
in most cases, it further enhances tunneling. As a result, if $\left|\delta S_{B}\right|/a\apprge1$---possible
even if $J\delta t\ll1$ for sufficiently large barrier heights $P/J$
and lengths $n_{\text{barr}}$---the effective tunneling amplitude
$\eta_{N,\text{eff}}$ is exponentially enhanced.

In order for WKBA to be applicable, $\eta_{N,\text{eff}}$ still has
to be smaller than all other energy scales. In the spirit of Eq.~(\ref{eq:underbarrier-action-change_direct_rough-estimation}),
the unperturbed under-barrier action $S_{B}$ is estimated as:
\begin{equation}
\frac{S_{B}}{a}\sim n_{\text{barr}}\times\begin{cases}
\sqrt{P/J}, & P/J\ll1,\\
P/J, & P/J\gg1.
\end{cases}
\label{eq:tunneling-action_rough-estimation}
\end{equation}
Requiring $\eta_{N,\text{eff}}\ll J,P,\Delta,...$ while still assuming
that the barrier is semiclassical, i.e., $n_{\text{barr}}\gg1$ and
$S_{B}/a\gg1$, leads to the following condition on $\delta t$:
\begin{equation}
(J\delta t)^{2}<\max\left\{ 1,J/P\right\} ,
\label{eq:semiclassical-condition_underbarrier-motion}
\end{equation}
which is nearly equivalent to the naive criteria~(\ref{eq:Suzuki-Trotter_naive-applicability-criteria}).
STA is thus capable of strongly enhancing tunneling provided that
the detuning effects described in \subsecref{Rabi-oscillations_detuning}
are irrelevant.

\section{Non-perturbative Spectrum Reconstruction at $\Lambda\delta t\sim2\pi$\label{sec:Large-Trotter-steps}}

\begin{figure}
\begin{centering}
\includegraphics[width=0.99\columnwidth]{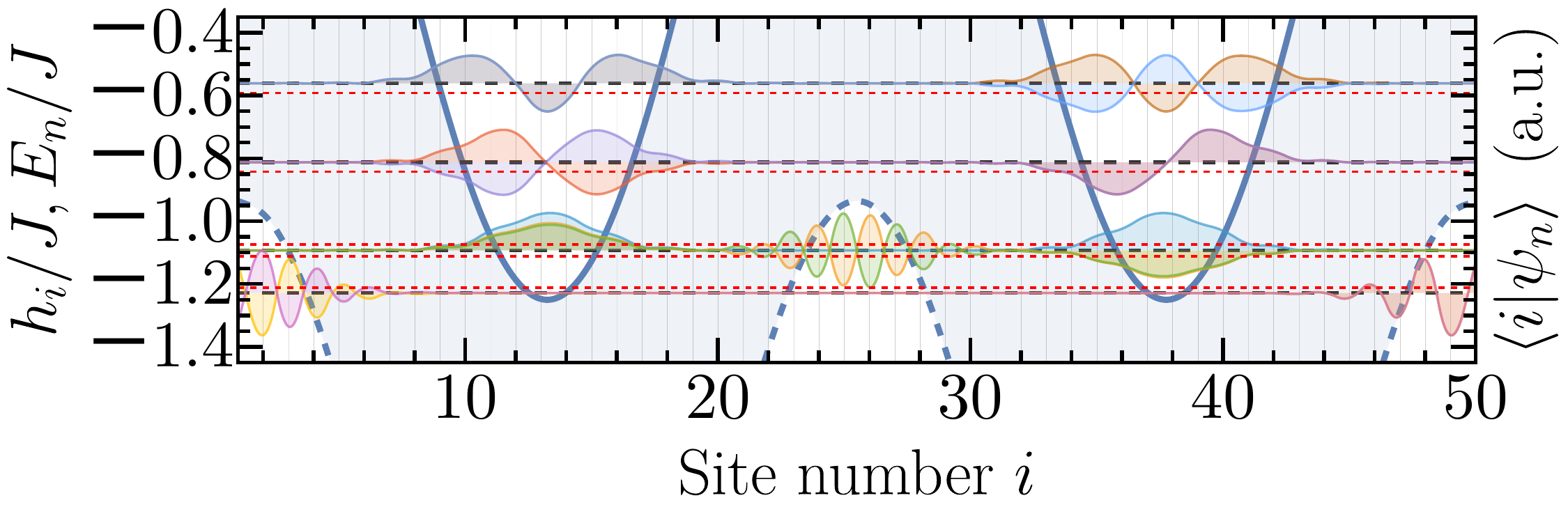}
\par\end{centering}
\caption{A portion of the energy levels (black dashed lines, shifted by $-J$
for clarity) and eigenvectors (color-filled curves) for STA for the
double-well experiment in \figref{tunneling-problem_potential-sketch}
with potential $h_{n}/J=5/4\,\cos4\pi\,(n-1)/(L-1)$ (thick blue line)
on $L=50$ sites with $J\delta t=1.5011125$. The dashed thick blue
line corresponds to the Floquet-translated band top $\left[h_{n}+2J-\Omega\right]/J$,
with the classically forbidden region shown in light blue. The red
dashed lines are the energy levels of the original Hamiltonian~(\ref{eq:single-particle_Hamiltonian}).
Red and black dashed lines are shifted by $-\Omega$ where necessary.
The tunnel splittings are too small to be visible. The central bound
state~$\left|B\right>$, with oscillating wave function $\psi_{n}\propto(-1)^{n}$,
is tuned into resonance with the lowest resonant subspace of the two
wells via a precise choice of $\delta t$. The parameters for model~(\ref{eq:effective-three-state-model})
are: $\eta/J=1.65\times10^{-4}$, $\varepsilon/J=-0.262\times10^{-4}$,
$\alpha\approx0.726$, with the eigenstates $\left|S'\right>$~(blue),
$\left|A'\right>$~(orange), and $\left|B'\right>$~(green). For
comparison, the tunnel splitting between $\left|A\right>$ and $\left|S\right>$
in the unperturbed problem is $\eta_{0}/J=1.97\times10^{-14}$. For
higher pairs of states, the splitting enhancement factors are $1.2\times10^{5}$
(purple/red) and $5.3\times10^{3}$ (light brown/light blue).\label{fig:low-energy-distortion_of-trotterized-evolution}}
\end{figure}

When the total spectral bandwidth $\Lambda\approx2J+P$ of the original
Hamiltonian approaches $\Omega=2\pi/\delta t$, processes involving
changes of energy by a multiple of $\Omega$ cease to be purely virtual,
invalidating the perturbative expansion in powers of $\delta t$.
According to the Floquet theorem, under periodic drive with period
$\delta t$, the energy spectrum is defined modulo $\Omega=2\pi/\delta t$,
which leads to mixing between high- and low-energy states of the original
Hamiltonian.

A nontrivial example of this reconstruction is shown in \figref{low-energy-distortion_of-trotterized-evolution}
(obtained numerically): By fine-tuning $\delta t$ around $\pi/J$,
a high-energy state $\left|B\right>$ with momentum close to $\pi$,
localized near the \emph{maximum} of the double-well potential, is
brought into resonance with the lowest resonant subspace. Due to the
matching conditions, $\left|B\right>$ hybridizes with the asymmetric
state $\left|A\right>$ of the resonant subspace. In contrast to Eq.~(\ref{eq:resonant-subspace_Hamiltonian}),
the resulting co-tunneling dynamics are now described by three eigenpairs
to account for the symmetric state $\left|S\right>$:
\begin{align}
E_{A} & =-\sqrt{\eta^{2}+(\varepsilon/2)^{2}}, &  & \left|A'\right>=\sqrt{1-\alpha^{2}}\left|A\right>-\alpha\left|B\right>,\nonumber \\
E_{S} & =-\varepsilon/2, &  & \left|S'\right>=\left|S\right>,
\label{eq:effective-three-state-model}\\
E_{B} & =\sqrt{\eta^{2}+(\varepsilon/2)^{2}}, &  & \left|B'\right>=\alpha\left|A\right>+\sqrt{1-\alpha^{2}}\left|B\right>,\nonumber 
\end{align}
where $E_{i}$ are the quasi-energies (defined up to $\Omega k,\,k\in\mathbb{Z}$),
$\eta$ is the tunneling amplitude between $B$ and either of the
wells, $\varepsilon$ is the detuning of $B$ from the resonant subspace,
and $0<\alpha=\eta/\sqrt{\eta^{2}+(\varepsilon/2)^{2}}<1$ characterizes
hybridization between $\left|A\right>$ and $\left|B\right>$. The
dynamics consist of three superimposed Rabi oscillations, demonstrating
a fully non-perturbative spectrum reconstruction. Importantly, the
tunnel splitting $\eta$ is many orders of magnitude larger than that
of the unperturbed problem due to the observed resonant co-tunneling
(see the caption of \figref{low-energy-distortion_of-trotterized-evolution}).

\begin{figure}
\begin{centering}
\includegraphics[width=0.99\columnwidth]{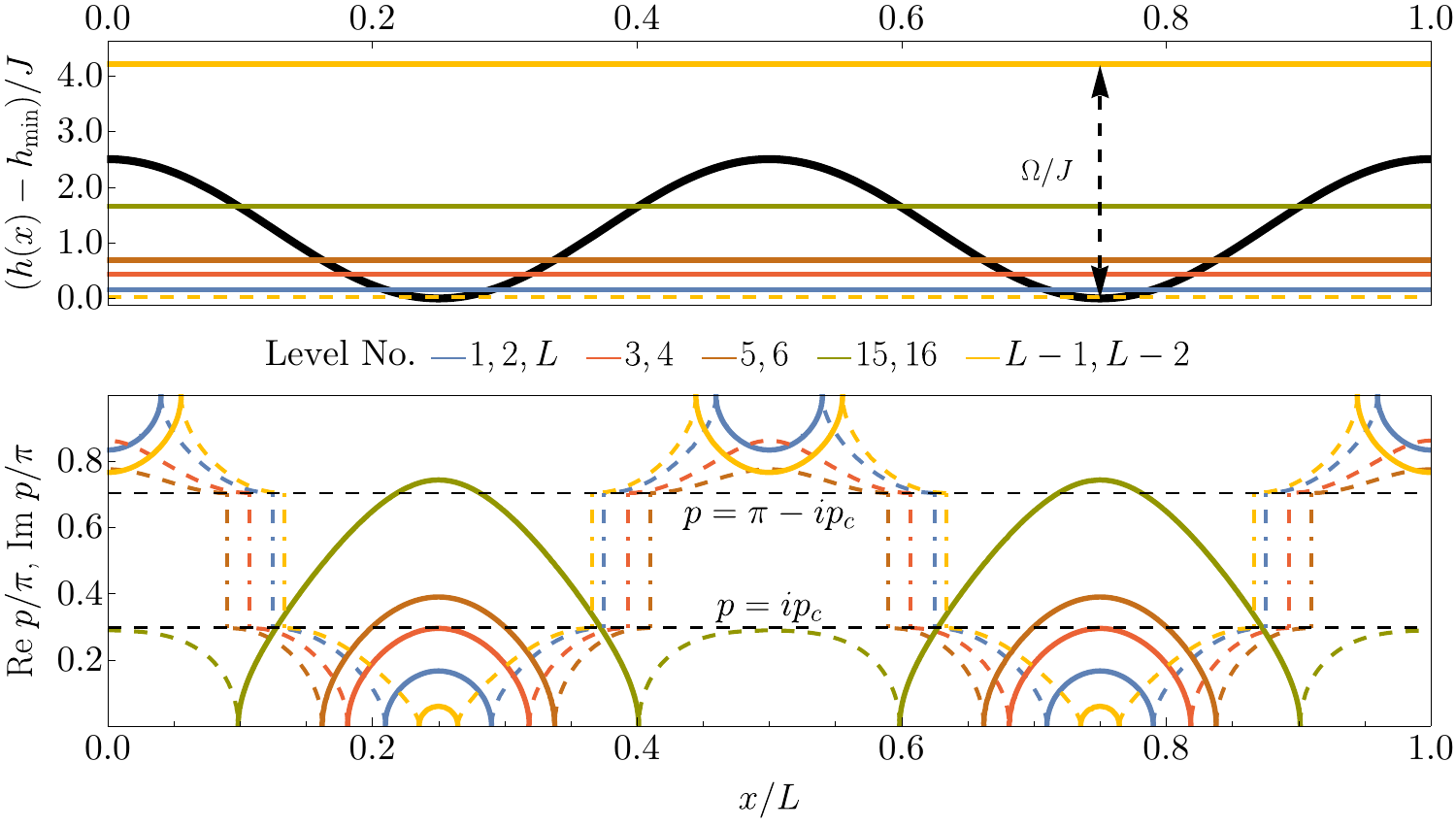}
\par\end{centering}
\caption{Phase portraits of the effective classical Hamiltonian~(\ref{eq:trotterized-classical-hamiltonian})
with the same parameters as in~\figref{low-energy-distortion_of-trotterized-evolution}
for the three lowest resonant subspaces (blue, red, brown), the 7th
subspace (olive), and the high-energy subspaces of states localized
near the boundary (yellow), colored in accordance with the representative
states in \figref{low-energy-distortion_of-trotterized-evolution}.
\emph{Top}: Potential profile (black curve) with quasi-energy levels
of \figref{low-energy-distortion_of-trotterized-evolution} (horizontal
lines) according to the legend below the panel. The tunnel splittings
are too small and not shown. Energy is counted from the bottom of
the potential in units of $J$. States $L-1,\,L-2$ are shown together
with a copy shifted by~$\Omega$ (dashed line). \emph{Bottom:} Solution
to Eqs.~(\ref{eq:classical-hamiltonian})~and~(\ref{eq:trotterized-classical-hamiltonian})
at the energies of the top panel: classical trajectories with real~$p$
(solid curves), tunneling trajectories with complex~$p$ (dashed
curves: $p_{r}+\text{Im}p(x)$, where $p_{r}=\text{Re}p(x)=\text{const}$
equals the value of $p$ at the turning point). Dot-dashed vertical
lines denote band-switching at $p=ip_{c},\,\pi-ip_{c}$ (indicated
by horizontal dashed lines), see Eq.~(\ref{eq:critical-momentum}).
\label{fig:Trotterized-evolution_phase-portraits}}
\end{figure}

To describe this regime semiclassically, one must use the full effective
Hamiltonian~(\ref{eq:trotterized-classical-hamiltonian}), since
criteria~(\ref{eq:Suzuki-Trotter_naive-applicability-criteria})
no longer justify the expansion in powers of~$\delta t$. The corresponding
equation of motion~(\ref{eq:classical-eq-of-motion}) manifests the
Floquet theorem in the form of allowed regions at $\left|h(x)-\Omega k-E\right|<J$
for integer $k$, with $k=0$ corresponding to the allowed regions
of the original Hamiltonian~(\ref{eq:single-particle_Hamiltonian}).
Each allowed region with $k\neq0$ also hosts bound states, which
formally correspond to \emph{high} energies $E\sim\Omega k$ for the
original Hamiltonian. This further implies that resonances can occur
between different allowed regions---a situation realized in \figref{low-energy-distortion_of-trotterized-evolution}.

The phase portrait of Hamiltonian~(\ref{eq:trotterized-classical-hamiltonian})
for the setup of \figref{low-energy-distortion_of-trotterized-evolution}
is shown in \figref{Trotterized-evolution_phase-portraits} with solid
lines. The position of each resonant subspace is correctly described
by quantization condition~(\ref{eq:Bohr-quantization-condition}).
In particular, the phase shift due to hard-wall boundary conditions
for states centered around $x=0,\,L$ causes the detuning of states
$L-1,\,L-2$ from state $L$. Tunneling is described by motion with
complex momenta, shown with dashed lines in \figref{Trotterized-evolution_phase-portraits}.
Importantly, tunneling is accompanied by a branch change as one passes
\begin{equation}
\text{Im}p=p_{c}=\text{arccosh}\frac{1}{\sin J\delta t/2},
\label{eq:critical-momentum}
\end{equation}
corresponding to the branch cut of the kinetic term in Eq.~(\ref{eq:trotterized-classical-hamiltonian})
in the complex $p$ plane. At this point, the momentum jumps between
$\text{Re}p=0$ and $\text{Re}p=\pm\pi$ to continue the under-barrier
motion. This is analogous to the Zener tunneling in indirect semiconductors~\citep{kriegerTheoryElectronTunneling1966}.
The corresponding part of the under-barrier trajectory is shown with
dot-dashed lines in \figref{Trotterized-evolution_phase-portraits}. 

Analytical computation of $\eta,\varepsilon$ in Eq.~(\ref{eq:effective-three-state-model})
requires a more accurate connection formula for the potential extrema~\citep{braun-1993_discrete-WKB}
or, equivalently, WKBA in the complex $x$~plane~\citep{berry-1972_semiclassical-approximation,complexMatrixWKBA}.
However, $\eta$ is still estimated by Eq.~(\ref{eq:semiclassical_tunneling-rate})
applied to the pair of states in either of the minima and the maximum
of the potential. In particular, for the parameters of \figref{low-energy-distortion_of-trotterized-evolution},
this equation yields a value that is just 6\% off the observed one.
Similarly, if $\delta t$ is changed slightly to destroy the resonance
between $\left|B\right>$ and the $\left|A\right>,\,\left|S\right>$
subspace, Eq.~(\ref{eq:semiclassical_tunneling-rate}) applied to
the original pair of states in the minima (with the allowed region
not contributing to the imaginary action) correctly estimates the
magnitude of the off-resonant co-tunneling splitting between $\left|A\right>$
and $\left|S\right>$, up to the factor of order unity associated
with the interference in the allowed region~\citep{complexMatrixWKBA}.
Finally, the tunnel splittings of higher resonant subspaces that do
not enter the allowed region under the barrier are correctly described
by Eq.~(\ref{eq:semiclassical_tunneling-rate}) as before.

\section{Discussion and Outlook\label{sec:Discussion}}

\subsection{Main Results\label{subsec:Main-qualitative-results}}

We analytically studied the dynamics of the Suzuki-Trotter algorithm
(STA), Eq.~(\ref{eq:Suzuki-Trotter_2nd-order}), for the unitary
evolution under Hamiltonian~(\ref{eq:Hamiltonian}) of an $XY$ spin-$1/2$
chain with a $z$-field that smoothly depends on position. Hamiltonian~(\ref{eq:Hamiltonian})
is mapped to a system of noninteracting fermions by the Jordan-Wigner
transformation~\citep{lieb-shultz-mattis_XY-model} described by
single-particle Hamiltonian~(\ref{eq:single-particle_Hamiltonian}).
We further restricted our analysis to the single-particle fermionic
subspace---the subspace with $z$-projection of total spin equal
to $-L/2+1$ (with $L$ being the length of the spin chain). We addressed
the tunneling problem realized by the double-well profile of the $z$-field,
\figref{tunneling-problem_potential-sketch},---a coherent, exponentially
long, nonlocal process.

We derived the effective time-independent single-particle Hamiltonian
$\mathcal{H}_{\text{eff}}$ that governs the evolution of the system
between discrete moments of time $t_{n}=n\delta t$, with $\delta t$
being the Trotter step. According to the Floquet theorem, the quasi-energies
of $\mathcal{H}_{\text{eff}}$ are defined up to a multiple of $\Omega=2\pi/\delta t$.
The naive STA applicability criterion is thus $\Lambda\delta t\ll1$,
where $\Lambda$ is the local energy scale of the simulated Hamiltonian
{[}see Eq.~(\ref{eq:Suzuki-Trotter_naive-applicability-criteria}){]}.
This criterion captures the evolution error in most cases, albeit
with the notable exception of tunneling that we address below. 

Under the assumption of a smooth potential profile, we developed the
Wentzel-Kramers-Brillouin approximation (WKBA)---a manifestly non-perturbative
in~$\delta t$ approach, enabling a faithful incorporation of the
Floquet theorem. We have shown that the main difference with the conventional
WKBA arises from non-commutativity of kinetic terms at neighboring
pairs of sites, modifying the kinetic energy of the semiclassical
Hamiltonian, Eq.~(\ref{eq:trotterized-classical-hamiltonian}).

For $\Lambda\delta t\ll1$, the main effect of STA on tunneling is
the energy shift caused by the perturbation in the classically allowed
region. The magnitude of this shift is given by Eq.~(\ref{eq:semiclassical-energy-shift})
and is estimated by Eq.~(\ref{eq:energy-shift-estimation}). Crucially,
this shift can detune the resonance and suppress tunneling. This effect
puts a limitation on $\delta t$, Eq.~(\ref{eq:criteria_absence-of-detuning}),
expressed in terms of the tunneling amplitude~$\eta$. Importantly,
$\eta$ is an exponentially small quantity, which makes Eq.~(\ref{eq:criteria_absence-of-detuning})
the most essential limitation on~$\delta t$. 

The situation changes significantly if the resonance is not spoiled,
e.g., due to a symmetry of the potential (which must be preserved
by the simulation protocol). In this case, the main effect of STA
is to alter~$\eta$ according to Eq.~(\ref{eq:effective-tunneling-amplitude}).
In most cases, the resulting effect accumulates along the barrier
length to an exponentially large \emph{increase} in the tunneling
rate even for $\Lambda\delta t\ll1$.

In contrast, for $\Lambda\delta t\sim1$, the unitary evolution under
STA is starkly different from the original Hamiltonian evolution.
\secref{Large-Trotter-steps} demonstrated this on the example of
elastic co-tunneling via the allowed region emerging within the potential
barrier due to the Floquet energy periodicity. Crucially, WKBA accurately
describes all qualitative spectral features, with quantitative agreement
achievable through a more refined application of WKBA that generalizes
relation~(\ref{eq:semiclassical_tunneling-rate}) to account for
interference in the allowed region~\citep{complexMatrixWKBA}. The
effective Hamiltonian~(\ref{eq:trotterized-classical-hamiltonian})
thus provides a comprehensive tool to construct and study tunneling
problems in Floquet dynamics.

\subsection{An Experimental Realization\label{subsec:Experimental-realization}}

We propose an experimental protocol to observe the described phenomena
on existing devices by monitoring the occupation number Rabi oscillations
in a series of device-tailored double-well setups. To better resolve
the oscillations, we employ the Fourier transform of the occupation
number, as done in Ref.~\citep{neill-2021_fermionic-ring-simulation}.
Due to the limitations of existing devices, several optimizations
are required: \emph{i)}~the energy of the initial state has to be
close to the top of the potential barrier to avoid too long tunneling
times; \emph{ii)}~the typical momentum of the initial state must
be close to $\pi/2$, to reduce the perturbative energy shift~(\ref{eq:semiclassical-energy-shift})
that may otherwise lift the target state above the barrier; \emph{iii)}~the
initial state has to be close to one of the resonant states to maximize
the visibility of the Rabi oscillations. The experimental design satisfying
these requirements for a system of $L=50$ qubits is presented in
\appref{Experimental-design}.

\begin{figure}
\begin{centering}
\includegraphics[width=0.95\columnwidth]{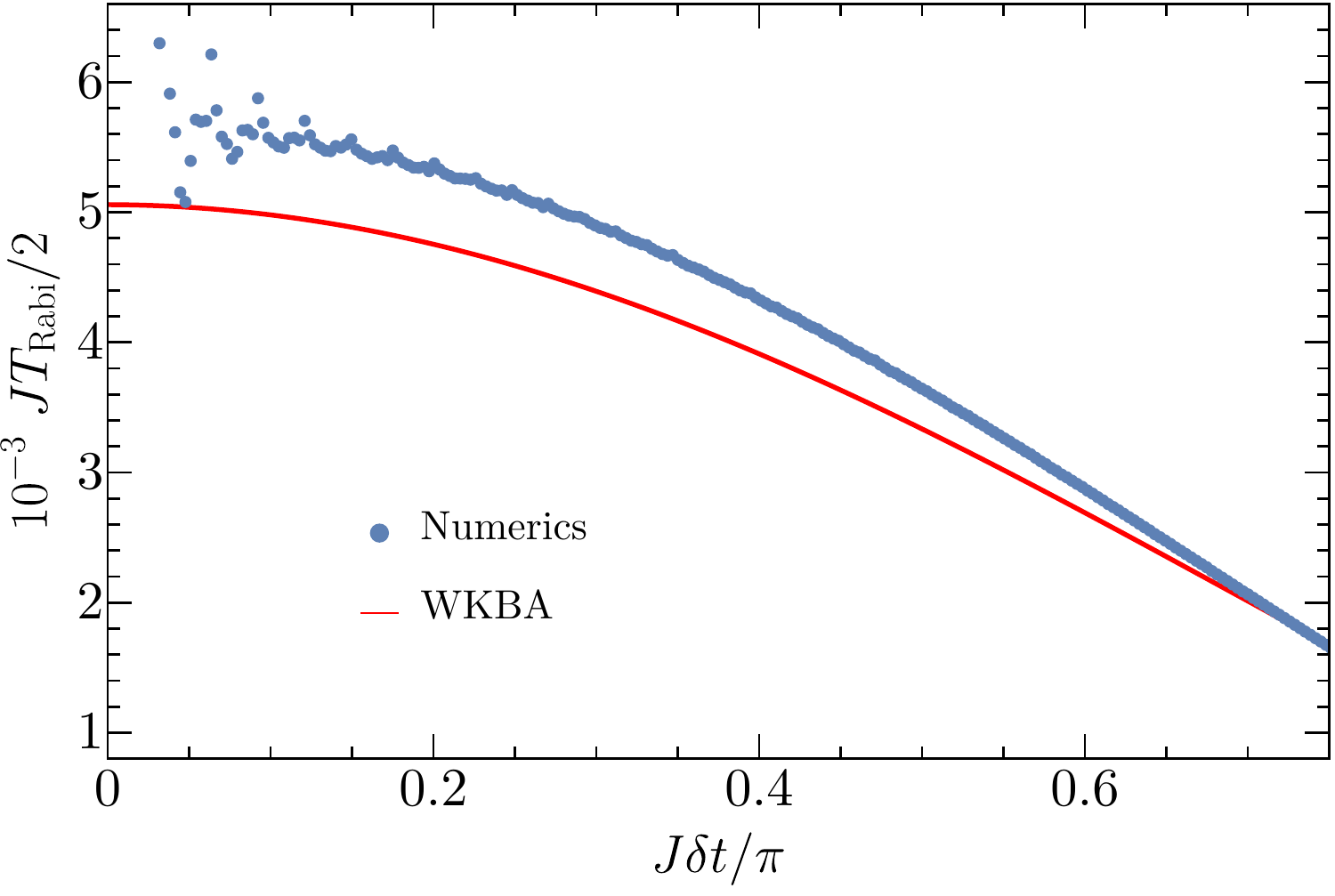}
\par\end{centering}
\caption{Dependence of the Rabi period on the Trotter step $\delta t$ for
the system described in \appref{Experimental-design}. The blue dots
are obtained by simulating the dynamics of the total occupation number
in one of the potential wells. The observed noise at small $\delta t$
is due to only one full period of oscillations fitting in the allowed
Trotter step count. The red curve is the WKBA description of the Rabi
oscillations according to \subsecref{Matrix-WKBA}, no fitting is
used. The remaining discrepancy between WKBA and the numerical simulation
originates from corrections to the semiclassical description.\label{fig:rabi-oscillation-period_vs_trotter-step}}
\end{figure}

Figure~\ref{fig:rabi-oscillation-period_vs_trotter-step} shows the
observed Rabi period as a function of~$\delta t$, alongside the
semiclassical prediction according to \subsecref{Matrix-WKBA}. Notably,
a fivefold acceleration of tunneling at large Trotter steps, whereas
the energy shift of the corresponding state is practically absent,
even falling below the tunnel splitting at large $\delta t$. The
detailed protocol for the numerical simulation is presented in \appref{Experimental-design},
which also contains additional simulations showcasing the main results
and the effect of operator ordering.

Experimentally reproducing \figref{rabi-oscillation-period_vs_trotter-step}
requires circuit depth of~$1.2\times10^{5}$. While Ref.~\citep{neill-2021_fermionic-ring-simulation}
performed spectral analysis of noisy circuits with depths up to $1.4\times10^{3}$
for a similar experiment, we expect depths $\sim10^{5}$ to be possible
in the near future. This anticipation is motivated by two reasons:
\emph{i)}~the problem is manifestly single-particle, which renders
error mitigation by post-selection efficient~\citep{neill-2021_fermionic-ring-simulation},
and \emph{ii)}~the coherence times of superconducting devices are
already in the range of $3\times10^{2}\,\mu s$~\citep{nature-2021_long-coherence-times,sivak-2023_nature_error-correction},
more than 10 times larger than that of the device in Ref.~\citep{neill-2021_fermionic-ring-simulation},
and further improvement is likely to occur. Moreover, due to design
optimizations, the proposed experiment can tolerate relative static
coherent error of order~$5\times10^{-3}$, which is well within the
capabilities of the device used in Ref.~\citep{neill-2021_fermionic-ring-simulation}.
As a result, we expect this experiment to be feasible in the near
future.

\subsection{Relation with the Rigorous Error Bounds\label{subsec:Relation-with-error-bounds}}

It is instructive to compare our results with the available rigorous
bounds and assess their saturation. A rigorous upper bound on the
spectral norm distance between the exact $U$ and approximate $U_{\text{appr}}$
evolution operators during one Trotter step is given by~\citep[Suppl. Sec. F.2.1, Eq. (87)]{childs_quantum-simulations-with-speedup}:
\begin{equation}
\left\Vert U(\delta t)-U_{\text{appr}}(\delta t)\right\Vert \le C\frac{x^{3}}{n}+\frac{x^{4}}{12}e^{x},
\label{eq:exact-upper-bound}
\end{equation}
where $n$ is the number of qubits, $\left\Vert O\right\Vert $ is
the spectral norm, $x=3n\Lambda\delta t$, $C<1/4$ is a numerical
constant, and $\Lambda=\max\{h,J\}$ is the maximum magnitude of local
terms in the Hamiltonian. This implies an upper bound on the eigenvalue
error:
\begin{equation}
\left|E_{N}-E_{N,\text{eff}}\right|\le3C'\Lambda\left(3n\Lambda\delta t\right)^{2},
\label{eq:detuning-upper-bound}
\end{equation}
where $C'=C\left[1+O(3n\Lambda\delta t)\right]$ can be made arbitrarily
close to $C$ for sufficiently small $3n\Lambda\delta t$. This upper
bound correctly reproduces the semiclassical estimate, Eq.~(\ref{eq:energy-shift-estimation}),
as a function of $\delta t$ and $\Lambda$, up to model-specific
details. The remaining discrepancy---notably, the $n$ (or, rather,
$n_{\text{cl}}$) dependence---originates from the fact that Eq.~(\ref{eq:exact-upper-bound})
does not contain information about the eigenstate $\left|N\right>$
itself.

Importantly, both Eqs.~(\ref{eq:energy-shift-estimation}) and (\ref{eq:detuning-upper-bound})
are governed by short-time dynamics, which is why they predictably
fail at large times. After $r=t/\delta t$ Trotter steps, the upper
bound on the evolution operator error reads~\citep[Suppl. Sec. F.2.1, Eq. (87)]{childs_quantum-simulations-with-speedup}
\begin{equation}
\left\Vert U(t)-\left[U_{\text{appr}}(\delta t)\right]^{t/\delta t}\right\Vert \le\frac{t}{\delta t}\left(C\frac{x^{3}}{n}+\frac{x^{4}}{12}e^{x}\right),
\end{equation}
which implies that at time scales of the order of the inverse detuning~(\ref{eq:detuning-upper-bound})
\begin{equation}
t_{\text{bound}}\sim\frac{1}{\delta E}\sim\frac{1}{\Lambda}\left(3n\Lambda\delta t\right)^{-2},
\end{equation}
the evolution error can attain order of unity. However, this cannot
be true for all physical observables and any required precision. To
illustrate this, consider the expectation value of the particle number
in the left well, $\mathcal{N}_{\text{left}}=2\sum^{L/2}_{i=1}S^{z}_{i}+L$,
for the double-well potential of \subsecref{Single-particle-tunneling-problem}.
The initial state is prepared as one of the metastable states in the
left well of the target Hamiltonian. The required precision $\epsilon\le1$
is assumed to be larger than the fidelity defect $\delta P_{N}$ of
the target metastable state, estimated as (see \appref{Probability-defect}):
\begin{equation}
\delta P_{N}:=1-\left|\left\langle N;0|N;\delta t\right\rangle \right|^{2}=C_{N}\left(J\delta t\right)^{4}+O\left(\delta t^{6}\right),
\label{eq:probability-defect-estimation}
\end{equation}
where $\left|N;\delta t\right>$ is the exact $N$-th metastable state
for Trotter step $\delta t$, $C_{N}\le n^{2}_{\text{cl}}$ is a state-dependent
constant available from WKBA, and $n_{\text{cl}}$ is the size of
the allowed region for the $N$-th level in the well. In an STA simulation,
the dynamics of the expectation value $\left\langle \mathcal{N}_{\text{left}}\right\rangle $
is approximately described as
\begin{equation}
\left\langle \mathcal{N}_{\text{left}}\left(t\right)\right\rangle \approx\frac{\varepsilon^{2}_{N,\text{eff}}/4}{\Omega^{2}_{N,\text{eff}}}+\left\{ 1-\frac{\varepsilon^{2}_{N,\text{eff}}/4}{\Omega^{2}_{N,\text{eff}}}\right\} \cos^{2}\Omega_{N,\text{eff}}t,
\label{eq:detuned-Rabi-oscillations}
\end{equation}
where $\eta_{N,\text{eff}},\,\varepsilon_{N,\text{eff}}$ are the
parameters of the resonant subspace for a given $\delta t$, and $\Omega_{N,\text{eff}}=\sqrt{\eta^{2}_{N,\text{eff}}+\varepsilon^{2}_{N,\text{eff}}/4}$
. In this expression the exponentially small contribution of states
in the right well and the total probability defect, Eq.~(\ref{eq:probability-defect-estimation}),
were neglected in comparison to the allowed error~$\epsilon$. The
dynamics of $\left\langle \mathcal{N}_{\text{left}}\right\rangle $
is thus strongly sensitive to whether the detuning effect from STA
is present, although in both cases the dominant error reveals itself
only at times much larger than $t_{\text{bound}}$. In the detuned
case, $\varepsilon_{N,\text{eff}}\gg\eta_{N}$, tunneling is suppressed,
but the difference between $\left\langle \mathcal{N}_{\text{left}}\right\rangle $
(now nearly constant) and its unperturbed counterpart exhibiting Rabi
oscillations reaches~$\epsilon$ at exponentially large times $t\sim\eta^{-1}_{N}\sqrt{\epsilon}\gg t_{\text{bound}}$.
In the case of no detuning, the Rabi oscillations are still observed
in the perturbed evolution, but the frequency of oscillations changes
according to Eq.~(\ref{eq:effective-tunneling-amplitude}), so error~$\epsilon$
is reached at slightly smaller times $t\sim\sqrt{\left|\eta^{-2}_{N,\text{eff}}-\eta^{-2}_{N}\right|}\sqrt{\epsilon}$,
still much larger than $t_{\text{bound}}$. This analysis, while missing
the short-time error from both detuning (if present) and fidelity
defect $\delta P_{N}$, clearly illustrates that rigorous upper bounds
overstate the actual magnitude of error at large times for certain
physical observables.

\subsection{Outlook\label{subsec:Outlook}}

\paragraph{Topological WKBA.}

An instance of phase interference influence on the tunnel splitting
is presented in \figref{low-energy-distortion_of-trotterized-evolution}.
One can further arrange for interference of multiple complex turning
points in the forbidden region. Equations~(\ref{eq:semiclassical_tunneling-rate})--(\ref{eq:underbarrier-action})
do not capture such effects, as explained in the end of \secref{Large-Trotter-steps}.
Instead, WKBA in the complex plane~\citep{berry-1972_semiclassical-approximation}
has to be extended to matrix Hamiltonians. The corresponding procedure
will be published elsewhere~\citep{complexMatrixWKBA}.

Notably, this enables the study of the interplay between tunneling
and topological properties of the system. In particular, our approach
is straightforwardly generalized to problems with complex hopping
amplitudes $Je^{i\phi_{n+1/2}}$, corresponding to a vector potential
$A_{n}=\phi_{n+1/2}-\phi_{n-1/2}$, by the Peierls substitution $p\to p-\phi$
in WKBA~\footnote{Cf. Eq. (4) of Ref.~\citep{vakulchykUniformAndersonLocalization2023}
or Eq. (1) of Ref.~\citep{neill-2021_fermionic-ring-simulation}.}. This allows one to study current response, as well as to observe
the effect of ring topology.

Further examples concern the topology of the Riemann surface of the
classical momentum and the associated Berry connection of the semiclassical
wave function. While closed classical trajectories of the system studied
in this work always have a trivial~\footnote{This can be verified using Eq.~(\ref{eq:trotterized-evolution_eigenfunction})
or concluded directly from the fact that the vector of components
of the matrix semiclassical Hamiltonian (see \appref{Matrix-WKB})
in the Pauli basis lies in the $y-z$-plane, thus sweeping solid angles
that are multiples of~$2\pi$.} Berry phase proportional to a multiple of~$2\pi$, adding a smooth
spatial modulation of the potential, $h_{n}=h(an)+(-1)^{n}w(an)$
with smooth $h(x),\,w(x)$, will produce a nontrivial Berry phase.
Notably, the resulting semiclassical description will depend on the
operator ordering within a single Trotter step. More importantly,
this topological phase will contribute to interference patterns in
the complex momentum plane (corresponding to tunneling) and, specifically,
around the branch point $p=\pm ip_{c}$ under the barrier. A detailed
analysis of the interplay between these topological properties and
tunneling is a subject of future work~\citep{complexMatrixWKBA}.

\paragraph{Higher-order STA.}

The present theoretical analysis can be generalized to higher-order
product formulas that are expected to yield $O\left(\left(J\delta t\right)^{2k}\right)$
naive accuracy of $H_{\text{eff}}-H$ in powers of $\delta t$~\citep{suzuki-1991}.
These corrections are more non-local and are thus more sensitive to
the potential inhomogeneities. However, we expect our analysis to
remain qualitatively valid: the effect of STA in the classically allowed
region can be studied perturbatively, capturing the content of the
rigorous error bounds, while the effects on tunneling can be described
by the leading-order correction to the classical tunneling action,
as long as the barrier is not too high. It also follows that the tunneling
events still impose the limitation $P\delta t\ll1$ regardless of
the order $k$ of the product formula due to the Floquet theorem reasoning.
Equivalently, it can be verified that the singularities in the kinetic
part of the effective semiclassical Hamiltonian are still located
at $p\sim\pm i\ln\frac{1}{J\delta t}$ (see Appendix~\subsecref{Tunneling_vs_large-distance-hopping}
for details).

\paragraph{Potential Disorder.}

Introducing potential disorder to the single-particle Hamiltonian~(\ref{eq:single-particle_Hamiltonian}),
similarly to Ref.~\citep{neill-2021_fermionic-ring-simulation},
constitutes another interesting extension. Importantly, arbitrarily
small disorder induces Anderson localization in one dimension~\citep{Scaling-theory-of-localization_1979}.
While our method is not straightforwardly applicable to local potential
disorder, we still expect that the key effect of STA for small disorder
amounts to smearing the potential disorder, causing only a renormalization
of the mean free path and, consequently, of the localization length.
This expectation is confirmed by Ref.~\citep{vakulchykUniformAndersonLocalization2023}
that studies the localization length for the first-order STA circuit.
Interestingly, Fig.~3 of this work indicates that the localization
length at weak disorder increases with typical momentum. Despite lower
order of STA, we expect their results to hold qualitatively for our
case.

\paragraph{Time-dependent Hamiltonians.}

Extending the analysis to time-dependent Hamiltonians is another interesting
challenge. While rigorous upper bounds on the STA error for time-dependent
Hamiltonians are also available~\citep{childs-2019_Nearly-optimal-simulation-by-product-formulas},
their saturation is understood even less, as it depends on the exact
dynamics arranged by the target time-dependent Hamiltonian. In particular,
a general time dependence renders the Floquet theorem inapplicable.

One particular example is the dynamic inelastic enhancement of tunneling
due to a weak periodic drive with frequencies $\eta\ll\omega\ll P$~\citep{ivlev_review},
where $\eta$ is the tunneling rate in the static problem, and $P$
is the potential barrier height. Interestingly, this problem can be
regarded as the opposite limit of STA from the point of view of criterion~(\ref{eq:Suzuki-Trotter_naive-applicability-criteria}).
However, a direct comparison is hindered by a different structure
of the periodic evolution. On the other hand, the method of Ref.~\citep{ivlev_review}
provides a direct way to generalize WKBA of this work to time-dependent
scenarios.

Another example of a time-dependent problem comes in the form of a
revival of the debate on the duration of a tunneling event~\citep{attoclock-atom-ionozation-experiments_review_2020}.
While in most experimental designs, the interpretation of the measurement
outcomes proves difficult, the existing superconducting qubit systems
provide a promising platform for an in-depth study of the issue, leveraging
the unique opportunity to perform simultaneous measurements at every
spatial point. Notably, this platform allows a straightforward implementation
of controlled periodic modulation of the barrier height---a convenient
way to probe the time spent under the barrier~\citep{Buttiker-1990_barrier-modulation}.
The experiment in \subsecref{Experimental-realization} can be tailored
to contribute to the tunneling time measurement. However, to provide
an accurate interpretation of the results, the magnitude and the structure
of the STA error for the resulting \emph{time-dependent} Hamiltonian
must be carefully analyzed.

\paragraph{Noise and Coupling to an External Bath.}

Decoherence and energy relaxation are inescapable consequences of
the coupling between the physical device implementing the qubit system
and the environment~\citep{neill-2021_fermionic-ring-simulation,miInformationScramblingQuantum2021,arute-2019_quantum-supremacy-paper,satzinger-2021_realizing-toric-code-on-quantum-proccessor}.
In addition, a particular implementation of two-qubit gates causes
leakage to non-computational subspace of the physical device, contributing
to decoherence and causing a violation of spin conservation.

Crucially, coupling to an external bath generally suppresses tunneling~\citep[Sec. 3.5]{kamenevFieldTheoryNonequilibrium2011},
allowing assessment of the underlying device coherence. However, to
this end, the interplay between noise and STA has to be studied in
more detail. Moreover, a bath induces finite level broadening, leading
to energy absorption from any periodic perturbation with frequency
larger than the single-particle level spacing~$\delta$~\citep[Ch. 8]{mahanManyParticlePhysics2000},
such as STA itself (see \appref{Trotterization_as-periodic-perturbation_and_thermalization}).
Studying these issues requires generalizing the present approach to
open quantum systems.

On the other hand, a monitored external bath can be employed to study
the tunneling process itself by examining the excitations emitted
by the tunneling particle. Implementing a controlled weakly anharmonic
bath (e.g., in the form of a qubit chain) coupled locally to a high-potential
region of the target Hamiltonian~(\ref{eq:Hamiltonian}) would allow
such an experiment.

\paragraph{Many-body Physics and Interaction.}

As soon as many-body dynamics of system~(\ref{eq:Hamiltonian}) are
considered, one is forced to study the interplay between fermionic
statistics and STA. One immediate model is the decay of large spin
density initially confined to a local potential minimum. More importantly,
while our model enjoys mapping to non-interacting fermions, all known
applications of existing quantum hardware necessarily invoke truly
many-body dynamics in an interacting system, taking advantage of the
large Hilbert space~\citep{smelyanskiy_2020_nonergodic,AQCReview,NISQReview}.
Analyzing the effect of STA in this class of problems is thus of great
practical significance. In addition, interaction also induces the
aforementioned heating by a periodic perturbation, albeit the corresponding
heating time~$\tau^{*}$ is exponentially large in the STA parameter,
$\ln\tau^{*}\sim\Omega/\max\left\{ P,J\right\} $~\citep{prethermalization_abanin-2017}.
However, experimentally simulating complex many-body physics is not
currently possible with existing noisy quantum devices~\citep{Preskill-Vidal_2022_many-body-spin-flip-evolution}
and requires instead implementing quantum error correction protocols~\citep{quantum-error-correction_2023,kitaev_2003}.
A deeper understanding of the interplay between noise and STA is thus
of practical importance.
\begin{acknowledgments}
The authors thank Mikhail V. Feigel'man for numerous fruitful discussions.
A.V.K. is also grateful to Laboratoire d’excellence LANEF in Grenoble
(ANR-10-LABX-51-01). At the time of revision, A.V.K. is affiliated
with CNRS, ENS de Lyon, LPENSL, UMR5672, 69342, Lyon cedex 07, France.
\end{acknowledgments}

\bibliographystyle{apsrev4-2_improved.bst}
\bibliography{Bibliography.bib}

\appendix
\onecolumngrid 

\counterwithin{figure}{section} 
\renewcommand\thefigure{\thesection\arabic{figure}} 

\let\secappref=\relax 
\newref{secapp}{ refcmd={Sec.~\ref{#1}} } 

\let\subsecappref=\relax 
\newref{subsecapp}{ refcmd={Sec.~\ref{#1}} } 

\section{Gate Ordering in the Product Formula\label{app:Gate-ordering}}

In this Appendix, we review the operator orderings in STA for Hamiltonian~(\ref{eq:Hamiltonian})
in quantum hardware with arbitrary local one- and two-qubit gates.
The two-qubit operators within each term in Eq.~(\ref{eq:Ham-non-commuting terms})
act on non-intersecting pairs of qubits and can thus be applied simultaneously.
As a result, one can apply the following evolution operators: $\exp\left\{ -iaK_{\text{even}}\right\} $,
$\exp\left\{ -ibK_{\text{odd}}\right\} $, and $\exp\left\{ -icP\right\} $
with real times $a,b,c$. 

Consider a Hamiltonian $H=A_{1}+..+A_{n}$, where each term $A_{k},\,k=1,...,n$
can be exponentiated individually as $U_{k}(\delta t)=e^{-i\delta tA_{k}}$,
but only one at a time. There are $n!$ distinct second-order product
formulae~(\ref{eq:Suzuki-Trotter_2nd-order}) corresponding to the
number of operator orderings, each yielding a different one-step unitary
$U(\delta t)$. However, for any unitary $S$, the spectra of $U_{\text{appr}}(\delta t)$
and $S^{\dagger}U_{\text{appr}}(\delta t)S$ coincide, allowing one
to identify all product formulae that differ by a unitary conjugation,
as far as spectral properties are concerned. To keep track of the
matrix elements, one also has to apply the same transformation to
observables: $O\to S^{\dagger}OS$.

It is useful to consider the following $S_{k+\alpha}$ with $k+\alpha\in[0,n]$:
\begin{equation}
S_{k+\alpha}=\begin{cases}
U_{n}(\delta t/2)\,U_{n-1}(\delta t/2)...U_{n-k}(\alpha\delta t/2), & k\in0,...,n-1,\,\,\,\alpha\in[0,1),\\
U_{n}(\alpha\delta t/2), & k=0,\,\,\,\alpha\in[0,1),\\
U_{n}(\delta t/2)\,U_{n-1}(\delta t/2)...U_{1}(\delta t/2), & k=n-1,\,\,\,\alpha=1.
\end{cases}
\end{equation}
Such $S_{k+\alpha}$ transforms the second-order product formula~(\ref{eq:Suzuki-Trotter_2nd-order})
by cyclically shifting the operator ordering and splitting the gate
$U_{k}$ for $\alpha\neq0$: 
\begin{equation}
U_{n}...U_{1}U_{1}...U_{n}\to U^{1-\alpha}_{k+1}U_{k}...U_{1}U_{1}...U_{n-1}U_{n}U_{n}U_{n-1}...U^{\alpha}_{k+1},
\end{equation}
where we denote $U_{k+1}(\alpha\delta t/2)=U^{\alpha}_{k+1}$ for
brevity. This is why we refer to $S_{k+\alpha}$ as the time translation
operator: it ``shifts'' the beginning of a Trotter step sequence
by $k+\alpha$ gates forward, splitting a gate if necessary.

Importantly, $S_{k+\alpha}$ for Hamiltonian~(\ref{eq:Hamiltonian})
contains at most two operations that act on pairs of qubits, and preserves
the $z$-projection of the total spin. In particular, it conserves
the single-particle subspace, in which it acts at most as a distance-two
operation. This suggests that nonlocal matrix elements, such as tunneling
amplitudes, are weakly affected by the local dressing induced by $S_{k+\alpha}$,
which additionally motivates our identification of product formulae
that differ by a cyclic shift of the underlying operator ordering.
In particular, we note that the first-order product formula for $n=2$
operators, $U^{(1)}_{\text{appr}}=U_{1}(\delta t)U_{2}(\delta t)$,
is connected to the second-order one, Eq.~(\ref{eq:Suzuki-Trotter_2nd-order}),
by $S_{k+\alpha}$ with $k=1,\,\alpha=0$. It also follows that for
$n=2$, the two possible orderings are identified with one another
by $S_{k+\alpha}$ with $k=2,\,\alpha=0$.

Another important point concerns specifically the two-qubit gate count.
Due to hardware limitations~\citep{neill-2021_fermionic-ring-simulation,miInformationScramblingQuantum2021,arute-2019_quantum-supremacy-paper,satzinger-2021_realizing-toric-code-on-quantum-proccessor},
two-qubit gates are typically more detrimental to device performance.
To reduce their gate count, one has to place the two-qubit gates at
the ends of the operator ordering. Indeed, adjacent $U_{k}(\delta t/2)$
can be merged as $e^{-i\delta tA_{k}/2}e^{-i\delta tA_{k}/2}=e^{-i\delta tA_{k}}$.
As a result, in an $M$-step STA sequence for operators $A_{1},...,A_{n}$,
operators $A_{2},...,A_{n-1}$ are applied $2M$ times, whereas operators
$A_{1},\,A_{n}$ are applied only $M$ and $M+1$ times, respectively.
For this reason, operator orderings $\left\{ K_{\text{even}},P,K_{\text{odd}}\right\} $
and $\left\{ K_{\text{odd}},P,K_{\text{even}}\right\} $ are preferred.

It is further worth considering the representation $H=A_{1}+A_{2}$
with $A_{1}=K_{\text{even}}+\beta P$ and $A_{2}=K_{\text{odd}}+(1-\beta)P$,
with $\beta\in[0,1]$. This construction eliminates a part of the
STA error originating from the noncommutativity of $P$ with $K_{\text{even}},\,K_{\text{odd}}$
(see \appref{trotter_perturb-corrections}), while requiring the same
number of two-qubit gates. Indeed, to represent the evolution operators
of $A_{1},\,A_{2}$, one uses the fact that a two-qubit gate acting
on qubits $i,\,i+1$ can be decomposed as
\begin{equation}
\exp\left\{ -i\delta t\left(K_{i,i+1}+x\left[P_{i}+P_{i+1}\right]\right)\right\} =\left(u_{i}\otimes u_{i+1}\right)\,U_{i,i+1}\,\left(v_{i}\otimes v_{i+1}\right),
\label{eq:two-qubit-gate-decomposition}
\end{equation}
where $U_{i,i+1}$ is a purely two-qubit gate, and $u_{i},\,v_{i}$
are the single-qubit gates acting on qubit~$i$, all depending on
$J_{i,i+1},\,\delta t,\,h_{i},\,x$. The evolution operator of Hamiltonian~(\ref{eq:Hamiltonian})
is then approximated by the product formula for just two operators~$A_{1},\,A_{2}$.
Upon merging the neighboring gates where possible, the implementation
of $M$ Trotter steps contains $2M+1$ layers of two-qubit gates interleaved
with $2M+2$ single-qubit layers. Provided that the resulting $U_{i,i+1}$
can be implemented as a single two-qubit gate in the underlying hardware,
this representation requires the same number of physical two-qubit
gates as, e.g., $\left\{ K_{\text{even}},P,K_{\text{odd}}\right\} $.
At the same time, this representation better suppresses corrections
to WKBA when the potential is not smooth enough (as is the case for
some simulations in \appref{Experimental-design}).

As a result, a three-term Hamiltonian~(\ref{eq:Ham-decomposition})
admits three different product formulae: $\left\{ K_{\text{even}},\,K_{\text{odd}},\,P\right\} $,
$\left\{ K_{\text{even}},\,P,\,K_{\text{odd}}\right\} $, and $\left\{ K_{\text{even}}+\beta P,\,K_{\text{odd}}+(1-\beta)P\right\} $
with $\beta\in[0,1]$. All other possibilities differ from these three
either by a time translation $S_{k+\alpha}$, or by interchanging
$K_{\text{even}}$ and $K_{\text{odd}}$ (which creates differences
only at system boundaries). For practical considerations, ordering
$\left\{ K_{\text{even}},\,P,\,K_{\text{odd}}\right\} $ produces
a gate sequence with the smallest two-qubit gate count, whereas ordering
$\left\{ K_{\text{even}}+\beta P,\,K_{\text{odd}}+(1-\beta)P\right\} $
reduces the Trotter error, while also yielding the near-optimal two-qubit
gate count in some hardware.

\section{Discrete Semiclassical Approximation\label{app:Discrete-Semiclassical-Approximation}}

In this Appendix, the semiclassical approximation for 1D systems is
reviewed. The derivation of the semiclassical ansatz is in line with
Ref.~\citep{maslov-2001-semiclassical-approx}, and the matching
procedure is done similarly to Ref.~\citep{berry-1972_semiclassical-approximation},
with modifications for discrete systems from Ref.~\citep{braun-1993_discrete-WKB}.

\subsection{The WKB ansatz for a discrete system}

We start from the exact Schrödinger equation:
\begin{equation}
\sum_{m}\mathcal{H}_{nm}\psi_{m}=E\psi_{n},
\label{eq:Stationary-Schroedinger-eq}
\end{equation}
where $\mathcal{H}_{nm}$ are the matrix elements of the Hamiltonian
in the coordinate (site position) basis. We assume that the matrix
elements~$\mathcal{H}_{nm}$ obey two conditions: \emph{i)}~\emph{locality}:
$\mathcal{H}_{nm}$ decay quickly (e.g., exponentially) with the difference
$\left|n-m\right|$, and \emph{ii)}~\emph{smoothness}: $\mathcal{H}_{nm}$
depend smoothly on $a(n+m)$ with some characteristic spatial scale
$a\ll1$ (playing the role of Planck's constant). For Hamiltonians
of the form~(\ref{eq:single-particle_Hamiltonian}), this implies
short-distance hopping terms, with both the hopping amplitudes and
the potential terms depending on site number~$n$ only via the combination
$an$.

We then search the wave function in the form of the following asymptotic
series:
\begin{equation}
\psi_{n}=\exp\left\{ \frac{iS(x)}{a}\right\} \left[\varphi_{0}(x)+a\,\varphi_{1}(x)+...\right],\,\,\,\,x=an,
\label{eq:semiclassical-series}
\end{equation}
where $S(x)$ and $\varphi_{i}(x)$ are slow functions of their arguments.
The action of the Hamiltonian on such wave functions can be expanded
in powers of~$a$:
\begin{align}
\sum_{m}\mathcal{H}_{nm}\psi_{m} & \approx\exp\left\{ \frac{iS(x)}{a}\right\} \mathcal{H}(x,p)\left[\varphi_{0}(x)+a\,\varphi_{1}(x)\right]\nonumber \\
 & +\left(-ia\right)\,\exp\left\{ \frac{iS(x)}{a}\right\} \,\left\{ \partial_{p}\mathcal{H}(x,p)\,\varphi_{0}'(x)+\frac{1}{2}\,S''(x)\,\partial^{2}_{p}\mathcal{H}(x,p)\,\varphi_{0}(x)-i\partial_{a}\mathcal{H}(x,p)\,\varphi_{0}(x)\right\} +O\left(a^{2}\right),
\label{eq:action-of-operator-on-semiclassical-wave-function}
\end{align}
where $x=an$, $p=dS/dx$ is the semiclassical momentum, and $\mathcal{H}(x,p)$
is the Maslov symbol of the Hamiltonian:
\begin{equation}
\mathcal{H}(x=an,p)=\sum_{m}\mathcal{H}_{nm}\exp\left\{ ip(m-n)\right\} \equiv\frac{\left\langle n\left|\mathcal{H}\right|p\right\rangle }{\left\langle n|p\right\rangle }.
\label{eq:maslov-symbol-def}
\end{equation}
According to the locality and smoothness conditions, this function
depends smoothly on both arguments, with $n$-dependence expressed
via the combination $x=an$. However, $\mathcal{H}$ may retain residual
regular dependence on~$a$, captured by the term $\propto\partial_{a}\mathcal{H}$
in Eq.~(\ref{eq:action-of-operator-on-semiclassical-wave-function}).
A relevant example is the position dependence of the hopping term:
\begin{equation}
\mathcal{H}_{\text{kin}}=-\frac{1}{2}\sum_{n}\left(J_{n+1/2}\left|n\left\rangle \right\langle n+1\right|+J^{*}_{n-1/2}\left|n\left\rangle \right\langle n-1\right|\right),
\label{eq:kin-term-example_position-dependent-hopping}
\end{equation}
The Maslov symbol then evaluates to
\begin{equation}
\mathcal{H}_{\text{kin}}(x,p;a)=-\frac{1}{2}\left[J(x+a/2)\,e^{ip}+J(x-a/2)\,e^{-ip}\right],
\end{equation}
where $J(x+a/2)\equiv J_{x/a+1/2}$. In this case, the $a$ derivatives
are given by:
\begin{equation}
\partial_{a}\equiv\left(-\frac{1}{2}i\partial_{x}\partial_{p}\right).
\end{equation}

The two leading orders in powers of $a$ of Eq.~(\ref{eq:action-of-operator-on-semiclassical-wave-function})
and the Schrödinger equation~(\ref{eq:Stationary-Schroedinger-eq})
imply the classical equation of motion,
\begin{equation}
\mathcal{H}(x,p)=E,
\label{eq:classical-eq-of-motion_app}
\end{equation}
as well as the transport equation for $\varphi_{0}$: 
\begin{equation}
\left\{ \partial_{p}\mathcal{H}\,\frac{d}{dx}+\frac{1}{2}\,S''\,\partial^{2}_{p}\mathcal{H}-i\partial_{a}\mathcal{H}\right\} \,\varphi_{0}(x)=0.
\label{eq:classical-transport-equation}
\end{equation}
Similar equations exist for higher orders $\varphi_{i},\,i\ge1$.
The transport equation can be solved explicitly, yielding
\begin{equation}
\varphi_{0}(x)=\frac{\text{const}}{\sqrt{\partial_{p}\mathcal{H}(x,p)}}\exp\left\{ \intop^{x}_{x_{0}}dx\,\left.\frac{\frac{1}{2}\partial_{x}\partial_{p}\mathcal{H}-i\partial_{a}\mathcal{H}}{\partial_{p}\mathcal{H}}\right|_{p=p(x)}\right\} ,
\label{eq:scalar-transpot-eq_solution}
\end{equation}
where $p(x)$ is the position-dependent classical momentum found from
the classical equation of motion, Eq.~(\ref{eq:classical-eq-of-motion_app}),
and the lower limit of integration $x_{0}$ can be set to any $x$-independent
constant (typically, the position of a classical turning point). 

In the allowed region, the equation of motion has two real solutions
with opposing signs of~$p$, corresponding to waves propagating in
different directions, according to the sign of the semiclassical velocity.
In the classically forbidden region, where the equation of motion
is solved by complex momenta, a finite imaginary part of $p$ means
that Eq.~(\ref{eq:semiclassical-series}) describes the two solutions
that grow or decay exponentially as one moves into the forbidden region.

In the special case when the Hamiltonian has the form $\mathcal{H}^{nm}_{\text{kin}}+V_{n}\delta^{nm}$,
with the kinetic term of the form~(\ref{eq:kin-term-example_position-dependent-hopping})
with $J_{n}=\text{const}$, the cross-term $\frac{1}{2}\partial_{x}\partial_{p}\mathcal{H}-i\partial_{a}\mathcal{H}$
vanishes, and one arrives at the standard form of the WKB ansatz:
\begin{equation}
\psi_{n}=\frac{\text{const}}{\sqrt{\left|v(x)\right|}}\exp\left\{ \frac{i}{a}\intop^{x}_{x_{0}}p(x)\:dx\right\} ,\,\,x=an,
\label{eq:wave-function_semiclassical-approximation}
\end{equation}
where $v(x)=\partial_{p}\mathcal{H}\left(x,p(x)\right)$ is the classical
velocity.

Importantly, Eq.~(\ref{eq:semiclassical-series}) is an asymptotic
series whose applicability is controlled by the two initial assumptions
of locality and smoothness of the Hamiltonian. \emph{A~posteriori},
these assumptions are expressed in the requirement for the de~Broglie
wavelength $\lambda(x)=2\pi a/p(x)$ to change slowly with $x=an$
at the scale of the wavelength itself:
\begin{equation}
\frac{d\lambda}{dx}\lambda\ll\lambda\Leftrightarrow a\,\frac{1}{p^{2}}\frac{\partial_{x}\mathcal{H}}{\partial_{p}\mathcal{H}}\ll1.
\label{eq:semiclassics_applicability-condition}
\end{equation}
While this condition is generally satisfied due to smallness of $a$,
it is clearly violated if either $p$ or $v=\partial_{p}\mathcal{H}$
approaches zero. The points where this happens are, by definition,
the classical turning points, where the corresponding classical particle
switches the direction of motion. In the vicinity of these points,
the semiclassical expansion~(\ref{eq:semiclassical-series}) is not
applicable, and one has to employ the matching procedure of Refs.~\citep{berry-1972_semiclassical-approximation,braun-1993_discrete-WKB}. 

\subsection{Matching conditions and quantization rules for the cosine spectrum\label{subsecapp:Matching-and-quantization-conditions}}

For the particular case of Hamiltonian~(\ref{eq:single-particle_Hamiltonian}),
the Maslov symbol is given by Eq.~(\ref{eq:classical-hamiltonian}),
and the positions of the classical turning points are described by
$E-h(x)=\pm J$. A turning point with $dp/dx\neq0$ splits the $x$-axis
into a classically allowed and classically forbidden region. At sufficiently
large distances from the turning point, the semiclassical solution
becomes applicable, but one still has to connect the amplitudes of
the semiclassical wave functions on either side of the turning point.
To this end, one needs to provide an analytical solution for the vicinity
of the turning point. These two solutions are then matched together
by their asymptotic forms. Importantly, there has to be an overlap
in the regions of applicability of the two approximations with spatial
sizes that fit many de~Broglie wavelengths, i.e., the phase $\frac{1}{a}\intop^{x}p(x)dx\propto\intop^{x}dx/\lambda(x)\gg1$
as one traverses the overlap region. 

The vicinity of the critical point can be analyzed in the important
case of a smooth potential that can be expanded as $h(x)\approx h_{0}+\alpha\left(x-x_{0}\right)$
in a sufficiently large neighborhood of the classical turning point
$x_{0}$. One then distinguishes two cases: a standard turning point
$p=0$, corresponding to $h_{0}=E+J$, and an anomalous turning point,
for which one has $h_{0}=E-J$. The existence of anomalous turning
points leads to a different structure of the classically allowed region
in comparison to the textbook case of quadratic dispersion law. In
particular, if the potential $h(x)$ has a local minimum at $h\left(x_{\min}\right)=h_{\min}$,
finite bandwidth of the kinetic term implies that at energies higher
than $J+h_{\min}$ one expects a forbidden region to appear in the
middle of the well. The quantization conditions involving anomalous
turning points are then required to accurately describe the energy
levels.

\subsubsection{A Standard Turning Point}

Let $\alpha>0$ be the potential gradient, and let $E$ satisfy $\alpha\,a\delta=E+J$
for some $\delta\in\left[0,1\right]$ (one can always shift the enumeration
of sites to force the exact position of the turning point to lie in
$[0,1]$). The exact Schrödinger equation~(\ref{eq:Stationary-Schroedinger-eq})
reads
\begin{equation}
-\frac{J}{2}\left[\psi_{n+1}+\psi_{n-1}\right]+\alpha\,an\,\psi_{n}=\left(\alpha\,a\delta-J\right)\psi_{n}.
\label{eq:linearized-potential_schroedinger-eq}
\end{equation}
Although this equation is exactly solvable in terms of Bessel functions,
we will resort to a controllable approximation that exploits the smoothness
of the potential and corresponds to an asymptotic expansion of the
exact solution in powers of the potential gradient $\alpha a\ll1$.
Near $n=0$, we expect a solution that is a smooth function of $x=an$,
as the relevant classical momenta are small due to the proximity to
the standard turning point $x_{0}=a\delta$ where $p\left(x_{0}\right)=0$.
One can then replace the discrete difference $\psi_{n+1}+\psi_{n-1}-2\psi_{n}$
by its continuous approximation $a^{2}d^{2}\psi/dx^{2}+O\left(a^{4}\right)$,
which is equivalent to replacing the true $p$-dependence of the Maslov
symbol~(\ref{eq:classical-hamiltonian}) with its quadratic approximation
near $p=0$. The criterion of applicability for such a replacement
is given by the smallness of the momentum that validates the quadratic
expansion:
\begin{equation}
\left|p\right|^{2}\approx\left|\frac{E+J-\alpha a\,x}{J}\right|\ll1\Leftrightarrow\frac{\alpha}{J}\left|x-a\delta\right|\ll1,
\label{eq:quadratic-approximation-condition}
\end{equation}
where the momentum is found from the energy conservation. Performing
the substitution, one obtains the following differential equation:
\begin{equation}
-\frac{Ja^{2}}{2}\,\psi''(x)+\alpha x\,\psi(x)=\alpha\,a\delta\,\psi(x),
\label{eq:Schroedinger-eq_continuos-appr_linear-potential}
\end{equation}
which is exactly solvable in terms of the Airy functions:
\begin{equation}
\psi(x)=C_{1}\text{Ai}\left(K(x-a\delta)\right)+C_{2}\text{Bi}\left(K(x-a\delta)\right),\,\,\,K=\left(\frac{2\alpha}{Ja^{2}}\right)^{1/3}\gg1.
\label{eq:approximate-solution}
\end{equation}
Despite condition~(\ref{eq:quadratic-approximation-condition}),
$K(x-a\delta)$ can attain large values, so it is legitimate to use
the asymptotic expansion of the Airy functions in the region $K^{-1}\ll\left|x-a\delta\right|\ll(\alpha/J)^{-1}$
:
\begin{equation}
\psi_{n}\approx\begin{cases}
z>0: & C_{1}\,\frac{1}{2\sqrt{\pi}}\sqrt[4]{\frac{1}{z}}\exp\left(-\frac{2}{3}z^{3/2}\right)+C_{2}\,\frac{1}{\sqrt{\pi}}\sqrt[4]{\frac{1}{z}}\exp\left(+\frac{2}{3}z^{3/2}\right),\\
z<0: & C_{1}\,\frac{1}{\sqrt{\pi}}\sqrt[4]{\frac{1}{\left|z\right|}}\cos\left(-\frac{2}{3}\left|z\right|^{3/2}+\frac{\pi}{4}\right)+C_{2}\,\frac{1}{\sqrt{\pi}}\sqrt[4]{\frac{1}{\left|z\right|}}\cos\left(\frac{2}{3}\left|z\right|^{3/2}+\frac{\pi}{4}\right),
\end{cases}\,\,z=K(x-a\delta).
\label{eq:exact-solution_asymptotics}
\end{equation}
On the other hand, WKBA can be applied in the region described by
condition~(\ref{eq:semiclassics_applicability-condition}), which
amounts to $K\left|x-a\delta\right|\gg1$. The semiclassical wave
functions then reads:
\begin{equation}
\psi\left(x\right)\approx\begin{cases}
\text{allowed region}: & \frac{1}{\sqrt{\left|v(x)\right|}}\exp\left\{ \pm\frac{i}{a}\intop^{x}_{a\delta}p(x)dx\right\} ,\\
\text{forbiden region}: & \frac{1}{\sqrt{\left|v(x)\right|}}\exp\left\{ \pm\frac{1}{a}\intop^{x}_{a\delta}\left|p(x)\right|dx\right\} ,
\end{cases}
\end{equation}
where $p(x)=\arccos\left(h(x)-E\right)/J$, and $v(x)=J\sin p(x)$.
Given condition~(\ref{eq:quadratic-approximation-condition}), one
can expand $\arccos1-u\approx\sqrt{2u}$, which is equivalent to replacing
the kinetic energy with its quadratic approximation, so the semiclassical
approximation for the wave function is given by:
\begin{equation}
\psi_{n}\approx\begin{cases}
z>0: & C_{\text{decaying}}\frac{1}{z^{1/4}}\exp\left\{ -\frac{2}{3}z^{3/2}\right\} +C_{\text{growing}}\frac{1}{z^{1/4}}\exp\left\{ +\frac{2}{3}z^{3/2}\right\} ,\\
z<0: & C_{\text{towards}}\frac{1}{\left|z\right|^{1/4}}\exp\left\{ +\frac{2i}{3}\left|z\right|^{3/2}\right\} +C_{\text{away}}\frac{1}{\left|z\right|^{1/4}}\exp\left\{ -\frac{2i}{3}\left|z\right|^{3/2}\right\} ,
\end{cases}\,\,\,\,z=K(x-a\delta),
\label{eq:semiclassical-approximation_asymptotics}
\end{equation}
where the labels of the $C$ coefficients denote the direction of
propagation relative to the turning point according to the sign of
the group velocity $v=J\sin p\approx Jp$ for $z>0$, and the behavior
with the distance from the turning point for $z<0$. Both expressions~(\ref{eq:exact-solution_asymptotics})~and~(\ref{eq:semiclassical-approximation_asymptotics})
are applicable in the same region, which contains $\frac{2}{3}\left(Ka\right)^{-1}$
de~Broglie wavelengths, so the matching procedure is valid if $Ka\ll1\Leftrightarrow\alpha a/J\ll1$,
which is yet another expression of the potential smoothness. Matching
the two asymptotic expressions then yields:
\begin{equation}
\begin{cases}
n>\delta: & C_{\text{decaying}}=C_{1}\,\frac{1}{2\sqrt{\pi}},\,\,\,C_{\text{growing}}=C_{2}\,\frac{1}{\sqrt{\pi}},\\
n<\delta: & C_{\text{towards}}=C_{1}\,\frac{1}{\sqrt{\pi}}\frac{e^{i\pi/4}}{2}+C_{2}\,\frac{1}{\sqrt{\pi}}\frac{e^{-i\pi/4}}{2},\,\,\,C_{\text{away}}=C_{1}\,\frac{1}{\sqrt{\pi}}\frac{e^{-i\pi/4}}{2}+C_{2}\,\frac{1}{\sqrt{\pi}}\frac{e^{i\pi/4}}{2}.
\end{cases}
\end{equation}
Excluding $C_{1},\,C_{2}$, one obtains the well-known~\citep{berry-1972_semiclassical-approximation}
connection between the semiclassical expansion coefficients on either
side of the turning point:
\begin{equation}
\begin{cases}
C_{\text{towards}}=C_{\text{decaying}}e^{i\pi/4}+C_{\text{growing}}\frac{e^{-i\pi/4}}{2},\\
C_{\text{away}}=C_{\text{decaying}}e^{-i\pi/4}+C_{\text{growing}}\frac{e^{i\pi/4}}{2}.
\end{cases}
\label{eq:standard-turning-point_matching-conditions}
\end{equation}
Although these relations are obtained for $\alpha>0$, they remain
identical for the case $\alpha<0$ upon identifying the growing and
decaying solutions in the forbidden region and the left- and right-propagating
waves in the allowed region.

\subsubsection{An Anomalous Turning Point}

Let $\alpha>0$ be the potential gradient, and let $E$ satisfy $\alpha\,a\delta=E-J$
for some $\delta\in\left[0,1\right]$. This corresponds to an anomalous
turning point $x_{0}=a\delta$ described by $p\left(x_{0}\right)=\pm\pi$,
with $x>a\delta$ being the classically allowed region and $x<a\delta$
being the classically forbidden one. To solve Eq.~(\ref{eq:Stationary-Schroedinger-eq})
for this case, we first perform the substitution $\psi_{n}=\eta_{n}(-1)^{n}$,
which maps the problem to the previous case, Eq.~(\ref{eq:linearized-potential_schroedinger-eq}),
with the inverted sign of $\alpha$:
\begin{equation}
-\frac{J}{2}\left[\psi_{n+1}+\psi_{n-1}\right]+\alpha\,an\,\psi_{n}=(\alpha\,a\delta+J)\psi_{n}\Leftrightarrow-\frac{J}{2}\left[\eta_{n+1}+\eta_{n-1}\right]-\alpha\,an\,\eta_{n}=(-\alpha\,a\delta-J)\eta_{n}.
\end{equation}
The latter equation is again solved by the Airy functions, so in the
region $K^{-1}\ll\left|x-a\delta\right|\ll\left(\alpha/J\right)^{-1}$
the asymptotic form of the solution is given by:
\begin{equation}
\eta_{n}=\psi_{n}(-1)^{n}=\begin{cases}
z<0: & C_{1}\,\frac{1}{2\sqrt{\pi}}\sqrt[4]{\frac{1}{\left|z\right|}}\exp\left(-\frac{2}{3}\left|z\right|^{3/2}\right)+C_{2}\,\frac{1}{\sqrt{\pi}}\sqrt[4]{\frac{1}{\left|z\right|}}\exp\left(\frac{2}{3}\left|z\right|^{3/2}\right),\\
z>0: & C_{1}\,\frac{1}{\sqrt{\pi}}\sqrt[4]{\frac{1}{z}}\cos\left(-\frac{2}{3}z^{3/2}+\frac{\pi}{4}\right)+C_{2}\,\frac{1}{\sqrt{\pi}}\sqrt[4]{\frac{1}{z}}\cos\left(\frac{2}{3}z^{3/2}+\frac{\pi}{4}\right),
\end{cases}\,\,\,z=-K(x-a\delta),
\label{eq:anomalous-turning-point-exact-solution_asymptotic}
\end{equation}
where $K=\left(2\alpha/Ja^{2}\right)^{1/3}$. On the other hand, the
semiclassical momentum is close to $\pm\pi$, so the semiclassical
wave function is described by:
\begin{equation}
\psi\left(x\right)\approx\begin{cases}
\text{allowed region}: & \frac{1}{\sqrt{\left|v(x)\right|}}\exp\left\{ \pm\frac{i}{a}\intop^{x}_{a\delta}\left[\pi-\delta p(x)\right]dx\right\} ,\\
\text{forbiden region}: & \frac{1}{\sqrt{\left|v(x)\right|}}\exp\left\{ \pm\frac{1}{a}\intop^{x}_{a\delta}\left[i\pi+\left|\delta p(x)\right|\right]dx\right\} ,
\end{cases}
\end{equation}
where $\delta p=\pi-\arccos\left\{ \left(h(x)-E+J\right)/J-1\right\} \approx\sqrt{2\left[E-J-h(x)\right]/J}$
(a positive quantity in the allowed region), while the group velocity
reads $v=J\sin\left(\pi-\delta p\right)\approx J\delta p$. The anomalous
turning point is characterized by the fact that the real contribution
to the phase $\pm i\pi(x-a\delta)/a=\pm i\pi(n-\delta)$ is present
on both sides of the turning point and causes oscillatory behavior
of the wave function. The semiclassical solution then reads
\begin{equation}
\psi_{n}\approx\begin{cases}
z<0: & C_{\text{decaying}}\frac{1}{\left|z\right|^{1/4}}\exp\left\{ +i\pi\frac{x-a\delta}{a}-\frac{2}{3}\left|z\right|^{3/2}\right\} +C_{\text{growing}}\frac{1}{\left|z\right|^{1/4}}\exp\left\{ -i\pi\frac{x-a\delta}{a}+\frac{2}{3}\left|z\right|^{3/2}\right\} ,\\
z>0: & C_{\text{towards}}\frac{1}{\left|z\right|^{1/4}}\exp\left\{ -i\pi\frac{x-a\delta}{a}+\frac{2i}{3}z^{3/2}\right\} +C_{\text{away}}\frac{1}{\left|z\right|^{1/4}}\exp\left\{ +i\pi\frac{x-a\delta}{a}-\frac{2i}{3}z^{3/2}\right\} ,
\end{cases}\,\,\,z=-K(x-a\delta),
\end{equation}
where the constant labels are again chosen to reflect the characteristic
behavior of the wave functions (e.g., the group velocity is positive
for the right-propagating wave moving away from the turning point).
Performing the matching procedure in the region $K^{-1}\ll\left|x-a\delta\right|\ll\left(\alpha/J\right)^{-1}$
yields:
\begin{equation}
\begin{cases}
C_{\text{towards}}=C_{\text{decaying}}\exp\left\{ -2i\pi\delta\right\} e^{-i\frac{\pi}{4}}+C_{\text{growing}}\frac{1}{2}e^{i\frac{\pi}{4}},\\
C_{\text{away}}=C_{\text{decaying}}e^{i\frac{\pi}{4}}+C_{\text{growing}}\exp\left\{ +2i\pi\delta\right\} \frac{1}{2}e^{-i\frac{\pi}{4}}.
\end{cases}
\label{eq:matching-conditions_anomalous-turning-point}
\end{equation}
where we have taken into account that $\exp\left\{ \pm i\pi x/a\right\} =\exp\left\{ \pm i\pi n\right\} =(-1)^{n}.$

One important feature of this matching condition in comparison to
the standard one, Eq.~(\ref{eq:standard-turning-point_matching-conditions}),
is that it includes the additional phase factor featuring the exact
position $\delta\in\left[0,1\right]$ of the turning point, which
has been a point of controversy~\citep{braun-1993_discrete-WKB}.
While one might argue that the potential is not well-defined at non-integer
coordinate values of $n=x/a$, it becomes well-defined after one requires
it to be a smooth function of the coordinate (e.g., by imposing a
condition on its Fourier spectrum). Due to the fact that the wave
function has momentum $p=\pm\pi$ at the turning point, this exact
coordinate becomes essential. It is important for an accurate description
of the quantization conditions.

\subsubsection{Quantization Conditions\label{subsecapp:Quantization-conditions}}

For a given energy $E$, consider a potential that arranges a bounded
classically allowed region $x_{1}<x<x_{2}$, where $x_{1,2}$ are
turning points. If one does not consider tunneling outside this region,
the matching conditions allow one to derive the quantization rules
for the bound states. A bound state is defined as a purely decaying
wave in the forbidden region close to both of the turning points $x_{i},\,i=1,2$:
\begin{equation}
C^{\left(i\right)}_{\text{grow}}=0\Rightarrow\frac{C^{\left(i\right)}_{\text{towards}}}{C^{\left(i\right)}_{\text{away}}}=\begin{cases}
e^{i\pi/2}, & p\left(x_{i}\right)=0,\\
e^{-i\pi/2}\exp\left\{ -2i\pi\delta^{\left(i\right)}\right\} , & p\left(x_{i}\right)=\pm\pi.
\end{cases}
\label{eq:coefficient-phase-relation_from-matching}
\end{equation}
where $\delta^{\left(i\right)}\in\left[0,1\right]$ is the exact position
of the anomalous turning point. Obviously, one has $\left|C^{\left(i\right)}_{\text{towards}}/C^{\left(i\right)}_{\text{away}}\right|=1$,
so the phase $\theta_{i}=\text{arg}C^{\left(i\right)}_{\text{towards}}/C^{\left(i\right)}_{\text{away}}$
defines the shape of the semiclassical wave function in the allowed
region up to an overall phase counted from the corresponding turning
point:
\begin{equation}
\psi^{\left(1\right)}\left(x\right)=\frac{C^{\left(1\right)}_{\text{away}}}{\sqrt{\left|v\left(x\right)\right|}}\exp\left\{ \frac{i}{a}\intop^{x}_{x_{1}}p\left(x\right)dx\right\} +\frac{C^{\left(1\right)}_{\text{towards}}}{\sqrt{\left|v\left(x\right)\right|}}\exp\left\{ -\frac{i}{a}\intop^{x}_{x_{1}}p\left(x\right)dx\right\} =\frac{2C^{\left(1\right)}_{\text{away}}e^{\theta_{1}/2}}{\sqrt{\left|v\left(x\right)\right|}}\cos\left\{ \frac{1}{a}\intop^{x}_{x_{1}}p\left(x\right)dx-\frac{\theta_{1}}{2}\right\} ,
\label{eq:semiclassical-wave-function_bound-state_left-turning-point}
\end{equation}
\begin{equation}
\psi^{\left(2\right)}\left(x\right)=\frac{C^{\left(2\right)}_{\text{towards}}}{\sqrt{\left|v\left(x\right)\right|}}\exp\left\{ \frac{i}{a}\intop^{x}_{x_{2}}p\left(x\right)dx\right\} +\frac{C^{\left(2\right)}_{\text{away}}}{\sqrt{\left|v\left(x\right)\right|}}\exp\left\{ -\frac{i}{a}\intop^{x}_{x_{2}}p\left(x\right)dx\right\} =\frac{2C^{\left(2\right)}_{\text{away}}e^{i\theta_{2}/2}}{\sqrt{\left|v\left(x\right)\right|}}\cos\left\{ \frac{1}{a}\intop^{x}_{x_{2}}p\left(x\right)dx+\frac{\theta_{2}}{2}\right\} 
\label{eq:semiclassical-wave-function_bound-state_right-turning-point}
\end{equation}
where we have taken into account $x_{1}<x<x_{2}$ , and $p$ is chosen
to be the positive root of the classical equation of motion~(\ref{eq:classical-eq-of-motion_app})
in the vicinity of the turning point $x_{i}$ (so the corresponding
group velocity is also positive). The two expressions for $i=1,2$
must yield the same wave function (up to an overall phase), which
is equivalent to requiring that the difference of phases of the cosines
is $\pi N$, yielding
\begin{equation}
\frac{1}{a}\intop^{x_{2}}_{x_{1}}p(x)dx-\frac{\theta_{1}}{2}-\frac{\theta_{2}}{2}=\pi N,
\label{eq:quantization-condition}
\end{equation}
which is equivalent to Eqs.~(\ref{eq:Bohr-quantization-condition})~and~(\ref{eq:anomalous-phases-def})
upon substitution of $\theta_{i}=\text{arg}C^{\left(i\right)}_{\text{towards}}/C^{\left(i\right)}_{\text{away}}$
from Eq.~(\ref{eq:coefficient-phase-relation_from-matching}). 

Let us illustrate this result for the exactly solvable case of a linear
potential $h_{n}=\alpha an$ with $a\ll1$. The exact eigenvalues
of the corresponding Schrödinger equation~(\ref{eq:Stationary-Schroedinger-eq})
are given by $E_{N}=\frac{J}{\alpha}+\alpha aN,\,\,N\in\mathbb{Z}$,
and the associated eigenfunctions $\left|N\right>$ read:
\begin{equation}
\left\langle n|N\right\rangle =J_{n-N}\left(\frac{J}{\alpha}\right),
\label{eq:linear-potential_exact-solution}
\end{equation}
where $J_{k}(x)$ is the Bessel function of order~$k$ (which reduces
to the special cases of Eqs.~(\ref{eq:exact-solution_asymptotics})~and~(\ref{eq:anomalous-turning-point-exact-solution_asymptotic})
with $C_{2}=0$ near the corresponding turning points for $a\ll1$).
From the semiclassical point of view, at energy $E$ the classically
allowed region of Hamiltonian $-J\cos p+\alpha x$ is given by $x_{1}<x<x_{2}$
with $x_{1,2}=\left(E\mp J\right)/\alpha$, and $x_{1}$ is the anomalous
turning point with $p\left(x_{1}\right)=\pm\pi$, while $x_{2}$ is
the standard turning point with $p\left(x_{2}\right)=0$. Applying
Eqs.~(\ref{eq:coefficient-phase-relation_from-matching})~and~(\ref{eq:quantization-condition})
then renders $\theta_{2}=+\pi/2$, and $\theta_{1}=-\pi/2-2\pi\left\{ x_{1}/a\right\} $,
with $\{n\}$ being the fractional part of $n$, so
\begin{equation}
\frac{1}{a}\intop^{x_{2}}_{x_{1}}\arccos\left\{ \frac{\alpha x-E}{J}\right\} dx+\pi\left\{ x_{1}/a\right\} =\pi N,
\end{equation}
The integral of momentum is $\pi J/\alpha$, so one obtains an equation
for the value of $E$:
\begin{equation}
\frac{J}{\alpha a}+\left\{ \frac{E}{\alpha a}-\frac{J}{\alpha a}\right\} \in\mathbb{Z},
\end{equation}
which is solved by the correct value $E=\alpha aN$. Crucially, the
correct quantization condition can be traced down to the presence
of the additional dependence of the matching conditions~(\ref{eq:coefficient-phase-relation_from-matching})
on~$\delta$.

\section{Semiclassical Description of Trotterization\label{app:Matrix-WKB}}

For finite~$\delta t$, the wave function is transformed by finite-time
unitary operators. To generalize the semiclassical approach to this
case, we will derive the generalization of the semiclassical effective
Hamiltonian and the transport equation. There are two qualitative
differences with the $\delta t\to0$ case: \emph{i)}~doubling of
the translation period of the uniform system, and \emph{ii)}~non-commutativity
of the unitary operators. This Appendix covers the technical details
of the semiclassical approximation for the matrix Hamiltonians following
the approach of Ref.~\citep[Ch. 11]{maslov-2001-semiclassical-approx}. 

\subsection{Matrix Semiclassical Description and Maslov Symbol\label{secapp:Matrix-semiclassical-description}}

\subsubsection{Motivation: The Kinetic Term}

The one-step Suzuki-Trotter evolution operator $\mathcal{U}_{\text{even}(\text{odd})}=\exp\left\{ -i\delta t\mathcal{K}_{\text{even}(\text{odd})}\right\} $
corresponding to the even (odd) part of the kinetic term $\mathcal{K}_{\text{even}(\text{odd})}$
has the following matrix elements:
\begin{align}
\mathcal{U}^{nm}_{\text{even}(\text{odd})} & =\cos\gamma\,\delta^{nm}+i\sin\gamma\,\left\{ \delta^{n+1,m}\,\delta_{n,\text{even}(\text{odd})}+\delta^{n-1,m}\,\delta_{n,\text{odd}(\text{even})}\right\} \nonumber \\
 & =\cos\gamma\,\delta^{nm}+i\sin\gamma\,\frac{1}{2}\left\{ \delta^{n+1,m}+\delta^{n-1,m}\right\} \mp(-1)^{n}\,i\sin\gamma\,\frac{1}{2}\left\{ \delta^{n-1,m}-\delta^{n+1,m}\right\} ,
\label{eq:evolution-operator_even/odd_kinetic-term}
\end{align}
where $\gamma=J\delta t/2$. The term $\propto(-1)^{n}$ in the last
expression instantiates the staggered structure of the evolution operators
that doubles the spatial translation period (see also \appref{trotter_perturb-corrections}).
Instead of a plain wave $\psi_{n}=e^{ipn}$, eigenfunctions now have
to be described by two components:
\begin{equation}
\psi_{n}=e^{ipn}\varphi_{+}+(-1)^{n}e^{ipn}\varphi_{-}\equiv e^{ipn}\,\begin{pmatrix}1 & (-1)^{n}\end{pmatrix}\cdot\overrightarrow{\varphi},\,\,\,\overrightarrow{\varphi}:=\begin{pmatrix}\varphi_{+}\\
\varphi_{-}
\end{pmatrix},
\end{equation}
where we introduced the component vector $\overrightarrow{\varphi}$.
$\mathcal{U}_{\text{even}(\text{odd})}$ acts on the components of
these functions by a $2\times2$ matrix:
\begin{equation}
\sum_{m}\mathcal{U}^{nm}_{\text{even}(\text{odd})}\psi_{m}=e^{ipn}\,\begin{pmatrix}1 & (-1)^{n}\end{pmatrix}\cdot\mathcal{\hat{U}}_{\text{even}(\text{odd})}(p)\cdot\overrightarrow{\varphi},
\end{equation}
\begin{equation}
\mathcal{\hat{U}}_{\text{even}(\text{odd})}(p)=\begin{pmatrix}\cos\gamma+i\sin\gamma\,\cos p & \pm\sin\gamma\,\sin p\\
\mp\sin\gamma\,\sin p & \cos\gamma-i\sin\gamma\,\cos p
\end{pmatrix}=\hat{\tau}_{0}\,\cos\gamma+i\hat{\tau}_{z}\sin\gamma\,\cos p\pm i\hat{\tau}_{y}\sin\gamma\,\sin p,
\label{eq:kinetic-term_Maslov-symbol}
\end{equation}
where we introduced the Pauli matrices $\hat{\tau}_{i}$ acting on
the component vectors. Note that the $\mathcal{\hat{U}}_{\text{even}(\text{odd})}(p)$
matrix is unitary for $p\in\mathbb{R}$. 

\subsubsection{The Semiclassical Series}

The example above dictates the proper generalization of the semiclassical
description. Let $\mathcal{O}$ be any operator with a staggered structure
of the matrix elements:
\begin{equation}
\mathcal{O}_{nm}=\mathcal{S}_{nm}+(-1)^{n}\mathcal{A}_{nm},
\end{equation}
with $\mathcal{S},\mathcal{A}$ being smooth functions of $(n+m)/2$
and rapidly decaying functions of $n-m$. The matrix generalization
of the Maslov symbol, Eq.~(\ref{eq:maslov-symbol-def}), is then
defined as:
\begin{equation}
\hat{\mathcal{O}}(x=an,p):=\begin{pmatrix}\mathcal{S}(x,p) & \mathcal{A}(x,p+\pi)\\
\mathcal{A}(x,p) & \mathcal{S}(x,p+\pi)
\end{pmatrix},
\label{eq:matrix-maslov-symbol_def}
\end{equation}
where $\mathcal{S}(x,p),\,\mathcal{A}(x,p)$ are the corresponding
scalar Maslov symbols, Eq.~(\ref{eq:maslov-symbol-def}). We will
henceforth use the hat to denote the \emph{(left) Maslov symbol} $\hat{\mathcal{O}}$
obtained in this way from operator $\mathcal{O}$. The semiclassical
wave function, Eq.~(\ref{eq:semiclassical-series}), is generalized
as
\begin{equation}
\psi_{n}=\exp\left\{ \frac{iS(an)}{a}\right\} \left(\varphi_{+}(an)+(-1)^{n}\,\varphi_{-}(an)\right)\equiv\exp\left\{ \frac{iS(an)}{a}\right\} \begin{pmatrix}1 & (-1)^{n}\end{pmatrix}\,\overrightarrow{\varphi}(x),\,\,\,\overrightarrow{\varphi}(x):=\begin{pmatrix}\varphi_{+}(an)\\
\varphi_{-}(an)
\end{pmatrix},
\label{eq:semiclassical_wave-vector}
\end{equation}
where $S(x),\,\overrightarrow{\varphi}(x)$ are assumed to be smooth
functions of $x$. The corresponding Brillouin zone is halved to $p=S'(x)\in\left[-\pi/2,\pi/2\right]$
to preserve the total number of states, and the redundancy of $\left|p\right|>\pi/2$
values is encoded in the gauge transform: $S\to S+\pi x$ is undone
by $\overrightarrow{\varphi}\to\hat{\tau}_{x}\overrightarrow{\varphi}$.
Maslov symbols of physical operators duly respect this symmetry: $\hat{\mathcal{O}}(x,p+\pi)=\hat{\tau}_{x}\hat{\mathcal{O}}(x,p)\hat{\tau}_{x}$. 

Acting with $\mathcal{O}$ on such a wave function results in a generalization
of Eq.~(\ref{eq:action-of-operator-on-semiclassical-wave-function}):
\begin{equation}
\sum_{m}\mathcal{O}_{nm}\psi_{m}=\exp\left\{ \frac{iS(an)}{a}\right\} \,\begin{pmatrix}1 & (-1)^{n}\end{pmatrix}\,\left[\hat{\mathcal{O}}-ia\left\{ \partial_{p}\hat{\mathcal{O}}\,\frac{d}{dx}+\frac{1}{2}\,\partial_{x}p\,\partial^{2}_{p}\hat{\mathcal{O}}+i\partial_{a}\hat{\mathcal{O}}\right\} +O\left(a^{2}\right)\right]\,\overrightarrow{\varphi}(x),
\label{eq:action-of-operator_with-oscillating-matrix-elems_on-semiclassical-wf}
\end{equation}
where we omit the arguments of $\hat{\mathcal{O}}$ for brevity, and
the derivative $\partial\hat{\mathcal{O}}/\partial a$ reflects possible
smooth dependence of $\hat{\mathcal{O}}$ on $a$. Crucially, this
yields back the semiclassical wave function with coefficients $\overrightarrow{\varphi}$
transformed as
\begin{equation}
\overrightarrow{\varphi}(x)\overset{\mathcal{O}}{\longmapsto}\left[\hat{\mathcal{O}}+a\mathcal{\hat{O}}^{(1)}+a\hat{\mathcal{O}}^{(x)}\frac{d}{dx}+O\left(a^{2}\right)\right]\,\overrightarrow{\varphi}(x).
\label{eq:matrix-Maslov-symbol_power-series_action}
\end{equation}
We will refer to the expression in front of $\overrightarrow{\varphi}$
in the right hand side as the \emph{semiclassical series} of operator~$\mathcal{O}$. 

The subleading matrix coefficients $\hat{\mathcal{O}}^{(x)},\,\hat{\mathcal{O}}^{(1)}$
are expressed in terms of $\hat{\mathcal{O}}(x,p)$ and its derivatives
with respect to $x,\,p,\,a$ according to Eq.~(\ref{eq:action-of-operator_with-oscillating-matrix-elems_on-semiclassical-wf}):
\begin{equation}
\mathcal{\hat{O}}^{(1)}:=-i\,\frac{1}{2}\,\partial_{x}p\,\partial^{2}_{p}\hat{\mathcal{O}}+\partial_{a}\hat{\mathcal{O}},\,\,\,\hat{\mathcal{O}}^{(x)}:=-i\partial_{p}\hat{\mathcal{O}}.
\label{eq:O-expansion-terms}
\end{equation}

The matrix dimension of the Maslov symbol depends only on the number
of qubits in the largest operator (always 2 qubits for this work);
it does not depend on the order of the Trotter approximation. 

\subsubsection{Conjugate Semiclassical Series}

The analysis is completed by repeating the calculation for the action
on bra states $\left<\psi\right|$. The counterpart of Eq.~(\ref{eq:action-of-operator_with-oscillating-matrix-elems_on-semiclassical-wf})
reads
\begin{equation}
\sum_{n}\psi^{*}_{n}\mathcal{O}_{nm}=\exp\left\{ -\frac{iS^{*}(am)}{a}\right\} \,\overrightarrow{\varphi}^{\dagger}(x)\,\left[\tilde{\mathcal{O}}+ia\,\left\{ \overleftarrow{\frac{d}{dx}}\,\partial_{p^{*}}\tilde{\mathcal{O}}+\frac{1}{2}\partial_{x}p\,\partial^{2}_{p^{*}}\tilde{\mathcal{O}}-i\partial_{a}\tilde{\mathcal{O}}\right\} +O\left(a^{2}\right)\right]\,\begin{pmatrix}1\\
(-1)^{n}
\end{pmatrix},
\label{eq:action-of-operator_conjugated}
\end{equation}
where $\overleftarrow{\frac{d}{dx}}$ acts on functions to the left
of it, $S^{*}{}'(x)=p^{*}$ (not necessarily equal to $p$ as it can
be imaginary), $\tilde{\mathcal{O}}(x,p^{*})$ is the \emph{right}
Maslov symbol defined as
\begin{equation}
\tilde{\mathcal{O}}(x,p^{*})=\begin{pmatrix}\tilde{\mathcal{S}}(x,p^{*}) & \tilde{\mathcal{A}}(x,p^{*}+\pi)\\
\tilde{\mathcal{A}}(x,p^{*}) & \tilde{\mathcal{S}}(x,p^{*}+\pi)
\end{pmatrix},
\end{equation}
and the right \emph{scalar} Maslov symbol is given by
\begin{equation}
\tilde{\mathcal{S}}(x=am,p^{*}):=\sum_{n}e^{-ip^{*}(n-m)}\mathcal{S}_{nm}\equiv\frac{\left\langle p^{*}\left|\mathcal{S}\right|m\right\rangle }{\left\langle p^{*}|m\right\rangle }.
\end{equation}
Importantly, right and left Maslov symbols---either scalar or matrix---do
not coincide in general. However, they are related by hermitian conjugation:
\begin{equation}
\left<\psi\right|\mathcal{O}^{\dagger}=\left(\mathcal{O}\left|\psi\right>\right)^{\dagger}\Rightarrow\tilde{\mathcal{O}}^{\dagger}=\hat{\mathcal{O}^{\dagger}},
\label{eq:left-right-symbol-relation}
\end{equation}
i.e., the matrix conjugate of right (tilded) Maslov symbol of $\mathcal{O}$
is given by the left (hatted) Maslov symbol of the hermitian conjugate
of $\mathcal{O}$. The two semiclassical series, Eqs.~(\ref{eq:action-of-operator_with-oscillating-matrix-elems_on-semiclassical-wf})
and~(\ref{eq:action-of-operator_conjugated}), are connected by the
same relation in all orders. In particular, the analogues of Eqs.~(\ref{eq:matrix-Maslov-symbol_power-series_action})--(\ref{eq:O-expansion-terms})
are given by
\begin{equation}
\overrightarrow{\varphi}^{\dagger}(x)\overset{\mathcal{O}}{\longmapsto}\overrightarrow{\varphi}^{\dagger}(x)\,\left[\tilde{\mathcal{O}}+a\tilde{\mathcal{O}}^{(1)}+a\overleftarrow{\frac{d}{dx}}\tilde{\mathcal{O}}^{(x)}+O\left(a^{2}\right)\right],
\label{eq:matrix-Maslov-symbol_power-series_action_conjugated}
\end{equation}
\begin{equation}
\tilde{\mathcal{O}}^{(1)}:=i\frac{1}{2}\partial_{x}p^{*}\,\partial^{2}_{p^{*}}\tilde{\mathcal{O}}+\partial_{a}\tilde{\mathcal{O}},\,\,\,\tilde{\mathcal{O}}^{(x)}:=i\,\partial_{p^{*}}\tilde{\mathcal{O}},
\label{eq:O-expansion-terms-conjugated}
\end{equation}
so one observes $\left[\tilde{\mathcal{O}}^{(1)}\right]^{\dagger}=\left[\hat{\mathcal{O}^{\dagger}}\right]^{(1)}$,
and similarly for $\tilde{\mathcal{O}}^{(x)}$, in accordance with
Eq.~(\ref{eq:left-right-symbol-relation}).

\subsection{Composition Rules for Matrix Maslov Symbols\label{secapp:Composition-rules}}

Composition of operators $\mathcal{A}\mathcal{B}$ induces composition
rules of semiclassical series in Eq.~(\ref{eq:matrix-Maslov-symbol_power-series_action}).
One can thus reduce operator products to products of corresponding
Maslov symbols and their derivatives. Directly computing the composition
yields the following values of the coefficients in Eq.~(\ref{eq:matrix-Maslov-symbol_power-series_action}):
\begin{equation}
\hat{\mathcal{A}\mathcal{B}}\equiv\hat{\mathcal{A}}\hat{\mathcal{B}},\,\,\,\left(\hat{\mathcal{A}\mathcal{B}}\right)^{(1)}=\left(\hat{\mathcal{A}}\hat{\mathcal{B}}\right)^{(1)}-i\left(\partial_{p}\hat{\mathcal{A}}\right)\left(\partial_{x}\hat{\mathcal{B}}\right),\,\,\,\left(\hat{\mathcal{A}\mathcal{B}}\right)^{(x)}=\left(\hat{\mathcal{A}}\hat{\mathcal{B}}\right)^{(x)},
\label{eq:semiclassic-composition-rules}
\end{equation}
where $\hat{\mathcal{A}\mathcal{B}}$ denotes the Maslov symbol for
the operator product $\mathcal{A}\mathcal{B}$, $\left(\hat{\mathcal{A}\mathcal{B}}\right)^{(1)},\,\left(\hat{\mathcal{A}\mathcal{B}}\right)^{(x)}$
are the associated coefficients in Eq.~(\ref{eq:matrix-Maslov-symbol_power-series_action}),
and $\hat{\mathcal{A}}\hat{\mathcal{B}}$ stands for the matrix product
of Maslov symbols, with $\left(\hat{\mathcal{A}}\hat{\mathcal{B}}\right)^{(1)},\,\left(\hat{\mathcal{A}}\hat{\mathcal{B}}\right)^{(x)}$
obtained by applying Eq.~(\ref{eq:O-expansion-terms}) to this product.
While the leading term is unsurprisingly given by the matrix product
of the constituting Maslov symbols, the coefficient $\left(\hat{\mathcal{A}\mathcal{B}}\right)^{(1)}$
differs from directly applying Eq.~(\ref{eq:O-expansion-terms})
to $\hat{\mathcal{A}}\hat{\mathcal{B}}$. This difference is responsible
for reproducing the quantum-classical correspondence between the commutator
and the Poisson bracket:
\begin{equation}
\hat{\left[\mathcal{A},\mathcal{B}\right]}=\left[\hat{\mathcal{A}},\hat{\mathcal{B}}\right],\,\,\,\left(\hat{\left[\mathcal{A},\mathcal{B}\right]}\right)^{(1)}=\left(\left[\hat{\mathcal{A}},\hat{\mathcal{B}}\right]\right)^{(1)}+i\left\{ \hat{\mathcal{A}},\hat{\mathcal{B}}\right\} ,
\end{equation}
\begin{equation}
\left\{ \hat{\mathcal{A}},\hat{\mathcal{B}}\right\} :=i\left[\left(\partial_{p}\hat{\mathcal{B}}\right)\left(\partial_{x}\hat{\mathcal{A}}\right)-\left(\partial_{p}\hat{\mathcal{A}}\right)\left(\partial_{x}\hat{\mathcal{B}}\right)\right].
\end{equation}
For scalar $\hat{\mathcal{A}},\hat{\mathcal{B}}$, this reduces to
the standard Poisson bracket.

\subsection{Logarithm of Matrix Maslov Symbols\label{secapp:operator-logarithm}}

Given an evolution operator $\mathcal{U}=\exp\left\{ -i\delta t\mathcal{H}\right\} $
and its semiclassical series~(\ref{eq:matrix-Maslov-symbol_power-series_action}),
the action of the corresponding Hamiltonian $\mathcal{H}$ on the
semiclassical wave function is also described a semiclassical series
of the form~(\ref{eq:matrix-Maslov-symbol_power-series_action}).
By definition, the two series are related as
\begin{equation}
\hat{\mathcal{U}}+a\mathcal{\hat{U}}^{(1)}+a\hat{\mathcal{U}}^{(x)}\,\frac{d}{dx}+O\left(a^{2}\right)=:\exp\left\{ -i\delta t\left[\hat{\mathcal{H}}+a\mathcal{\hat{H}}^{(1)}+a\hat{\mathcal{H}}^{(x)}\,\frac{d}{dx}+O\left(a^{2}\right)\right]\right\} .
\end{equation}
A straightforward calculation, using Eq.~(\ref{eq:O-expansion-terms}),
yields the leading expansion coefficients:
\begin{equation}
\hat{\mathcal{H}}=\frac{i}{\delta t}\ln\hat{\mathcal{U}},\,\,\,\,\hat{\mathcal{H}}^{(1)}=-i\,\frac{1}{2}\,\partial_{x}p\,\partial^{2}_{p}\hat{\mathcal{H}}+\partial_{a}\hat{\mathcal{H}}+\delta\mathcal{H}^{(1)},\,\,\,\hat{\mathcal{H}}^{(x)}=-i\,\partial_{p}\hat{\mathcal{H}},
\label{eq:matrix-Maslov-symbol-logarithm}
\end{equation}
\begin{equation}
\delta\hat{\mathcal{H}}^{(1)}=\partial_{p}\hat{\mathcal{H}}\,\left(i\,\partial_{x}\hat{\mathcal{U}}\right)\hat{\mathcal{U}}^{-1}+\hat{\mathcal{U}}i\,\left(\partial_{\delta t}\left\{ \hat{\mathcal{U}}^{-1}\delta\mathcal{\hat{U}}^{(1)}\right\} \right)\,\hat{\mathcal{U}}^{-1},
\label{eq:hamiltonian-subleading-term-correction}
\end{equation}
where $\partial_{\delta t}$ acts on the $\delta t$ dependence of
the expression inside $\{...\}$, and $\delta\mathcal{\hat{U}}^{(1)}=\mathcal{U}^{(1)}-\left\{ -i\,\frac{1}{2}\,\partial_{x}p\,\partial^{2}_{p}\hat{\mathcal{U}}+\partial_{a}\hat{\mathcal{U}}\right\} $
is the difference between the actual coefficient in the $\mathcal{U}$
series and the one computed by Eq.~(\ref{eq:O-expansion-terms})
from the Maslov symbol $\hat{\mathcal{U}}$. This term is present
when $\mathcal{U}$ itself is composed of several operators, according
to Eq.~(\ref{eq:semiclassic-composition-rules}). 

Equations~(\ref{eq:matrix-Maslov-symbol_power-series_action})--(\ref{eq:hamiltonian-subleading-term-correction})
are a generalization of Eq.~(\ref{eq:action-of-operator-on-semiclassical-wave-function})
for finite $\delta t$. Importantly, the leading term $\hat{\mathcal{H}}$
is expressed by the matrix logarithm of~$\hat{\mathcal{U}}$. In
particular, if $\hat{\mathcal{U}}$ is unitary for real $x,p$, $\hat{\mathcal{H}}$
is hermitian for real $x,p$. In addition, the entire extra term $\delta\hat{\mathcal{H}}^{(1)}$
is $O(\delta t)$ in the limit $\delta t\to0$, in which case Eq.~(\ref{eq:matrix-Maslov-symbol-logarithm})
reproduces the scalar case, Eq.~(\ref{eq:action-of-operator-on-semiclassical-wave-function}),
as expected from the logic of STA.

\subsection{Semiclassical Hamiltonian for a Smooth External Potential \label{secapp:Full-effective-Hamiltonian}}

We can now compute the Maslov symbols of interest. For the unitary
operator $\mathcal{U}^{nm}_{\text{pot}}=\exp\left\{ -i\delta th_{n}\right\} \delta^{nm}$,
the Maslov symbol is given by a scalar matrix:
\begin{equation}
\hat{\mathcal{U}}_{\text{pot}}(x,p)=\exp\left\{ -i\delta th(x)\right\} \,\hat{1},\,\,\,h(x)=h_{x/a},
\end{equation}
where $\hat{1}$ is the identity matrix. The Maslov symbol for the
unitary evolution corresponding to even and odd parts of the kinetic
term, Eq.~(\ref{eq:evolution-operator_even/odd_kinetic-term}), is
given by Eq.~(\ref{eq:kinetic-term_Maslov-symbol}). 

The Maslov symbol of the full evolution operator $\hat{\mathcal{U}}_{\text{tot}}$
is restored by composing the Maslov symbols of constituent operators
according to Eq.~(\ref{eq:semiclassic-composition-rules}). Crucially,
because the potential operator is a scalar matrix, only the relative
order of the kinetic terms matters for the Maslov symbol. The two
distinct options are given by:
\begin{align}
\hat{\mathcal{U}}_{\text{tot},1}= & \hat{\mathcal{U}}_{\text{pot}}(\delta t)\,\hat{\mathcal{U}}_{\text{even}}(\delta t/2)\,\hat{\mathcal{U}}_{\text{odd}}(\delta t)\,\hat{\mathcal{U}}_{\text{even}}(\delta t/2)=\exp\left\{ -i\delta th(x)-i\delta t\hat{\mathcal{K}}_{-}(p)\right\} ,\\
\hat{\mathcal{U}}_{\text{tot},2}= & \hat{\mathcal{U}}_{\text{pot}}(\delta t)\,\hat{\mathcal{U}}_{\text{odd}}(\delta t/2)\,\hat{\mathcal{U}}_{\text{even}}(\delta t)\,\hat{\mathcal{U}}_{\text{odd}}(\delta t/2)=\exp\left\{ -i\delta th(x)-i\delta t\hat{\mathcal{K}}_{+}(p)\right\} ,
\end{align}
where the matrices $\hat{\mathcal{K}}_{1,2}(p)$ differ only in the
sign of the term $\propto\hat{\tau}_{y}$:
\begin{equation}
\hat{\mathcal{K}}_{1,2}(p)=T(p)\,\frac{\mp\hat{\tau}_{y}\,\sin^{2}(\gamma/2)\,\sin2p+\hat{\tau}_{z}\left(1-2\sin^{2}(\gamma/2)\,\cos^{2}p\right)}{\sqrt{1-(\sin\gamma\,\cos p)^{2}}},
\label{eq:maslov-symbol_kinetic-term}
\end{equation}
\begin{equation}
T(p)=-\frac{2}{\delta t}\arcsin\left\{ \cos p\sin\gamma\right\} .
\label{eq:kinetic-suzuki-trotter_energy-spectrum}
\end{equation}
This sign can be absorbed into a gauge transform of the semiclassical
wave function, $\overrightarrow{\varphi}(x)\mapsto\hat{\tau}_{z}\overrightarrow{\varphi}(x)$,
which implies that $\hat{\mathcal{K}}_{1,2}$ lead to identical physics
at the semiclassical level. The Maslov symbol of the Hamiltonian is
calculated according to Eq.~(\ref{eq:matrix-Maslov-symbol-logarithm})
and reads
\begin{equation}
\hat{\mathcal{H}}_{\text{eff}}(x,p)=\hat{\mathcal{K}}_{1,2}(p)+h(x)\,\hat{1}.
\label{eq:semiclassical-matrix-Hamiltonian_smooth-potential}
\end{equation}
Importantly, unlike the leading term $\hat{\mathcal{H}}_{\text{eff}}$
and the $\hat{\mathcal{H}}^{(x)}_{\text{eff}}$ coefficient, the subleading
coefficient $\hat{\mathcal{H}}^{(1)}_{\text{eff}}$ depends on the
operator ordering, except in the $\delta t\to0$ limit (see \secappref{operator-logarithm}).

\subsection{Equation of Motion and Transport Equation\label{secapp:matrix_eq-of-motion_transport-eq}}

\subsubsection{Derivation}

Following Ref.~\citep{maslov-2001-semiclassical-approx}, we seek
the eigenfunctions of $\mathcal{H}$ (or, equivalently, $\mathcal{U}$)
in the semiclassical form~(\ref{eq:semiclassical_wave-vector}),
with $\overrightarrow{\varphi}$ as a formal power series of~$a$:
\begin{equation}
\overrightarrow{\varphi}(x)=\overrightarrow{\varphi}_{0}(x)+a\overrightarrow{\varphi}_{1}(x)+O\left(a^{2}\right).
\label{eq:semiclassical-envelope-series}
\end{equation}
By means of Eq.~(\ref{eq:matrix-Maslov-symbol_power-series_action}),
the stationary Schrödinger equation~(\ref{eq:Stationary-Schroedinger-eq})
then amounts to equality between two power series in $a$, with the
leading coefficients obeying:
\begin{equation}
\left(E\,\hat{1}-\hat{\mathcal{H}}_{\text{eff}}\right)\overrightarrow{\varphi}_{0}=0,
\label{eq:classical-matrix-eq-of-motion}
\end{equation}
\begin{equation}
\left(E\,\hat{1}-\hat{\mathcal{H}}_{\text{eff}}\right)\overrightarrow{\varphi}_{1}=\left\{ \hat{\mathcal{H}}^{(x)}_{\text{eff}}\frac{d}{dx}+\hat{\mathcal{H}}^{(1)}_{\text{eff}}\right\} \overrightarrow{\varphi}_{0},
\label{eq:matrix-subleading-order}
\end{equation}
where $\hat{\mathcal{H}}_{\text{eff}},\,\hat{\mathcal{H}}^{(x)}_{\text{eff}},\,\hat{\mathcal{H}}^{(1)}_{\text{eff}}$
are the leading coefficients in the semiclassical series~(\ref{eq:matrix-Maslov-symbol_power-series_action}).
To possess a nontrivial right eigenvector $\overrightarrow{\varphi}_{0}$,
system~(\ref{eq:classical-matrix-eq-of-motion}) has to be degenerate---a
requirement that yields a matrix generalization of the semiclassical
equation of motion~(\ref{eq:classical-eq-of-motion_app}):
\begin{equation}
\det\left\{ E\,\hat{1}-\hat{\mathcal{H}}_{\text{eff}}(x,p)\right\} =0.
\label{eq:classical-eq-of-motion_matrix-case}
\end{equation}
$E$ is then an eigenvalue of $\hat{\mathcal{H}}_{\text{eff}}$, whereas
$\overrightarrow{\varphi}_{0}$ is the corresponding eigenvector.
Eq.~(\ref{eq:matrix-subleading-order}) further yields a condition
that defines the spatial behavior of the \emph{normalization} of $\overrightarrow{\varphi}_{0}$:
since $E-\hat{\mathcal{H}}_{\text{eff}}$ is a degenerate matrix,
the linear system~(\ref{eq:matrix-subleading-order}) has a solution
for $\overrightarrow{\varphi}_{1}$ only if the right-hand side of
Eq.~(\ref{eq:matrix-subleading-order}) is orthogonal to the kernel
of $E-\hat{\mathcal{H}}_{\text{eff}}$, which implies 
\begin{equation}
\overrightarrow{\varphi}^{\dagger}_{0}\left\{ -i\,\partial_{p}\hat{\mathcal{H}}\frac{d}{dx}+\hat{\mathcal{H}}^{(1)}\right\} \overrightarrow{\varphi}_{0}=0,
\label{eq:transport-equation_matrix-cse}
\end{equation}
where $\overrightarrow{\varphi}^{\dagger}_{0}$ is the \emph{left}
eigenvector of $\hat{\mathcal{H}}_{\text{eff}}$ corresponding to
eigenvalue $E$, and we have used the explicit form of $\hat{\mathcal{H}}^{(x)}$,
Eq.~(\ref{eq:matrix-Maslov-symbol-logarithm}). This is the generalization
of the transport equation~(\ref{eq:classical-transport-equation})
that completely specifies the leading term in Eq.~(\ref{eq:semiclassical-envelope-series}):
If $\overrightarrow{f}_{0},\,\overrightarrow{f}^{\dagger}_{0}$ are
some smooth right and left eigenvectors of $\hat{\mathcal{H}}_{\text{eff}}$,
then $\overrightarrow{\varphi}_{0}=\sigma(x)\,\overrightarrow{f}_{0}$
for a smooth $\sigma(x)$ which obeys
\begin{equation}
\frac{1}{\sigma}\frac{d\sigma}{dx}=-\frac{\overrightarrow{f}^{\dagger}_{0}\partial_{p}\hat{\mathcal{H}}\frac{d\overrightarrow{f}_{0}}{dx}+i\,\overrightarrow{f}^{\dagger}_{0}\hat{\mathcal{H}}^{(1)}\overrightarrow{f}_{0}}{\overrightarrow{f}^{\dagger}_{0}\partial_{p}\hat{\mathcal{H}}\overrightarrow{f}_{0}}.
\end{equation}
Importantly, $\overrightarrow{\varphi}^{\dagger}_{0}$ is not necessarily
the hermitian conjugate of $\overrightarrow{\varphi}_{0}$, since
the Maslov symbol $\hat{\mathcal{H}}_{\text{eff}}$ is not guaranteed
to be hermitian for imaginary $p$, realized in the classically forbidden
region (consider the term $\propto\hat{\tau}_{y}$ in Eq.~(\ref{eq:semiclassical-matrix-Hamiltonian_smooth-potential})
as an example).

\subsubsection{Reproducing the $\delta t\to0$ Limit}

It is instructive to observe how the generalized description reduces
to the scalar case of \appref{Discrete-Semiclassical-Approximation}.
In the limit $\delta t\to0$, the matrix semiclassical Hamiltonian~(\ref{eq:semiclassical-matrix-Hamiltonian_smooth-potential})
reduces to
\begin{equation}
\hat{\mathcal{H}}(x,p)=-J\cos p\,\hat{\tau}_{z}+h(x)\,\hat{1}.
\label{eq:semiclassical-matrix_no-alternating-perturbation}
\end{equation}
The classical equation of motion~(\ref{eq:classical-eq-of-motion_matrix-case})
then reads
\begin{equation}
\begin{cases}
\mathcal{H}(x,p)=E,\,\,\,\overrightarrow{\varphi}_{0}=\begin{pmatrix}\varphi_{+}(x)\\
0
\end{pmatrix},\,\,\,\overrightarrow{\varphi}^{\dagger}_{0}=\left(\varphi^{*}_{+}(x),0\right),\\
\mathcal{H}(x,p+\pi)=E,\,\,\,\overrightarrow{\varphi}_{0}=\begin{pmatrix}0\\
\varphi_{-}(x)
\end{pmatrix},\,\,\,\overrightarrow{\varphi}^{\dagger}_{0}=\left(0,\varphi^{*}_{-}(x)\right),
\end{cases}
\label{eq:leading-order-wave-vectors_no-alternating-matrix-elements}
\end{equation}
and the semiclassical wave function~(\ref{eq:semiclassical_wave-vector})
is given by
\begin{equation}
\psi_{n}=\begin{cases}
\mathcal{H}(x,p)=E: & e^{iS(x)/a}\,\varphi_{+}(x),\\
\mathcal{H}(x,p+\pi)=E: & e^{iS(x)/a}\,(-1)^{n}\varphi_{-}(x)
\end{cases}\equiv e^{iS(x)/a}\,\varphi(x),\,\,\,\mathcal{H}(x,p)=E,
\end{equation}
where we have used the fact that the second alternative reduces to
the first one upon replacing $p\mapsto p+\pi$. The transport equation~(\ref{eq:transport-equation_matrix-cse})
for both alternatives in Eq.~(\ref{eq:leading-order-wave-vectors_no-alternating-matrix-elements})
coincides with the scalar one, Eq.~(\ref{eq:classical-transport-equation}).

\subsubsection{General Case: the Semiclassical Hamiltonian}

For finite $\delta t$, the classical equation of motion~(\ref{eq:classical-eq-of-motion_matrix-case})
for $\hat{\mathcal{H}}_{\text{eff}}$ reads
\begin{equation}
\pm T(p)+E-h(x)=0,
\label{eq:large-t-suzuki-trotter_equation-of-motion}
\end{equation}
where $\pm T(p)$ represents the eigenvalues of the kinetic term and
is given by Eq.~(\ref{eq:kinetic-suzuki-trotter_energy-spectrum}).
The corresponding eigenvector $\overrightarrow{\varphi}_{0}$ of $\hat{\mathcal{H}}_{\text{eff}}$
is given by the appropriate eigenfunction of Eq.~(\ref{eq:maslov-symbol_kinetic-term}),
resulting in Eqs.~(\ref{eq:trotterized-evolution_eigenfunction})--(\ref{eq:trotterized-evolution_oscillating-component})
of the main text. Because $T(p+\pi)=-T(p)$, we are further free to
choose the positive sign, in which case Eq.~(\ref{eq:large-t-suzuki-trotter_equation-of-motion})
formally corresponds to the equation of motion~(\ref{eq:classical-eq-of-motion_app})
for the \emph{scalar} classical Hamiltonian $T(p)+h(x)$ with $p\in[-\pi,\pi]$
as in the $\delta t\to0$ case. This concludes the derivation of Eqs.~(\ref{eq:trotterized-evolution_eigenvalue})--(\ref{eq:trotterized-classical-hamiltonian})
of the main text. The wave function normalization is a cumbersome
expression because of the complex form of the correction~$\delta\hat{\mathcal{H}}^{(1)}$
arising from the non-commutativity of the operators entering STA,
Eq.~(\ref{eq:hamiltonian-subleading-term-correction}), which does
not seem to admit significant simplifications. 

\subsubsection{Conditions of Applicability}

Similarly to the standard semiclassical treatment of \appref{Discrete-Semiclassical-Approximation},
one should \emph{a~posteriori} verify the smoothness of the de~Broglie
wavelength $\lambda(x)=2\pi a/p(x)$ and the semiclassical coefficients~$\overrightarrow{\varphi}_{0}$.
The results similarly indicate that semiclassical description breaks
down close to the classical turning points defined by vanishing of
the group velocity,
\begin{equation}
\text{turning point: }p(x)=0\,\,\,\text{or}\,\,\,\overrightarrow{\varphi}^{\dagger}_{0}\partial_{p}\hat{\mathcal{H}}_{\text{eff}}\overrightarrow{\varphi}_{0}=0.
\end{equation}

The matrix structure of Eq.~(\ref{eq:classical-matrix-eq-of-motion})
further requires the Hamiltonian eigenvalues to be non-degenerate---a
condition that is violated by Hamiltonian~(\ref{eq:semiclassical-matrix-Hamiltonian_smooth-potential})
at $p=\pm\pi/2$. However, the projection of $\hat{\mathcal{H}}^{(x)}=-i\partial_{p}\hat{\mathcal{H}}$
on this subspace is non-degenerate, allowing one to simply ignore
this singularity in the semiclassical description by allowing $|p|$
to grow past~$\pi/2$. The result is verified by performing the matching
of the semiclassical wave function to the exact solution for a linearized
potential around $x_{\pi/2}$ such that $p(x_{\pi/2})=\pm\pi/2$.

\subsection{Scattering Phases and Quantization Conditions}

The quantization conditions are obtained by the same procedure as
that of \subsecappref{Matching-and-quantization-conditions}, resulting
in Eq.~(\ref{eq:quantization-condition}). Importantly, the scattering
phases $\theta_{i}$ arguably constitute a topological property of
the problem~\citep{maslov-2001-semiclassical-approx} and thus preserve
their values, as we verified perturbatively in $\delta t$ (see \subsecref{Change-in-the-matching-conditions}). 

However, the global matching conditions including other critical points,
e.g., $p=\pm ip_{c},\,\pm\pi\pm ip_{c}$ {[}see Eq.~(\ref{eq:critical-momentum}){]},
are more complicated, as evident from \figref{low-energy-distortion_of-trotterized-evolution}:
while the smooth component $\varphi_{+}$ is dominant in both potential
wells, the region within the barrier is characterized by the dominant
oscillating component~$\varphi_{-}$. However, a complete analysis
of the matching problem, as determined by the underlying topological
properties of the Riemann surface of the semiclassical wave function
$\psi(x)$, will be presented elsewhere~\citep{complexMatrixWKBA}.

In addition, the system ends at $n=1,L$ impose boundary conditions
on the semiclassical wave function, relating the two counter-propagating
solutions, resulting in the middle line of Eq.~(\ref{eq:anomalous-phases-def}).
For the original tight-binding model, Eq.~(\ref{eq:single-particle_Hamiltonian}),
this manifests as a scattering phase $\theta=\pi$ in the quantization
condition. For finite $\delta t$, the scattering phase $\theta(J\delta t,p)$
acquires a $p$-dependence that must be determined from the scattering
problem off the boundary of a translationally invariant system, where
WKBA exactly describes the wave functions away from the boundary.
Below we briefly describe the procedure.

Let $\mathcal{U}_{nm}$ be the matrix elements of the product formula
for a homogeneous Hamiltonian of a semi-infinite system ($n,m\ge0$).
WKBA describes eigenfunctions of $\mathcal{U}$ for $n$ such that
$\mathcal{U}_{nm}$ is described by the translationally invariant
part, $n\ge t$, where $t$ is the radius of the region perturbed
by the boundary (for our case, $t=3$ because $\mathcal{U}$ is a
band matrix with size 3, as explained in \secappref{general-form_large-distance_matrix-elements}).
The eigenfunction equation yields a $t\times t$ inhomogeneous linear
system on $\psi_{m}$ for $0\le m<t$. Via the Fredholm alternative,
the existence of a solution translates to a linear system of equations
for the coefficients of incoming and outgoing waves for each $p$.
The scattering phase is then fixed by the choice of normalization
of the incoming and outgoing (or, in the forbidden region, decaying
and growing) waves, e.g., in the form of the lower integration limit
$x_{0}$ in Eq.~(\ref{eq:scalar-transpot-eq_solution}). By extension
of the limiting case $\delta t\to0$, it is convenient to place the
reference point $x_{0}=-a$, i.e., one site past the last physical
site of the system.

\begin{figure}
\begin{centering}
\includegraphics[width=0.7\textwidth]{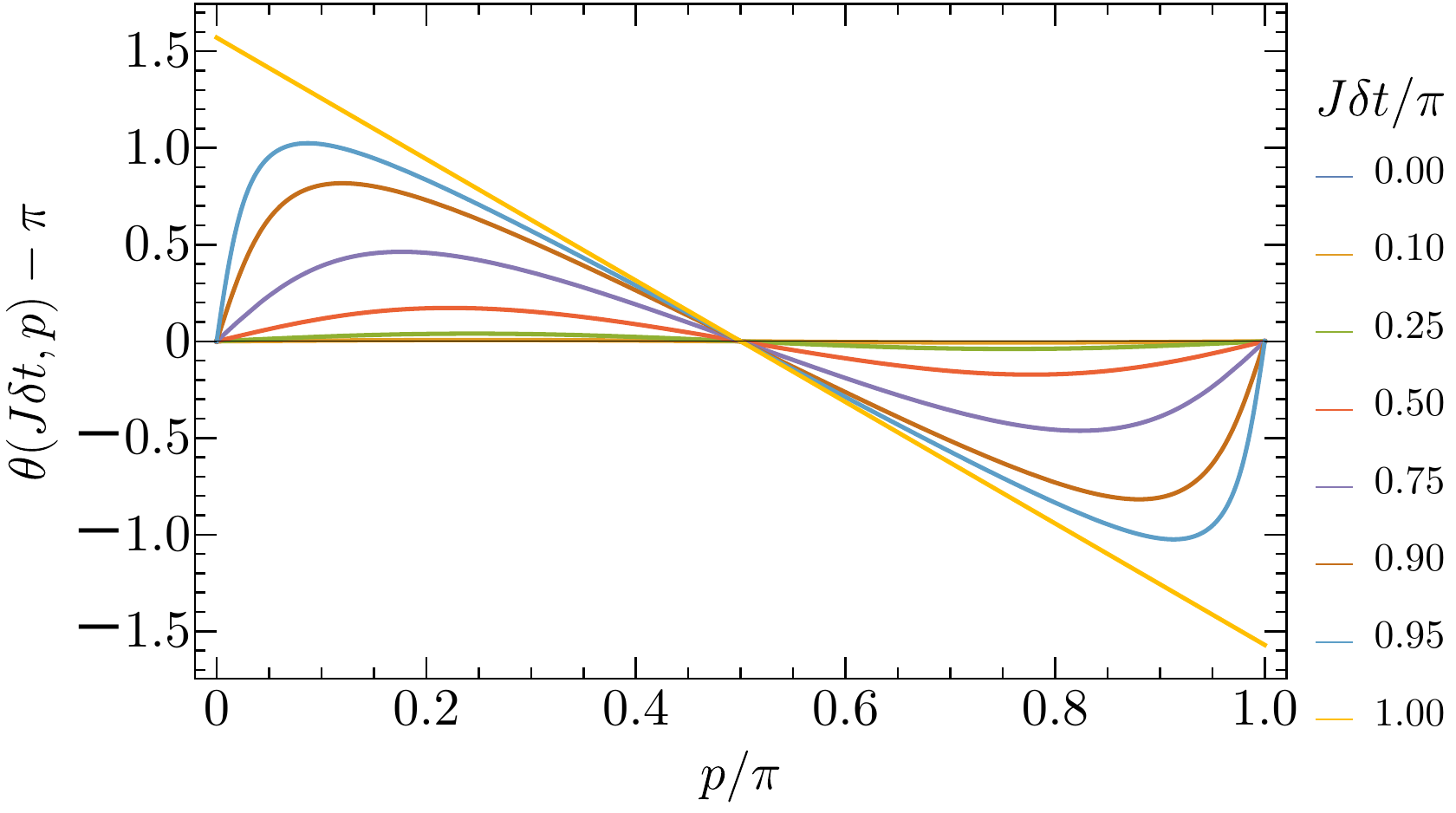}
\par\end{centering}
\caption{The scattering phase $\theta(J\delta t,p)$ as a function of $p$
for various $J\delta t$. The limiting cases are given by Eq.~(\ref{eq:scattering-phase_limitting-cases}).
\label{fig:scattering-phase}}

\end{figure}

Importantly, the two possible choices of the operator ordering in
a translationally invariant problem differ by a gauge transform that
does not affect the scattering phase. In particular, the quantity
$\tau=a\,\partial_{p}\theta\,\partial_{E}p=a\,\partial_{p}\theta/v$
is the physically observable wave packet scattering delay that is
gauge-invariant. \figref{scattering-phase} illustrates the resulting
$\theta(J\delta t,p)$. Unfortunately, the analytical expression occupies
many lines and does not seem to admit significant simplifications,
except in two limiting cases:
\begin{equation}
\theta(J\delta t=0,p)=\pi,\,\,\,\theta(J\delta t=\pi,p)=3\pi/2-p.
\label{eq:scattering-phase_limitting-cases}
\end{equation}

\section{Trotterization Error in the Perturbative Regime\label{app:trotter_perturb-corrections}}

\subsection{The Leading Correction to the Effective Hamiltonian\label{secapp:general-form_for-2nd-order-correction}}

Equation~(\ref{eq:Suzuki-Trotter_2nd-order}) for STA with $n$ operators
$A_{k}$ is described by the following recursive relation:

\begin{equation}
U_{k}(\delta t)=e^{-iA_{k}\delta t/2}U_{k-1}(\delta t)e^{-iA_{k}\delta t/2},\,\,\,U_{1}=e^{-iA_{1}\delta t}.
\label{eq:Suzuki-Trotter_evolution-operator_recursion}
\end{equation}
Due to the symmetry $U_{k}(\delta t)^{-1}=U_{k}(-\delta t)$, the
effective Hamiltonian at each step can be expanded in \emph{even}
powers of $\delta t$:
\begin{equation}
H^{(k)}_{\text{eff}}=-\frac{1}{i\delta t}\ln U_{k}(\delta t)=H^{(k)}_{\text{ }}+\delta t^{2}D^{(k)}+O\left(\delta t^{4}\right),\,\,\,H^{(k)}_{\text{ }}=\sum^{k}_{j=1}A_{j}.
\label{eq:effective-Hamiltonian-form}
\end{equation}
To extract the leading correction $D^{\left(k\right)}$, we apply
the second order of the Baker--Campbell--Hausdorff formula:
\begin{equation}
\ln\left\{ e^{-itA}e^{-itB}\right\} =-it\left[A+B+\frac{t}{2i}\left[A,B\right]+\frac{t^{2}}{12}\left(\left[A,\left[A,B\right]\right]+\left[B,\left[B,A\right]\right]\right)+O\left(t^{3}\right)\right].
\label{eq:CBH_leading-order}
\end{equation}
Applying this twice to Eq.~(\ref{eq:Suzuki-Trotter_evolution-operator_recursion})
and substituting into Eq.~(\ref{eq:effective-Hamiltonian-form})
yields:
\begin{equation}
D^{(k+1)}=D^{(k)}-\frac{1}{12}\left\{ \left[H^{(k)}_{\text{ }},\left[H^{(k)}_{\text{ }},A_{k+1}\right]\right]-\frac{1}{2}\left[A_{k+1},\left[A_{k+1},H^{(k)}_{\text{ }}\right]\right]\right\} .
\label{eq:Suzuki-Trotter_leading-correction_recursion}
\end{equation}
In general, $D^{(n)}$ is sensitive to the order of $A_{k}$. For
example, if the target Hamiltonian can be decomposed as $H=A_{1}+A_{2}+A_{3}$,
with $A_{i}$ ordered by their index according to Eq.~(\ref{eq:Suzuki-Trotter_evolution-operator_recursion}),
one obtains
\begin{equation}
H_{\text{eff}}=H+\delta t^{2}D^{(3)}+O\left(\delta t^{4}\right),
\end{equation}
\begin{equation}
D^{(3)}=-\frac{1}{12}\left\{ \left[A_{1},\left[A_{1},A_{2}\right]\right]-\frac{1}{2}\left[A_{2},\left[A_{2},A_{1}\right]\right]+\left[A_{1}+A_{2},\left[A_{1}+A_{2},A_{3}\right]\right]-\frac{1}{2}\left[A_{3},\left[A_{3},A_{1}+A_{2}\right]\right]\right\} .
\label{eq:Suzuki-Trotter-3-terms_leading-correction_explicit-form}
\end{equation}

\begin{figure}
\begin{centering}
\includegraphics[width=0.45\textwidth]{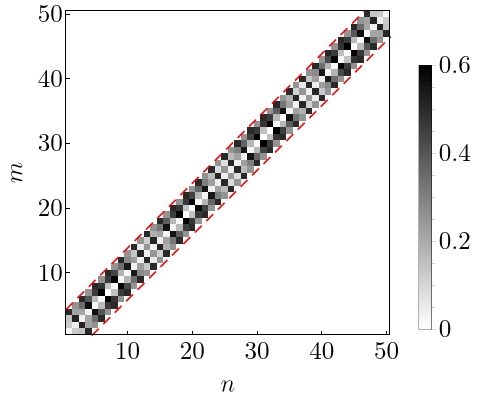}~~\includegraphics[width=0.49\textwidth]{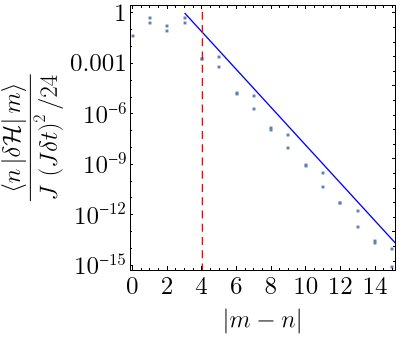}
\par\end{centering}
\caption{Matrix elements of the Suzuki-Trotter error $\left\langle n\left|\delta\mathcal{H}\right|m\right\rangle $
normalized by $J\,\left(J\delta t\right)^{2}/24$ with the same parameters
as \figref{low-energy-distortion_of-trotterized-evolution} but $J\delta t=0.3$.
\emph{Left:}~color plot of the matrix elements\emph{.} \emph{Right:}~matrix
element as a function of distance between sites $\left|m-n\right|$
for fixed $n=L/2$, with the blue line corresponding to $\left(J\delta t/4\right)^{\left|m-n\right|-3}$,
according to estimation~(\ref{app_effective-Ham_large-distance-matr-elems-estimation}).
On both plots, red dashed line denotes $\left|m-n\right|=4$. \label{fig:effective-Hamiltonian_matrix-elements}}
\end{figure}

Equations~(\ref{eq:Suzuki-Trotter_evolution-operator_recursion})--(\ref{eq:Suzuki-Trotter-3-terms_leading-correction_explicit-form})
translate to single-particle operators by replacing roman operator
symbols with calligraphic ones. The matrix elements of the difference
$\delta\mathcal{H}=\mathcal{H}_{\text{eff}}-\mathcal{H}$ between
the single-particle effective Hamiltonian~(\ref{eq:effective-Hamiltonian-def})
of STA and the simulated Hamiltonian~(\ref{eq:single-particle_Hamiltonian})
are shown in \figref{effective-Hamiltonian_matrix-elements},~\emph{left}.
They are dominated by short-range hopping of the leading correction
$D^{(3)}$, given by Eq.~(\ref{eq:Suzuki-Trotter-3-terms_leading-correction_explicit-form}).
However, \figref{effective-Hamiltonian_matrix-elements},~\emph{right},
shows that large-distance matrix elements are also present. To describe
these, one must examine higher orders of expansion~(\ref{eq:effective-Hamiltonian-form})
in powers of~$\delta t$.

\subsection{Estimation of Large-Distance Matrix Elements of the Effective Hamiltonian\label{secapp:general-form_large-distance_matrix-elements}}

The purpose of STA, Eq.~(\ref{eq:Suzuki-Trotter_2nd-order}), is
to obtain an approximation to the true evolution operator that involves
only local one- and two-qubit gates. Reflecting this, the potential
terms are diagonal, and the kinetic terms have nonzero matrix elements
only at distances at most one: $\left\langle i\left|\mathcal{U}_{k}\right|j\right\rangle =0$
for $\left|i-j\right|>2$ {[}see, e.g., Eq.~(\ref{eq:evolution-operator_even/odd_kinetic-term}){]}.
Moreover, off-diagonal matrix elements of each constituting operator
scale as $O\left(\left[J\delta t/4\right]\right)$, with $J$ being
the characteristic scale of the kinetic energy. In other words, the
following estimate for the matrix elements of each term in Eq.~(\ref{eq:Suzuki-Trotter_2nd-order})
is valid in the limit $\delta t\to0$:
\begin{equation}
\left\langle i\left|\mathcal{U}_{k}\right|j\right\rangle =O\left(\left[J\delta t/4\right]^{\left|i-j\right|}\right)\,\text{Id}\left(d_{k}\ge\left|i-j\right|\right),\,\,\,d_{k}=\begin{cases}
0, & k\in\text{potential terms},\\
1, & k\in\text{kinetic terms},
\end{cases}
\label{eq:local-gate-matrix-element-estimation}
\end{equation}
where we assume $h\sim J$, and $\text{Id}(\text{condition})$ is
the indicator function, equal to 1 if the condition in the argument
is satisfied, and 0 otherwise. Estimate~(\ref{eq:local-gate-matrix-element-estimation})
is preserved under composition:
\begin{equation}
\left\langle i\left|\prod^{m}_{k=1}\mathcal{U}_{k}\right|j\right\rangle =O\left(\left[J\delta t/4\right]^{\left|i-j\right|}\right)\text{Id}\left(d_{\text{appr}}\ge\left|i-j\right|\right),\,\,\,d_{\text{appr}}:=\sum^{m}_{k=1}d_{k},
\label{eq:matrix-element-estimate_composition}
\end{equation}
which is proven by induction on $m$. This implies the following structure
of the matrix elements of the STA evolution operator $\mathcal{U}_{\text{appr}}$:
\begin{equation}
\mathcal{U}^{ij}_{\text{appr}}=\begin{cases}
O\left(\left(J\delta t/4\right)^{\left|i-j\right|}\right), & \left|i-j\right|\le d_{\max},\\
0, & \left|i-j\right|>d_{\max}.
\end{cases}
\end{equation}

The second-order STA operator $\mathcal{U}_{\text{appr}}$, Eq.~(\ref{eq:Suzuki-Trotter_2nd-order}),
is composed of $m=2n-1$ local gates, where $n$ is the number of
terms in the decomposition of $\mathcal{H}$ into executable terms.
The $-1$ contribution arises because there are two identical operators
$e^{-i\mathcal{A}_{1}\delta t/2}$ in the middle of Eq.~(\ref{eq:Suzuki-Trotter_2nd-order})
that are merged together. Moreover, two of the constituting operators
are diagonal, as they represent the potential term. As a result, the
three different orderings of \appref{Gate-ordering} are characterized
by the following finite distances:
\begin{align}
\left\{ K_{\text{even}},\,K_{\text{odd}},\,P\right\} : & d_{\max}=4,\nonumber \\
\left\{ K_{\text{even}},\,P,\,K_{\text{odd}}\right\} : & d_{\max}=3,\\
\left\{ K_{\text{even}}+\beta P,\,K_{\text{odd}}+(1-\beta)P\right\} : & d_{\max}=3.\nonumber 
\end{align}

Using Eq.~(\ref{eq:matrix-element-estimate_composition}) on the
direct expansion of the operator logarithm in Eq.~(\ref{eq:effective-Hamiltonian-def})
in powers of $\delta t$, one obtains that large-distance matrix elements
of the effective Hamiltonian $\mathcal{H}_{\text{eff}}$ decay exponentially
with distance:
\begin{equation}
\left\langle i\left|\mathcal{H}_{\text{eff}}\right|j\right\rangle \equiv\frac{1}{i\delta t}\sum^{\infty}_{s=1}\frac{(-1)^{s}}{s}\left\langle i\left|\mathcal{U}^{s}_{\text{appr}}\right|j\right\rangle =\begin{cases}
J\,O\left(\left[J\delta t/4\right]^{\left|i-j\right|-1}\right), & \left|i-j\right|>0,\\
J\,O(1), & \left|i-j\right|=0.
\end{cases}\label{app_effective-Ham_large-distance-matr-elems-estimation}
\end{equation}
It is crucial that $J\delta t/4\ll1$ for the resulting Hamiltonian
to be truly local, which is one of the applicability criteria~(\ref{eq:Suzuki-Trotter_naive-applicability-criteria}).
This estimate is also valid for higher-order product formulae of Ref.~\citep{suzuki-1991},
as they are too composed of operators obeying Eq.~(\ref{eq:local-gate-matrix-element-estimation}).

\subsection{Competition Between the Potential Barrier and Large-Distance Hopping\label{subsec:Tunneling_vs_large-distance-hopping}}

The enhancement of tunneling described in \subsecref{Rabi-oscillations_no-detuning}
is a manifestation of next-to-nearest neighbor hopping present in
the effective Hamiltonian {[}see Eq.~(\ref{app_effective-Ham_large-distance-matr-elems-estimation}){]}.
The effect of large-distance hopping becomes prominent for large barriers.
From Eq.~(\ref{app_effective-Ham_large-distance-matr-elems-estimation}),
it follows that the sum over $k$ in the Maslov symbol, Eq.~(\ref{eq:Maslov-symbol}),
diverges for $\text{Im}p>p_{c}$, where
\begin{equation}
p_{c}\approx\pm i\ln\frac{4}{J\delta t},
\label{eq:kinetic-energy_approx-singularity-position}
\end{equation}
implying the presence of a singularity of $\mathcal{H}(x,p)$ in the
complex $p$-plane. The true position of this singularity is given
by Eq.~(\ref{eq:critical-momentum}), which reduces to Eq.~(\ref{eq:kinetic-energy_approx-singularity-position})
in the limit of small~$\delta t$.

Singularities of $\mathcal{H}(x,p)$ in the complex $p$ plane have
important implications for tunneling processes, as they correspond
to the motion with imaginary momentum. Estimating the maximum momentum
in the forbidden region by its unperturbed dynamics, Eqs.~(\ref{eq:classical-eq-of-motion})--(\ref{eq:Maslov-symbol}),
we obtain that the tunneling trajectory does not approach this singularity
if
\begin{equation}
\left(h(x)-E\right)\delta t\ll1,
\label{eq:underbarrier-singularity-condition}
\end{equation}
which is the second part of the STA applicability criterion~(\ref{eq:Suzuki-Trotter_naive-applicability-criteria}).
If condition~(\ref{eq:underbarrier-singularity-condition}) is violated,
large-distance hopping becomes more favorable than tunneling through
the barrier. Moreover, as evident from Eq.~(\ref{eq:underbarrier-singularity-condition}),
the low-energy subspace is the first to be affected, with the most
prominent effect occurring at the point with the highest potential.
This also elucidates the observed difference in the scaling of the
under-barrier action~(\ref{eq:tunneling-action_rough-estimation})
and the correction~(\ref{eq:underbarrier-action-change_direct_rough-estimation})
with the potential scale~$P$.

\subsection{How the Matrix WKBA is Connected to the Expansion in Powers of $\delta t$?\label{secapp:Correction-to-semiclassical-Hamiltonian}}

Another perspective on the machinery of \appref{Matrix-WKB} is obtained
by considering higher orders of expansion of the exact effective Hamiltonian~(\ref{eq:effective-Hamiltonian-def})
in powers of $\delta t$, which contain more complicated combinations
of nested commutators. Eq.~(\ref{eq:semiclassical-matrix-Hamiltonian_smooth-potential})
then amounts to observing that each time $\mathcal{P}$ is commuted
with other operators, the result is proportional to $a\ll1$. One
can thus retain only the contributions that do not involve commutation
with $\mathcal{P}$. This corresponds to computing the full effective
Hamiltonian in a translationally invariant system, yielding Eq.~(\ref{eq:trotterized-classical-hamiltonian}).
This perspective explains why expanding Eq.~(\ref{eq:trotterized-classical-hamiltonian})
in powers of $\delta t$ coincides with Eq.~(\ref{eq:effective-Hamiltonian_classical-correction}):
the latter corresponds to the leading orders in powers of $\delta t$.

\subsection{Change in the Matching Conditions\label{subsec:Change-in-the-matching-conditions}}

In the presence of the matrix structure of the semiclassical Hamiltonian~(\ref{secapp:Full-effective-Hamiltonian}),
one has to re-examine the matching conditions. To this end, we repeat
the calculation of \subsecappref{Matching-and-quantization-conditions}
for STA with operators ordered as~$K_{\text{even}},K_{\text{odd}},P$,
with other orderings yielding the same results. Using Eq.~(\ref{eq:Suzuki-Trotter_leading-correction_recursion})
for the problem with linear potential $h_{n}=\alpha\,an$, one obtains
the following structure of the matrix elements of the effective Hamiltonian:
\begin{equation}
\mathcal{H}^{nm}_{\text{eff}}=\mathcal{H}^{nm}-\frac{(J\delta t)^{2}}{24}\frac{J}{2}\left\{ \left[\left(\frac{\alpha a}{J}\right)^{2}-\frac{1}{4}+(-1)^{n}\frac{3}{4}\right]\,\,\delta_{n+1,m}+\,\left[\frac{1}{4}+(-1)^{n}\frac{3}{4}\right]\,\,\delta_{n+3,m}\right\} +(n\leftrightarrow m)+O\left(\delta t^{4}\right),
\label{eq:effective-hamiltonian_leading-order_matrix-elements}
\end{equation}
where $\mathcal{H}^{nm}$ are the matrix elements of the target Hamiltonian~(\ref{eq:single-particle_Hamiltonian}),
and the third term $(n\leftrightarrow m)$ is equal to the second
one with indices $n,m$ interchanged. While the solution of the corresponding
stationary Schrödinger equation (with $O\left(\delta t^{4}\right)$
terms neglected) is still expressible in terms of integral representation
(which reproduces the Bessel function solution of Eq.~(\ref{eq:linearized-potential_schroedinger-eq})
for $\delta t=0$), we will once again resort to a controllable approximate
solution that exploits the potential smoothness. While we limit the
analysis to precision $O\left(\delta t^{4}\right)$, higher powers
of $\delta t$ are expected to yield the same results for the connection
formula due to its topological origin~\citep{maslov-2001-semiclassical-approx}.

Just as with the semiclassical approximation, we first remove the
oscillating terms in $\mathcal{H}_{\text{eff}}$, which is done by
introducing the oscillating component to the wave function:
\begin{equation}
\psi_{n}=\varphi(x)+(-1)^{n}\eta(x),\,\,\,x=an,
\end{equation}
with both $\varphi,\,\eta$ being smooth functions of $x$, so one
can replace the discrete differences as
\begin{equation}
\varphi_{n+k}+\varphi_{n-k}=2\varphi(x)+k^{2}a^{2}\,\varphi''(x)+O\left(a^{4}\right),\,\,\,\varphi_{n+k}-\varphi_{n-k}=2ka\,\varphi'(x)+O\left(a^{3}\right),
\end{equation}
and similarly for $\eta$, with the criteria of applicability identical
to Eq.~(\ref{eq:quadratic-approximation-condition}). The stationary
Schrödinger equation is then described by a pair of coupled second-order
differential equations that can be written in the matrix form as:
\begin{equation}
\left(E-\left[\left\{ -J+\frac{J}{2}\hat{p}^{2}+\frac{\left(J\delta t\right)^{2}}{24}J\hat{p}^{2}\right\} \hat{\tau}_{z}+\frac{(J\delta t)^{2}}{8}J\,\hat{p}\,\hat{\tau}_{y}+\alpha x\,\hat{1}\right]+\frac{(\alpha a\delta t)^{2}}{24}\right)\begin{pmatrix}\varphi(x)\\
\eta(x)
\end{pmatrix}=O\left(\hat{p}^{3},\delta t^{4}\right),\,\,\,\hat{p}:=ia\frac{d}{dx}.
\label{eq:Shroedinger-eq_continuous-app_linear-potential_Suzuki-Trotter-corrected}
\end{equation}
In this equation, we have grouped terms in such a way that the term
in the square brackets of the left-hand side coincides with the expansion
of the semiclassical matrix Hamiltonian, Eq.~(\ref{eq:semiclassical-matrix-Hamiltonian_smooth-potential}),
around $p=0$ up to the second order, with subsequent replacement
$\hat{p}\mapsto ia\frac{d}{dx}$. The last term in the left-hand side
of Eq.~(\ref{eq:Shroedinger-eq_continuous-app_linear-potential_Suzuki-Trotter-corrected})
is beyond WKBA, as discussed in \secappref{Correction-to-semiclassical-Hamiltonian}.

\subsubsection{A Standard Turning Point $p=0$}

This case corresponds to $E+J=\alpha\,a\delta,\,\,\,\delta\in[0,1]$,
with $p$ vanishing at the turning point. For $\delta t=0$, the solution
to Eq.~(\ref{eq:Shroedinger-eq_continuous-app_linear-potential_Suzuki-Trotter-corrected})
is described by $\eta=0$ as there should be no oscillating component
in the system without oscillations, and the $\varphi$ component is
then given by $\psi(x)$ from Eq.~(\ref{eq:approximate-solution}).
At nonzero $\delta t$, one can treat the term $\propto\hat{\tau}_{y}$
in Eq.~(\ref{eq:Shroedinger-eq_continuous-app_linear-potential_Suzuki-Trotter-corrected})
as a perturbation inducing a finite $\eta(x)$:
\begin{equation}
\eta(x)=-i\left[\frac{(J\delta t)^{2}}{16}\hat{p}+O\left(a^{3}\right)\right]\varphi(x),\,\,\,\hat{p}=ia\frac{d}{dx},
\end{equation}
as found from the second component of Eq.~(\ref{eq:Shroedinger-eq_continuous-app_linear-potential_Suzuki-Trotter-corrected}).
Substituting this into the first component of Eq.~(\ref{eq:Shroedinger-eq_continuous-app_linear-potential_Suzuki-Trotter-corrected})
then shows that the resulting contribution is beyond the precision
of expansion in powers of $\delta t$, so the equation for $\varphi$
reads
\begin{equation}
\left(-J+\alpha\,(x-a\delta)-\left\{ 1+\frac{\left(J\delta t\right)^{2}}{12}\right\} \frac{J}{2}\hat{p}^{2}+\frac{\left(\alpha a\delta t\right)^{2}}{24}\right)\varphi=O\left(\delta t^{4},a^{3}\right),\,\,\,\hat{p}=ia\frac{d}{dx}.
\end{equation}
This equation coincides with Eq.~(\ref{eq:Schroedinger-eq_continuos-appr_linear-potential})
up to additive corrections to $K$ and $\delta$, with the difference
being scaling as $O\left(K\left(J\delta t\right)^{2}\right)$ and
$O\left(\alpha\,a^{2}\delta t^{2}\right)$, respectively. The matching
conditions, being independent on both $J$ and $\delta$ for this
case, are thus identical to~(\ref{eq:standard-turning-point_matching-conditions}). 

\subsubsection{An Anomalous Turning Point $p=\pm\pi$}

In this case, $E-J=\alpha\,a\delta$, and the solution for Eq.~(\ref{eq:Shroedinger-eq_continuous-app_linear-potential_Suzuki-Trotter-corrected})
is described by $\varphi(x)=0$ since the only oscillating solution
is present near the anomalous turning point, while the $\eta$ component
is described by Airy functions, with asymptotic behavior given by
Eq.~(\ref{eq:anomalous-turning-point-exact-solution_asymptotic}).
Similarly to the previous case, the presence of the term $\propto\hat{\tau}_{y}$
at nonzero $\delta t$ can be neglected, rendering 
\begin{equation}
\left(-\alpha\,(x-a\delta)+\left\{ 1+\frac{\left(J\delta t\right)^{2}}{12}\right\} \frac{J}{2}\hat{p}^{2}+\frac{\left(\alpha a\delta t\right)^{2}}{24}\right)\eta(x)\approx0,\,\,\,\hat{p}=ia\frac{d}{dx}.
\end{equation}
This equation is still solved by~(\ref{eq:anomalous-turning-point-exact-solution_asymptotic})
with the values of $K,\,\delta$ shifted by $O\left(K\left(J\delta t\right)^{2}\right)$
and $O\left(\alpha\,a^{2}\delta t^{2}\right)$, respectively. While
the matching conditions~(\ref{eq:matching-conditions_anomalous-turning-point})
for this case do depend on $\delta$, the shift in the latter has
additional smallness of $O\left(a^{2}\right)$ and thus does not alter
the WKBA.

\subsection{Change in the classical period of motion\label{secapp:Change-in-classical-period}}

In this section, we estimate the change of the classical period of
motion~$T(E)$ of a given bound state {[}entering the sub-exponential
prefactors in Eq.~(\ref{eq:semiclassical_tunneling-rate}){]} due
to perturbation~(\ref{eq:effective-Hamiltonian_classical-correction})
of the semiclassical Hamiltonian. This change is given by:
\begin{equation}
\delta T_{1}\left(E\right)=\frac{\partial\delta S_{12}}{\delta E}\approx-\intop^{x_{2}}_{x_{1}}dx\,\left[\frac{1}{v(p)}\,\frac{\partial}{\partial p}\left\{ \frac{\delta\mathcal{H}_{\text{eff}}(x,p)}{v(p)}\right\} \right]_{p=p(E;x)},
\label{eq:classical-period_fixed-energy_change-due-to-perturbation}
\end{equation}
where $x_{1,2}$ are the turning points of the unperturbed classical
trajectory, $\delta\mathcal{H}_{\text{eff}}(x,p)$ is given by Eq.~(\ref{eq:effective-Hamiltonian_classical-correction}),
and the momentum $p$ is found from the classical equation of motion~(\ref{eq:classical-eq-of-motion}).
The integral is convergent because $v\left(p(E;x)\right)$ vanishes
at $x_{1,2}$ as $\sqrt{\left|x-x_{i}\right|}$, and the quantity
$\delta\mathcal{H}_{\text{eff}}/v$ is a smooth function of $p$ with
no singularities at real momenta. We thus arrive at the following
estimate:
\begin{equation}
\left|\frac{\delta T_{1}}{T}\right|=\left|-\frac{1}{T}\intop^{t_{2}}_{t_{1}}dt\,\left[\frac{\partial}{\partial p}\left\{ \frac{\delta\mathcal{H}_{\text{eff}}(x,p)}{v(p)}\right\} \right]_{p=p(t),x=x(t)}\right|\le\max_{x}\left|\frac{\partial}{\partial p}\left\{ \frac{\delta\mathcal{H}_{\text{eff}}(x,p)}{v(p)}\right\} \right|\le\frac{\left[J\delta t\right]^{2}}{24},
\end{equation}
where we used that $t\equiv\intop dx/v$, with $x(t),p(t)$ describing
the classical trajectory. There is also a change of $T$ due to the
perturbative shift of the energy: 
\begin{equation}
\left|\frac{\delta T_{2}(E)}{T}\right|=\left|\frac{T(E+\delta E)-T(E)}{T}\right|\approx\left|\frac{1}{T}\frac{\partial T}{\partial E}\delta E\right|\le\frac{(J\delta t)^{2}}{24}\,\left|\frac{1}{T^{2}}\frac{\partial T}{\partial E}\right|\,4n_{\text{cl}}a
\label{eq:classical-period_change-due-to-shift-of-energy}
\end{equation}
where we have used estimate~(\ref{eq:energy-shift-estimation}) for
the energy shift $\delta E$ and expression~(\ref{eq:single-particle_level-spacing})
for the single-particle level spacing. The quantity $T^{-2}\partial T/\partial E$
strongly depends on the potential profile, but it typically attains
large values only if $E$ approaches local potential maxima above
which the classical trajectories are characterized by a qualitatively
different shape of the allowed region. On the other hand, estimate~(\ref{eq:classical-period_change-due-to-shift-of-energy})
contains~$a\ll1$, so the change of $T$ due to the energy shift
is typically much smaller than that due to the deformation of the
semiclassical Hamiltonian.

The total relative change $\delta T_{1}/T+\delta T_{2}/T$ is thus
typically a small quantity, although large values can be attained
at energies that are close to local maxima of the potential (notably,
the semiclassical expression~(\ref{eq:semiclassical_tunneling-rate})
for the tunnel splitting ceases to be applicable in the vicinity of
such points, requiring a more accurate treatment of the potential
maximum instead). However, in all practical cases, the effect of the
relative change of $T$ on the tunneling rates is negligible compared
to the exponentially strong change in the tunneling action, Eq.~(\ref{eq:effective-tunneling-amplitude}).

\section{Probability Defect\label{app:Probability-defect}}

In this Appendix, we examine the total probability defect $\delta P=1-\left|\left\langle N_{\text{eff}}|N\right\rangle \right|^{2}$
between a given exact bound state $\left|N\right\rangle $ of a potential
well and its counterpart $\left|N_{\text{eff}}\right\rangle $ in
the STA simulation in the perturbative regime, $\Lambda\delta t\ll1$.
For tunneling processes, the overlap with high-energy states is of
particular interest, as these states have much smaller tunneling exponents
or even no tunneling barriers at all, thus introducing a completely
different type of evolution. Estimating $\delta P$ thus provides
an upper limit on this effect.

We start by estimating the overlap $\left\langle N|N_{\text{eff}}\right\rangle $
via the dominant contribution from the allowed region, since both
wave functions decay into the forbidden region. Although both wave
functions are strongly oscillating, their product contains a slowly
varying component due to close classical momenta, providing the dominant
contribution:
\begin{equation}
\left\langle N|N_{\text{eff}}\right\rangle \approx\sum_{n\in\text{allowed}}\left\langle N|n\right\rangle \left\langle n|N_{\text{eff}}\right\rangle ,
\label{eq:eigenfunction-overlap}
\end{equation}
where the summation goes only over the allowed region of either of
the state. The condition $\Lambda\delta t\ll1$ justifies neglecting
the difference in the allowed regions of the two states.

The semiclassical wave function of a bound state is given by (see
\appref{Discrete-Semiclassical-Approximation}):
\begin{equation}
\left\langle n|N\right\rangle =\psi_{N}(x=an),\,\,\,\psi_{N}(x)=\frac{\sqrt{C_{N}}}{\sqrt{v\left(E_{N};x\right)}}\cos\left\{ \frac{iS\left(E_{N},x_{1};x\right)}{a}-\frac{\theta_{1}}{2}\right\} ,
\end{equation}
where $v(E;x)$ is the semiclassical velocity, $S\left(E_{N},x_{1};x\right)$
is the action of the classical trajectory counted from the turning
point $x_{1}$, and $\theta_{1}$ is the corresponding reflection
phase (see \appref{Discrete-Semiclassical-Approximation}). The normalization
constant $C_{N}$ is conventionally found as~\citep[par. 48]{landau-lifshitz_quantum-mechanics-2013}:
\begin{equation}
1\approx\sum_{n\in\text{allowed}}\left\langle N|n\right\rangle \left\langle n|N\right\rangle \approx\left|C_{N}\right|^{2}\sum_{n}\frac{1/2}{v\left(E_{N};x=an\right)}\approx\frac{\left|C_{N}\right|^{2}}{2a}\intop_{\text{allowed}}\frac{dx}{v\left(E_{N};x\right)}\equiv\frac{\left|C_{N}\right|^{2}}{4a}T\left(E_{N}\right),
\label{eq:semiclassical-normalization}
\end{equation}
where the oscillatory term of the squared cosine was dropped, the
replacement of the sum $\sum_{n}$ with the integral $a^{-1}\intop dx$
is justified by the smoothness of $v\left(E;x\right)$, and the final
result is expressed in terms of the classical period of motion at
energy $E_{N}$. 

The overlap~(\ref{eq:eigenfunction-overlap}) is then given by
\begin{equation}
\left\langle N|N_{\text{eff}}\right\rangle \approx C^{*}_{N}C_{N,\text{eff}}\sum_{n\in\text{allowed}}\frac{\cos\left\{ \left[S_{N,\text{eff}}(an)-S_{N}(an)\right]/a\right\} }{2\sqrt{v_{N}(an)}\sqrt{v_{N,\text{eff}}(an)}}\approx\frac{C^{*}_{N}C_{N,\text{eff}}}{a}\intop_{\text{allowed}}dx\,\frac{\cos\left\{ \left[S_{N,\text{eff}}(x)-S_{N}(x)\right]/a\right\} }{2\sqrt{v_{N}(x)}\sqrt{v_{N,\text{eff}}(x)}}.
\label{eq:eigenfunction-overlap_semiclassical}
\end{equation}
The derivation steps of the Eq.~(\ref{eq:eigenfunction-overlap_semiclassical})
above are completely identical to those in Eq.~(\ref{eq:semiclassical-normalization}).
While in the perturbed system, the wave function has an additional
contribution $\propto(-1)^{n}$ (see \appref{Matrix-WKB}), this correction
is small as $O\left(\left(J\delta t\right)^{2}\right)$ (see also
\subsecref{Change-in-the-matching-conditions}). Moreover, the overlap
of this correction is further suppressed for states $\left|N\right>$
whose classical momenta are far from satisfying $2p+\pi=0\mod{2\pi}$,
as will become apparent momentarily.

One can further neglect the difference between $v_{N}$ and $v_{N,\text{eff}}$,
and between $C_{N}$ and $C_{N,\text{eff}}$ in Eq.~(\ref{eq:eigenfunction-overlap_semiclassical}),
which yields the following expression:
\begin{equation}
\left\langle N|N_{\text{eff}}\right\rangle \approx\frac{2}{T\left(E_{N}\right)}\intop_{\text{allowed}}\frac{dx}{v\left(E_{N},x\right)}\,\cos\left\{ \left[S_{N,\text{eff}}(x)-S_{N}(x)\right]/a\right\} .
\label{eq:overlap_via-integral-over-trajectory}
\end{equation}
The difference in actions consists of the direct perturbation and
the effect of the change in the energy level:
\begin{equation}
S_{N,\text{eff}}(x)-S_{N}(x)\approx\delta S\left(E_{N},x_{1};x\right)+\frac{\partial S\left(E_{N},x_{1};x\right)}{\partial E}\,\delta E_{N}.
\end{equation}
Because $S$ is the action on the classical trajectory, Eq.~(\ref{eq:overlap_via-integral-over-trajectory})
can be rewritten as
\begin{equation}
\left\langle N|N_{\text{eff}}\right\rangle \approx\frac{1}{T\left(E_{N}\right)}\intop^{T}_{0}dt\,\cos\left\{ \frac{1}{a}\left(\delta S\left(E_{N};t\right)-t\frac{\delta S\left(E_{N};T\right)}{T}\right)\right\} ,
\label{eq:overlap_via-integral-over-time}
\end{equation}
where $\delta S\left(E_{N};t\right)\equiv\delta S\left(E_{N},x_{1};x(t)\right)$,
$t=\partial S\left(E_{N},x_{1};x\right)/\partial E$ is the time along
the classical trajectory $x(t)$ measured from $x=x_{1}$, and we
have used Eq.~(\ref{eq:semiclassical-energy-shift}) for $\delta E_{N}$,
with $\delta S\left(E_{N};T\right)\equiv2\delta S_{12}\left(E_{N}\right)$
being the total additional action along the trajectory. In Eq.~(\ref{eq:overlap_via-integral-over-time}),
the argument of the cosine is small, which justifies the Taylor expansion:
\begin{equation}
\delta P_{N}=1-\left|\left\langle N|N_{\text{eff}}\right\rangle \right|^{2}\approx\frac{1}{T\left(E_{N}\right)}\intop^{T}_{0}dt\,\left[\frac{1}{a}\left(\delta S\left(E_{N};t\right)-t\frac{\delta S\left(E_{N};T\right)}{T}\right)\right]^{2}.
\label{eq:probability-defect_WKBA}
\end{equation}
This equation is applicable as long as the result is small, and the
classically allowed region is the dominant one, i.e., $\delta P_{N}\ll1$
and $n_{\text{cl}}=\left(x_{2}-x_{1}\right)/a\gg1$. Compared to exact
numerical diagonalization, Eq.~(\ref{eq:probability-defect_WKBA})
gives the correct result away from $N\sim1$ (with $N=1$ being the
ground state), while for $N\sim1$ it overestimates the result by
a factor of order unity.%

Based on Eq.~(\ref{eq:probability-defect_WKBA}) and on the time-reversal
symmetry of the STA ($\delta t\to-\delta t$), one arrives at Eq.~(\ref{eq:probability-defect-estimation})
of the main text:
\begin{equation}
\delta P_{N}=C_{N}\left(J\delta t\right)^{4}\left[1+O\left(\delta t^{2}\right)\right],
\label{eq:probability-defect}
\end{equation}
where $C_{N}$ is a positive constant that can be estimated by Eq.~(\ref{eq:probability-defect_WKBA})
away from $N=1$:
\begin{equation}
\left(J\delta t\right)^{4}C_{N}=\frac{1}{T\left(E_{N}\right)}\intop^{T}_{0}dt\,\left[\frac{1}{a}\left(\delta S\left(E_{N};t\right)-t\frac{\delta S\left(E_{N};T\right)}{T}\right)\right]^{2}.
\end{equation}
$C_{N}$ is bounded from above by $n^{2}_{\text{cl}}$ (with $n_{\text{cl}}$
being the number of sites in the allowed region) but typically of
order~1. Indeed, using the exact form of the action correction, Eqs.~(\ref{eq:effective-Hamiltonian_classical-correction})--(\ref{eq:correction-to-action}),
one obtains:
\begin{equation}
\left(J\delta t\right)^{4}C_{N}\le\max_{t\in\left[0,T\right]}\left[\frac{1}{a}\left(\delta S\left(E_{N};t\right)-t\frac{\delta S\left(E_{N};T\right)}{T}\right)\right]^{2}\le\left(\frac{2}{a}\,\delta S\left(E_{N};T\right)\right)^{2}=\left(\frac{2}{a}\,\delta E_{N}\,T\right)^{2}\le\left(\frac{\left(J\delta t\right)^{2}}{3}n_{\text{cl}}\right)^{2},
\end{equation}
where estimation~(\ref{eq:energy-shift-estimation}) for $\delta E_{N}$
was used in the last step. Therefore, the probability defect~(\ref{eq:probability-defect})
is guaranteed to be small if
\begin{equation}
\left(J\delta t\right)^{2}n_{\text{cl}}\ll1.
\label{eq:criterion_small-probability-defect}
\end{equation}
Criterion~(\ref{eq:criterion_small-probability-defect}) can be derived
from energy conservation: If the energy shift of state~$N$, Eq.~(\ref{eq:semiclassical-energy-shift}),
approaches the level spacing~$\Delta$, i.e., $\delta E_{N}/\Delta\sim\left(J\delta t\right)^{2}n_{\text{cl}}\sim1$,
then this state must be smeared across other states with close energies.

\begin{figure}
\begin{centering}
\includegraphics[width=0.8\textwidth]{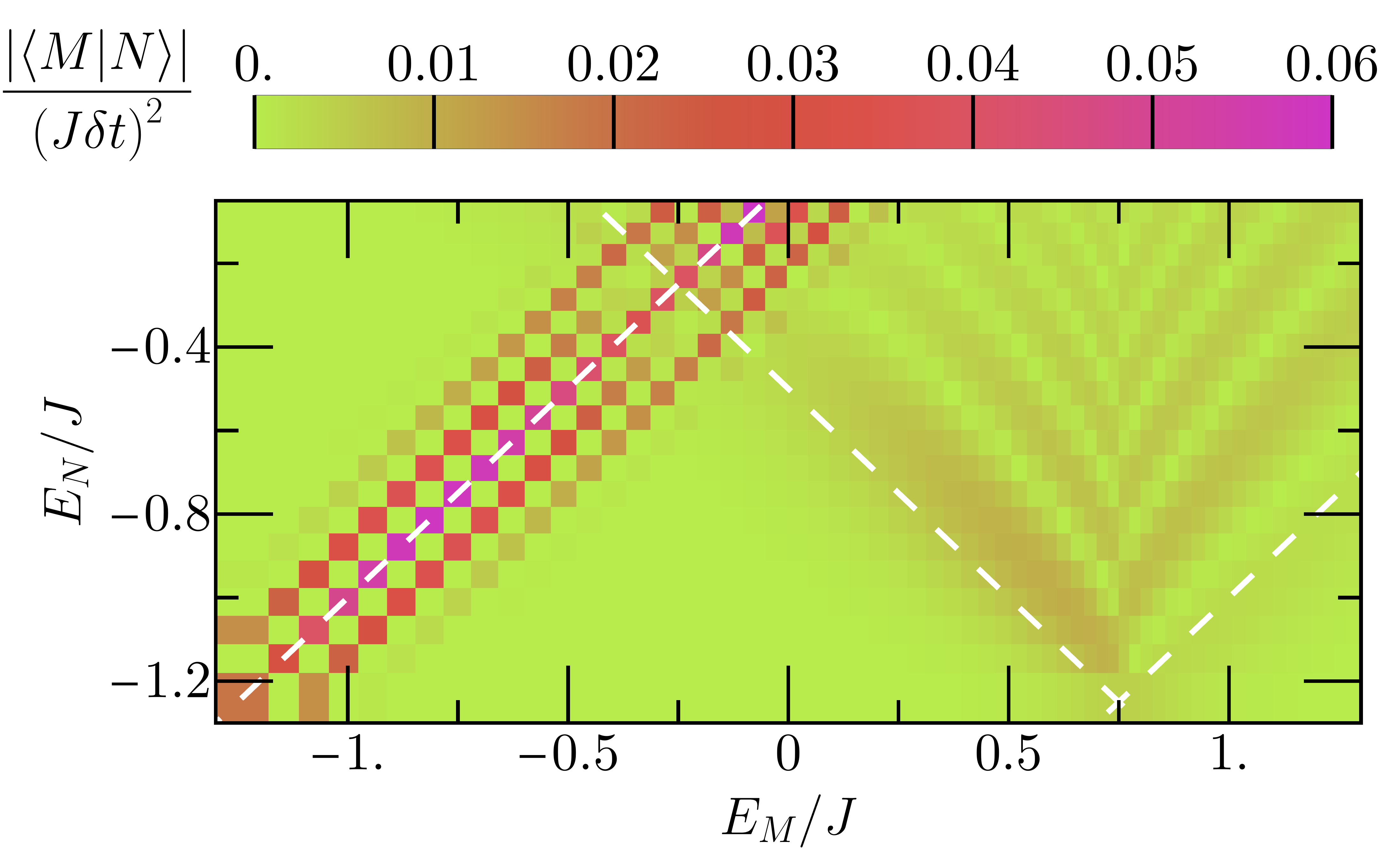}
\par\end{centering}
\caption{Color plot of the absolute value of the overlaps $\left|\left\langle M|N\right\rangle \right|$
between the metastable levels $\left|N\right\rangle $ in the left
well of the double-well Hamiltonian and the metastable levels $\left|M\right\rangle $
of the effective Hamiltonian with $\delta tJ=0.2$ with operator ordering
$\left\{ P,\,K_{\text{even}},\,K_{\text{odd}}\right\} $. The potential
profile is given by $h_{n}=P\,\cos\left\{ 4\pi\left(n-1\right)/\left(L-1\right)\right\} $,
$P=1.25\,J$. For $N=M$, the quantity $\sqrt{\delta P_{N}}\equiv\sqrt{\sum_{M\protect\neq N}\left|\left\langle M|N\right\rangle \right|^{2}}$
is plotted. All values are normalized by $\left(J\delta t\right)^{2}$.
All energies are shifted by $J$ to simplify comparison with the potential
profile. The diagonal white dashed lines correspond to (from left
to right): \emph{i)}~$E_{M}=E_{N}$, \emph{ii)}~$E_{M}=2J-2P-E_{N}$,
\emph{iii)}~$E_{M}=2J+E_{N}$. The first two indicate, respectively,
\emph{i)}~the smearing of the target state over states with close
energies, and \emph{ii)}~the overlap between the states with typical
momenta $p_{N}$ and $p_{M}\sim\pi-p_{N}$. Computing $\left|\left\langle M|N\right\rangle \right|$
for $M\protect\neq N$ with the first-order perturbation theory, $\left\langle M|N\right\rangle \approx\left\langle M|\delta\mathcal{H}|N\right\rangle /\left(E_{M}-E_{N}\right)$,
yields indistinguishable results. \label{fig:Perturbation-matrix-elements}}
\end{figure}

Importantly, there are also significant overlaps $\left\langle M|N_{\text{eff}}\right\rangle $
between states with momenta $\sim p$ and $\sim\pm\pi+p$, facilitated
by the oscillating term in the effective Hamiltonian (see \figref{effective-Hamiltonian_matrix-elements},~\emph{left}),
as shown in \figref{Perturbation-matrix-elements}. While not directly
relevant to the single-particle problem, these phenomena might be
important for the many-body problem via the orthogonality catastrophe~\citep{anderson-1967_orthogonality-catastrophe}.

\section{Proposal on Experimental Observation of Tunneling\label{app:Experimental-design}}

\subsection{The Experimental Design}

In this Appendix, we present a detailed experimental setup to demonstrate
the results reported in the main text. We consider the system discussed
in \secref{Solvable-model} with $L=50$ spins, within the capabilities
of existing devices~\citep{neill-2021_fermionic-ring-simulation}.
We will limit ourselves to the maximum depth of $10^{5}$ two-qubit
layers. Although this is roughly 60 times better than what was reported
in Ref.~\citep{neill-2021_fermionic-ring-simulation}, our choice
is motivated by the recent advances in the coherence times~\citep{nature-2021_long-coherence-times},
as well as the possibility of post-selection of $S^{z}_{\text{total}}$,
as done in Ref.~\citep{neill-2021_fermionic-ring-simulation}. Unless
otherwise stated, all simulations use the product formula, Eq.~(\ref{eq:Suzuki-Trotter_2nd-order}),
with ordering $\left\{ K_{\text{even}},\,P,\,K_{\text{odd}}\right\} $,
which delivers up to $M=5\times10^{4}$ Trotter steps (see \appref{Gate-ordering}).
All other orderings introduce only limited quantitative changes that
are suppressed by the smoothness of the potential. In general, the
choice of the operator ordering depends on the hardware (see \appref{Gate-ordering}).

The experiment aims to observe tunneling in the double-well potential
(see \figref{tunneling-problem_potential-sketch}). To this end, one
initializes the system in a state localized in one of the wells and
observes the dynamics of the total number of particles in one half
of the system. Given the limitations of existing devices, we must
carefully design the double-well potential according to the following
requirements:
\begin{enumerate}
\item The well has to be deep enough to be amenable to the semiclassical
description, i.e., to accommodate many de~Broglie wavelengths at
the target energy.
\item There must be bound states that are close to the maximum of the potential
profile, such that the tunnel splitting is not too small to be resolved
within the maximum number of Trotter steps. On the other hand, the
barrier has to be sufficiently long and high for the semiclassical
description to be applicable.
\item Since the barrier transparency strongly depends on energy, one wishes
to minimize the trivial effect of the energy shift of the bound states
inside the well due to STA. According to Eqs.~(\ref{eq:effective-Hamiltonian_classical-correction})--(\ref{eq:semiclassical-energy-shift}),
one should choose initial states with momenta close to $\pi/2$, i.e.,
at energies close to $J$ above the bottom of the potential well. 
\item To maximize the visibility of the Rabi oscillations, one has to maximize
the overlap of the initial state with the target bound state. As demonstrated
below, the accuracy of the initial state preparation is not essential:
even a 10\% error in the amplitude of the wave function at each site
yields good visibility of the oscillations. One can thus use the semiclassical
approximation~(\ref{eq:semiclassical-wave-function}) for the wave
function in the well. Such a state can be created by flipping one
spin at the end of the system and evolving it with a sequence of $n_{\text{cl}}$
purely two-qubit gates, where $n_{\text{cl}}$ is the number of sites
in the classically allowed region. Furthermore, according to Eqs.~(\ref{eq:trotterized-evolution_eigenfunction})--(\ref{eq:trotterized-evolution_WKB-ansatz}),
the unperturbed semiclassical wave function for $p\sim\pi/2$ is qualitatively
correct for any $J\delta t/2$ since the oscillating component $\sim(-1)^{n}$
is nearly absent.
\item The potential should be smooth enough, such that semiclassical treatment
is applicable away from the classical turning points, i.e., criterion~(\ref{eq:semiclassics_applicability-criteria})
should be satisfied.
\end{enumerate}
Based on these requirements, we employ the following potential profile:
\begin{align}
h_{n} & =P\,\frac{f\left(n,\delta n\right)}{f\left(n_{0},\delta n\right)}+\alpha\,\left(\frac{n-(L+1)/2}{L}\right),\,\,n_{0}=\left\lfloor (L+1)/2\right\rfloor ,\nonumber \\
f\left(n,\delta n\right) & =\exp\left\{ -\frac{1}{2}\left(\frac{n-\delta n-(L+1)/2}{w}\right)^{4}-\frac{L/2}{\sqrt{(n-1)(L-n)}}\right\} .
\label{eq:experimental-potential-profile}
\end{align}
Examples of this potential profile for various parameters are shown
in Figs.~\ref{fig:experimental-eigensystem_symmetric}~and~\ref{fig:experimental-eigensystem_asymmetric}.
Potential~(\ref{eq:experimental-potential-profile}) includes two
terms: the first term governs the principal profile, while the second
term enables the tuning accidental resonance in asymmetric potentials.
The parameters $P,w$ in Eq.~(\ref{eq:experimental-potential-profile})
control the height and width of the barrier. The second term in the
exponent of Eq.~(\ref{eq:experimental-potential-profile}) is needed
to force $h_{n}\rightarrow0$ close to the ends of the system. Tuning
$\delta n$ in Eq.~(\ref{eq:experimental-potential-profile}) away
from zero breaks the mirror symmetry of the potential, such that two
wells are of different spatial sizes. One then manually tunes $\alpha$
away from zero to create accidental resonance between a chosen pair
of localized levels in the two wells. The resulting eigenlevels for
various choices of the parameters are also shown in Figs.~\ref{fig:experimental-eigensystem_symmetric}~and~\ref{fig:experimental-eigensystem_asymmetric}.

For simplicity, we choose the initial state to be the superposition
of exact left and right states that maximizes the population in the
left half of the system (for instance, in the case of a symmetric
potential the superposition is equally weighted with properly chosen
signs). To imitate the state preparation inaccuracies, we further
add 10\% multiplicative noise to the amplitude of the wave function
independently at each site as $\psi_{i}\mapsto\psi_{i}\left(1+\epsilon_{i}\right)$,
$\epsilon_{i}\in\left[-0.1,0.1\right]$. We then truncate the wave
function to the left half of the system by setting $\psi_{i}=0$ for
$i>(L+1)/2$, after which the state is normalized.

The wave function is then evolved for $M=60000$ Trotter steps. The
total number of particles in the left half of the system,
\begin{equation}
\mathcal{N}_{\text{left}}(t)=\sum^{L/2}_{i=1}\left|\left\langle i|\psi(t)\right\rangle \right|^{2},
\label{eq:left-well_occupation-number}
\end{equation}
is recorded every $\delta M=250$ Trotter steps. The Fourier transform
$n_{\omega}=\frac{\delta M}{M}\sum_{t}\mathcal{N}_{\text{left}}(t)e^{i\omega t}$
of the resulting time series is calculated for $\omega_{n}=\frac{2\pi}{\delta t\,\delta M}\frac{n-1}{M/\delta M-1}$,
$n\in\left\{ 0,\,M/\delta M-1\right\} $. The Rabi oscillations of
the form~(\ref{eq:detuned-Rabi-oscillations}) manifest as a sharp
peak in $\left|n_{\omega}\right|^{2}$ at the Rabi frequency. The
density plots of $\left|n_{\omega}\right|$ as a function of both
$\omega$ and $\delta t$ are shown in Figs.~\ref{fig:particle-number_density-plot}~and~\ref{fig:particle-number_density-plot_with-detuning}
for different sets of parameters. Given the Fourier transform $n_{\omega}$,
one then extracts the observed period of Rabi oscillations as $T_{\text{Rabi}}=2\pi/\omega_{\text{max}}$,
where $\omega_{\text{max}}$ is the angular frequency where the maximum
of $n_{\omega}$ is achieved. The resulting dependencies of $T_{\text{Rabi}}$
on the Trotter step $\delta t$ are shown in Figs.~\ref{fig:rabi-oscillation-period_vs_trotter-step},~\ref{fig:Rabi-period_vs_Trotter-step_with-energy-shift},
and~\ref{fig:particle-number_density-plot_with-detuning} for different
sets of parameters. The parameters are chosen to demonstrate the effects
discussed in the main text, as explained below.

The semiclassical description of the system is constructed according
to \secref{Semiclassical-Approach-to-Trotterization}. The action
integrals, Eqs.~(\ref{eq:Bohr-quantization-condition}) and~(\ref{eq:underbarrier-action}),
and their energy derivatives are calculated numerically. The semiclassical
energy levels are found by numerically solving the exact quantization
condition, Eq.~(\ref{eq:Bohr-quantization-condition}).

\subsection{Demonstration of Tunneling and its Enhancement\label{subsec:Demonstration-of-tunneling-enhancement}}

\begin{figure}
\begin{centering}
\includegraphics[width=0.49\textwidth]{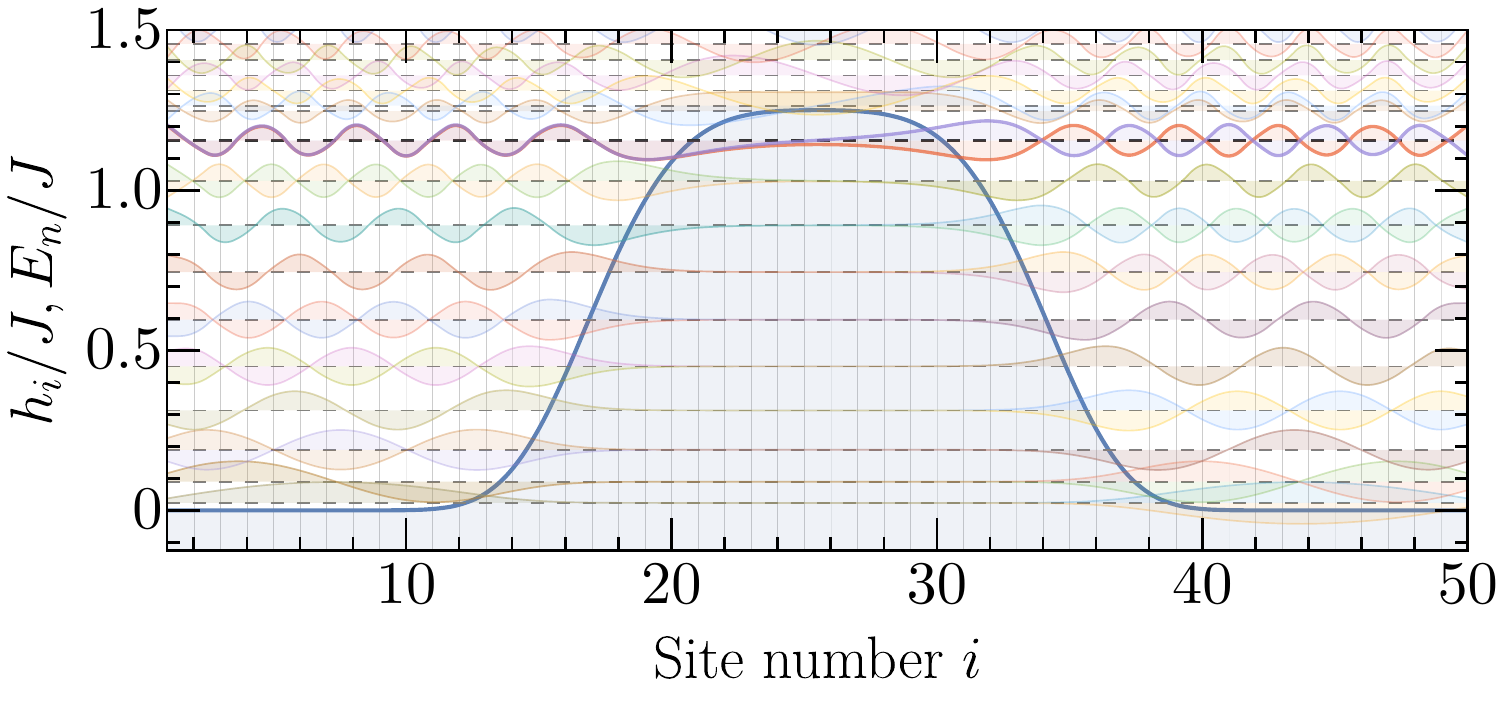}\hfill{}\includegraphics[width=0.49\textwidth]{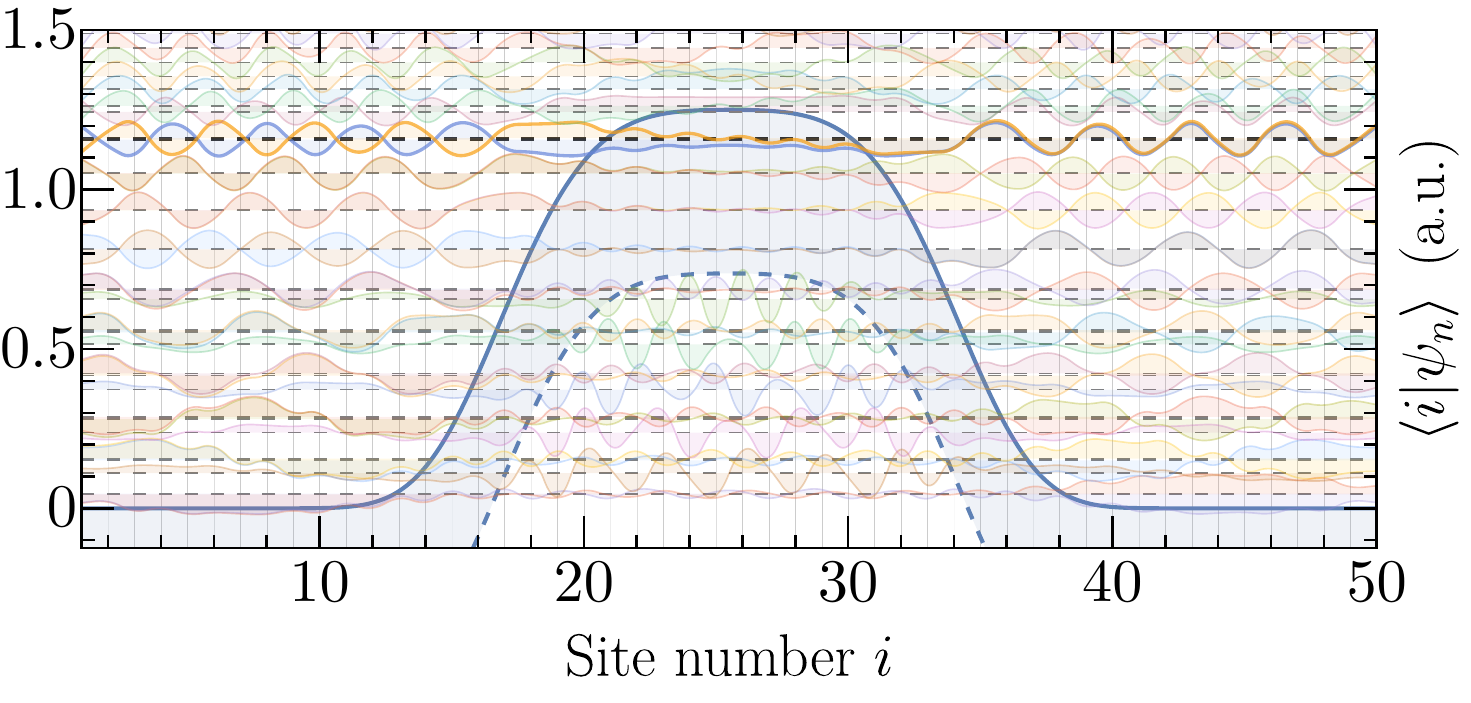}
\par\end{centering}
\caption{Visualization of the eigensystem for the potential profile~(\ref{eq:experimental-potential-profile})
with $P=1.25J$, $w=8$, $\delta n=0$, $\alpha=0$ (\emph{left})
and for the STA with $J\delta t/\pi=0.795$ (\emph{right}). The dashed
thick blue line in the right panel corresponds to the Floquet-translated
band top $\left[h_{n}+2J-\Omega\right]/J$, with the classically forbidden
region shown in light blue. All energies are shifted by $J$ for clarity.
The energies of the highlighted pair of levels are close to $-J+1.156J$
(\emph{left}) and $-J+1.159J$ (\emph{right}). The level splitting~$\eta$
of the highlighted pair is $1.102\times10^{-3}\,J$ (\emph{left})
and $4.82\times10^{-3}\,J$ (\emph{right}). Note the strong distortion
of the wave functions at low energies in the right panel. \label{fig:experimental-eigensystem_symmetric}}
\end{figure}

We start by demonstrating the enhancement of tunneling in the case
of a symmetric potential with $P=1.25$, $w=8$, $\delta n=0$, $\alpha=0$.
The 10th pair of energy levels is situated at energy $E_{10}\approx-J+1.156J$,
fulfilling all the requirements listed above. \figref{experimental-eigensystem_symmetric}
shows the corresponding eigenpair and its counterpart for a large
Trotter step, $J\delta t/\pi\approx0.795$. Two details are worth
pointing out: \emph{i)}~the energy shift due to STA is nearly absent
due to the particular choice of the energy level, but \emph{ii)}~the
eigenstates at low energies are strongly distorted by STA, as described
in \secref{Large-Trotter-steps}.

\begin{figure}
\begin{centering}
\includegraphics[width=0.7\textwidth]{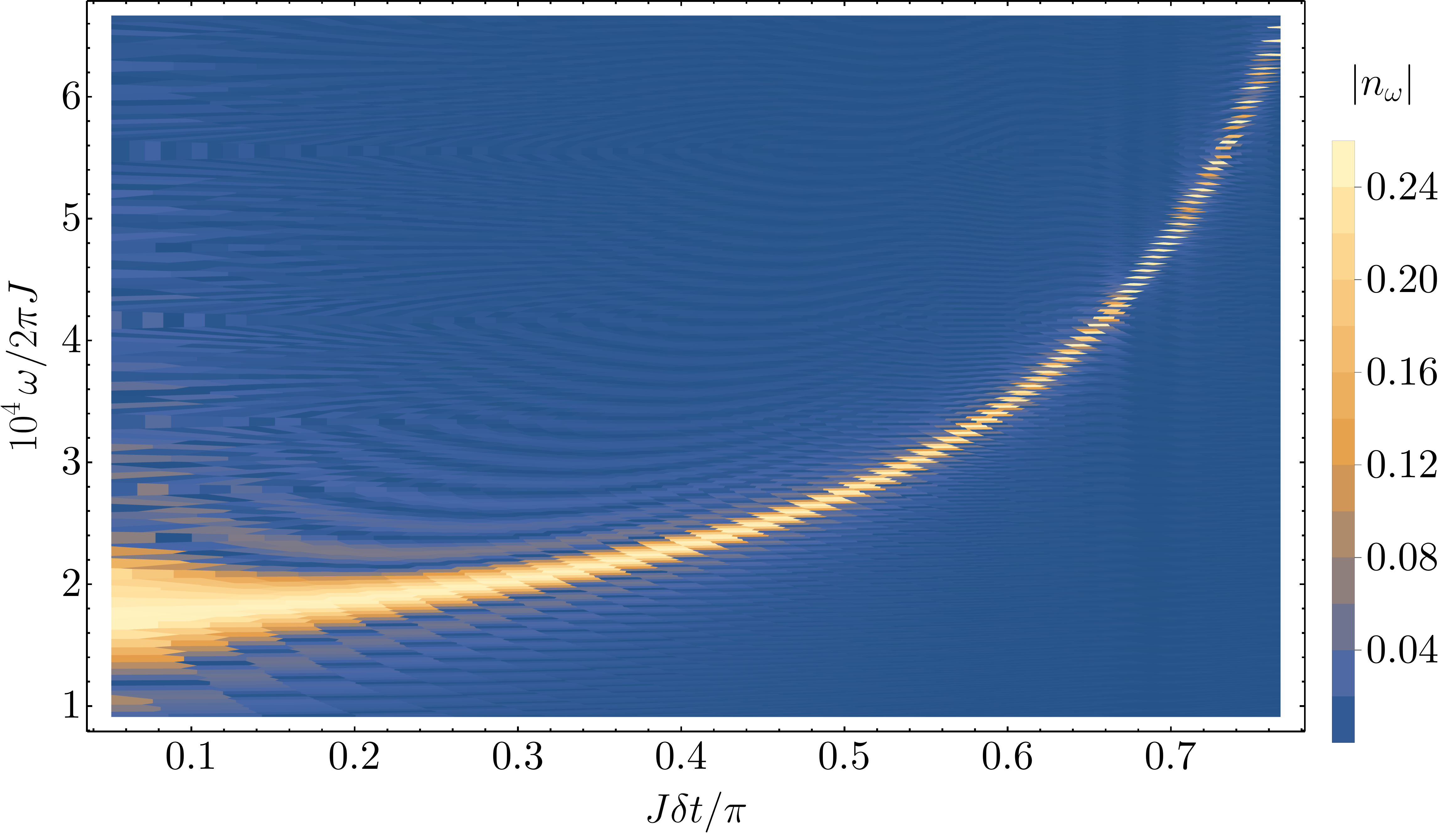}
\par\end{centering}
\caption{Density map of the absolute value of the Fourier harmonic $\left|n_{\omega}\right|$
of the occupation number in the left well, Eq.~~(\ref{eq:left-well_occupation-number}),
as a function of the angular frequency $\omega$ and Trotter step
$\delta t$. The parameters of the unperturbed system are the same
as for \figref{experimental-eigensystem_symmetric}. The position
$\omega_{\max}$ of the maximum amplitude $\left|n_{\omega}\right|$
for each $\delta t$ corresponds to a point $T_{\text{Rabi}}=2\pi/\omega_{\max}$
in \figref{rabi-oscillation-period_vs_trotter-step} of the main text.
\label{fig:particle-number_density-plot}}
\end{figure}

One then applies the protocol described above to probe the period
of the Rabi oscillations. The density map of the occupation number
harmonic $\left|n_{\omega}\right|$ as a function of angular frequency
$\omega$ and Trotter step $\delta t$ is shown on \figref{particle-number_density-plot}.
The corresponding dependence of the Rabi oscillation period $T_{\text{Rabi}}$
on $\delta t$ is shown in~\figref{rabi-oscillation-period_vs_trotter-step}
of the main text. Crucially, the period of Rabi oscillations decreases
in agreement with the semiclassical prediction, despite the lack of
energy shift of the corresponding eigenpair, as verified both by semiclassical
approximation and by exact diagonalization.

\subsection{Additional Enhancement of Tunneling due to the Energy Shift}

\begin{figure}
\begin{centering}
\includegraphics[width=0.7\textwidth]{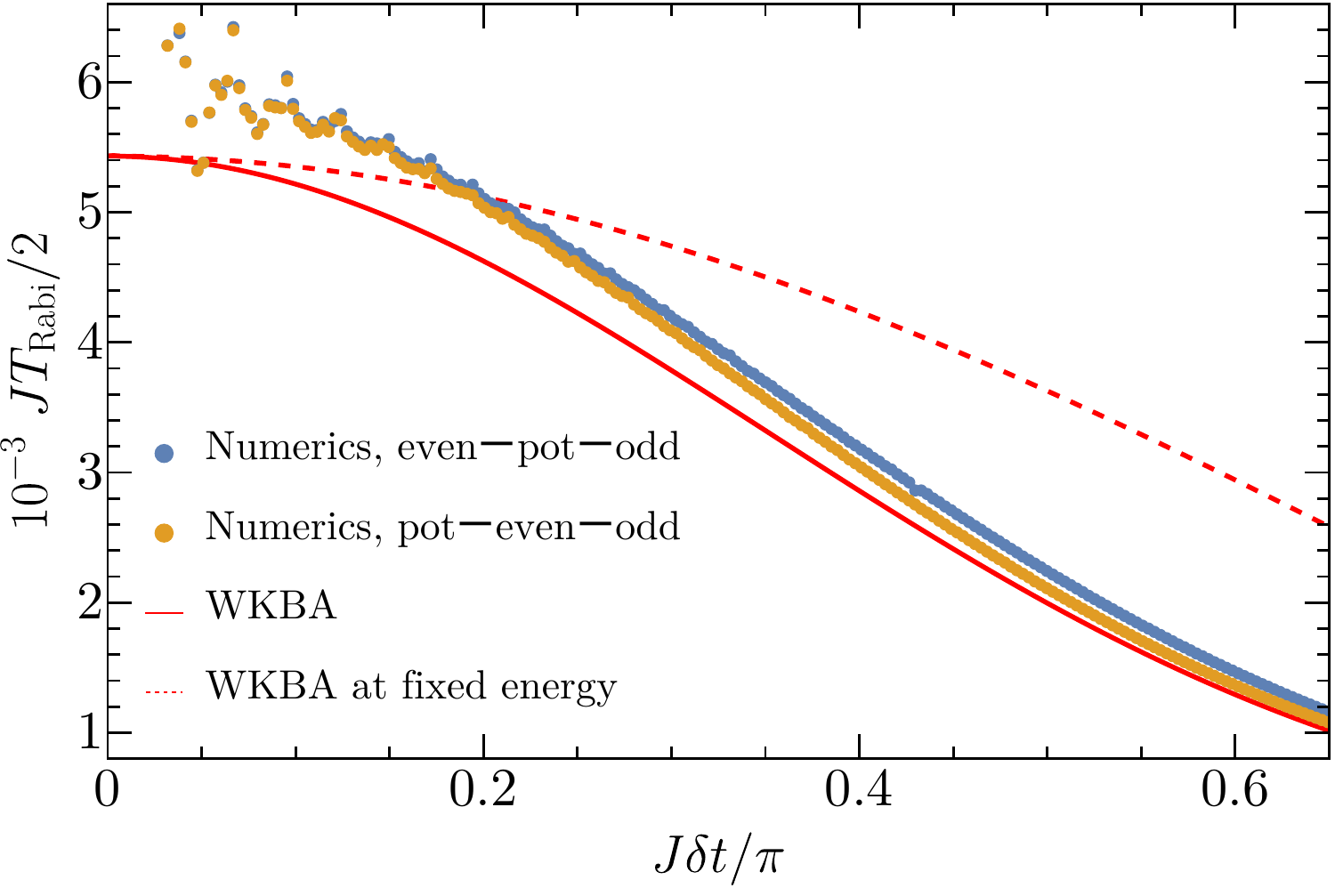}
\par\end{centering}
\caption{The dependence of the Rabi oscillation period $T_{\text{Rabi}}$
on the Trotter step~$\delta t$. The evolution is performed according
to potential~(\ref{eq:experimental-potential-profile}) with parameters
$P=1.14,\,w=15$, $\delta n=0,\,\alpha=0$. The initial state is based
on the 7th eigenpair. The dots correspond to the numerical simulation
of STA with operator orderings $\left\{ K_{\text{even}},\,P,\,K_{\text{odd}}\right\} $
(blue) and $\left\{ P,\,K_{\text{even}},\,K_{\text{odd}}\right\} $
(orange). The red curves correspond to the semiclassical description
of the Rabi oscillations according to \secref{Semiclassical-Approach-to-Trotterization}.
The dashed line shows the semiclassical prediction calculated at the
energy of the eigenpair for $\delta t=0$. The solid line is the full
semiclassical prediction at the energy found from the correct quantization
condition for each $\delta t$. In contrast to \figref{rabi-oscillation-period_vs_trotter-step},
the two approaches yield manifestly different results due to the perturbation-induced
energy shift. The remaining discrepancy between the solid line and
direct simulation is due to corrections beyond the semiclassical description
(the potential is not sufficiently smooth). \label{fig:Rabi-period_vs_Trotter-step_with-energy-shift}}
\end{figure}

In this subsection, we demonstrate how the shift of the bound state
energies additionally enhances tunneling. To this end, we consider
the symmetric version of potential~(\ref{eq:experimental-potential-profile})
with $P=1.14J$, $w=15$, $\delta n=0$, $\alpha=0$. The 7th eigenpair
is located at $E\approx-J+1.087J$, and thus experiences a noticeable
shift due to STA. The resulting period of Rabi oscillations as a function
of $\delta t$ is shown in \figref{Rabi-period_vs_Trotter-step_with-energy-shift},
along with the corresponding semiclassical predictions. This figure
also shows the same curve for STA with the operator ordering $\left\{ P,\,K_{\text{even}},\,K_{\text{odd}}\right\} $.
To further illustrate the effect of the energy shift, the dashed red
curve in \figref{Rabi-period_vs_Trotter-step_with-energy-shift} shows
the semiclassical period of Rabi oscillations calculated for the effective
Hamiltonian~(\ref{eq:trotterized-classical-hamiltonian}) at finite
$\delta t$ but at the \emph{fixed} energy of the level at $\delta t=0$,
as opposed to the actual energy level, which is itself a function
of the Trotter step via the quantization condition~(\ref{eq:Bohr-quantization-condition}).
In contrast to \figref{rabi-oscillation-period_vs_trotter-step} of
the main text, the difference due to this energy shift is well-pronounced.

\subsection{Suppression of Tunneling by the Resonance Detuning}

\begin{figure}
\begin{centering}
\includegraphics[width=0.49\textwidth]{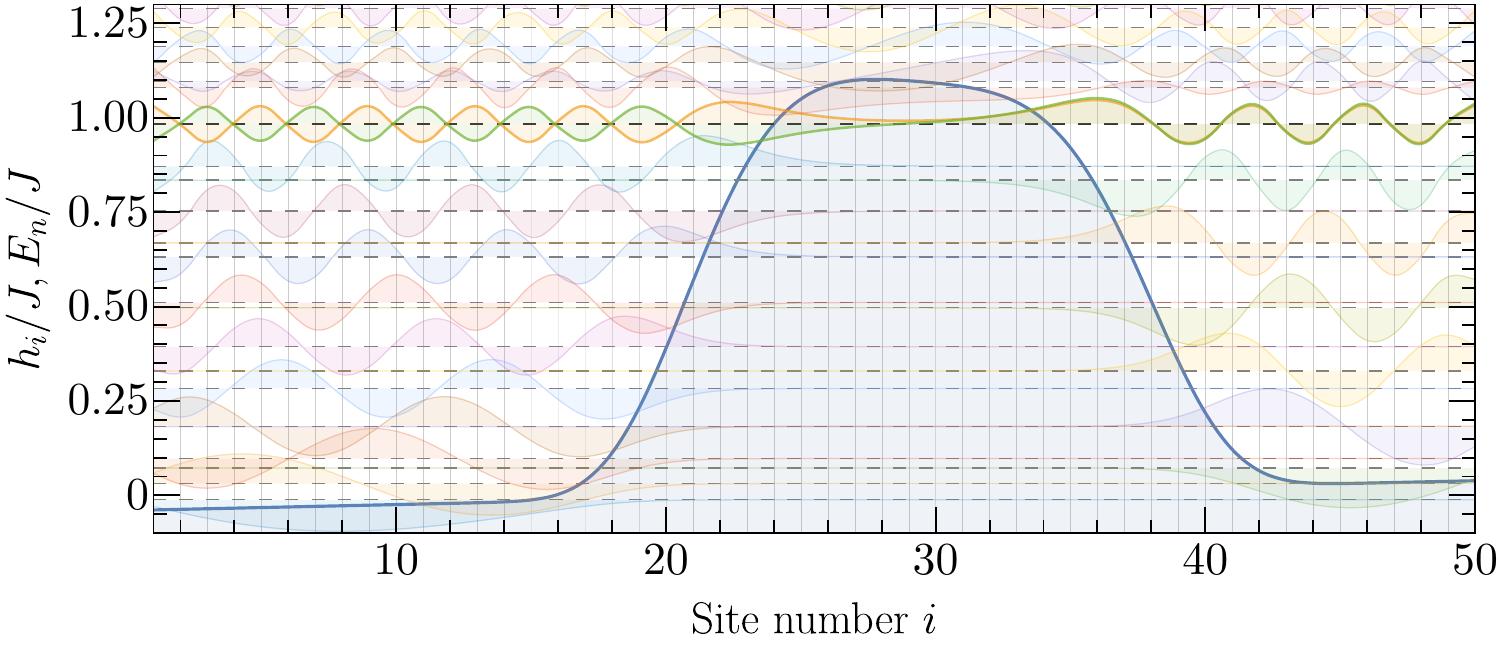}\hfill{}\includegraphics[width=0.49\textwidth]{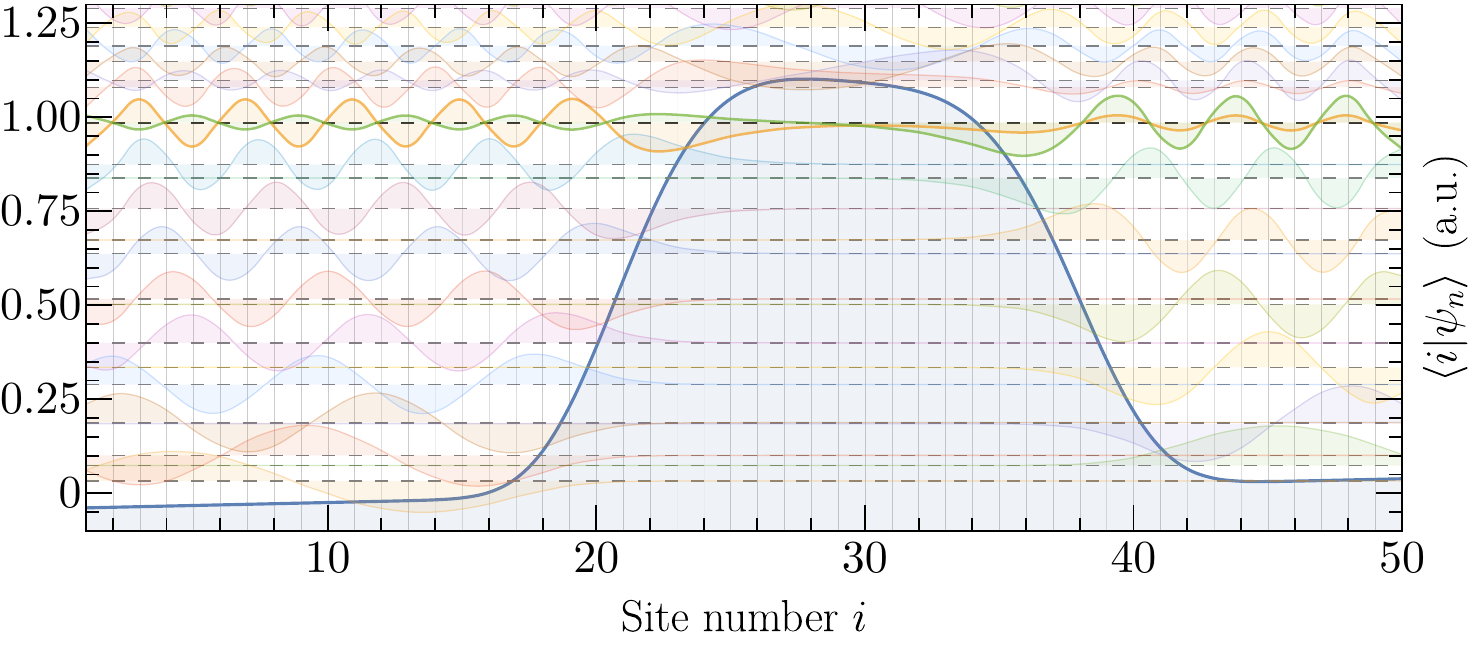}
\par\end{centering}
\caption{Visualization of the eigensystem for potential~(\ref{eq:experimental-potential-profile})
with $P=1.05,\,w=8,\,\delta n=4$, and $\alpha\approx0.0394$, found
with high precision by maximizing the Rabi period for the 9th pair
of eigenstates (highlighted, $E\approx-J+0.9837J$). \emph{Left:}~$J\delta t=0$.
\emph{Right:}~$J\delta t/\pi\approx0.223$. In contrast to \figref{experimental-eigensystem_symmetric},
the value of $J\delta t$ in the right panel is not large enough to
cause noticeable distortion of the low-energy wave functions. However,
due to the potential asymmetry, the hybridization between the two
levels is destroyed, which causes them to be predominantly localized
in only one of the two wells.\label{fig:experimental-eigensystem_asymmetric}}
\end{figure}

We then turn to the effect of resonance detuning, which is present
in asymmetric potentials. To this end, we set $P=1.05J,\,\,w=8,\,\,\delta n=4,$
and find $\alpha\approx0.0394$ from maximizing the period of Rabi
oscillations for the 9th pair of states located close to $E=-J+0.9837J$.
The eigensystem of the original system and its STA are shown in \figref{experimental-eigensystem_asymmetric}.
In full agreement with \subsecref{Rabi-oscillations_detuning}, one
can clearly observe that STA introduces shifts in the bound state
energies that are different for the two wells. As a result, the hybridization
of the eigenpair is destroyed, such that each state is mostly localized
in only one of the two wells. 

\begin{figure}
\begin{centering}
\includegraphics[width=0.49\textwidth]{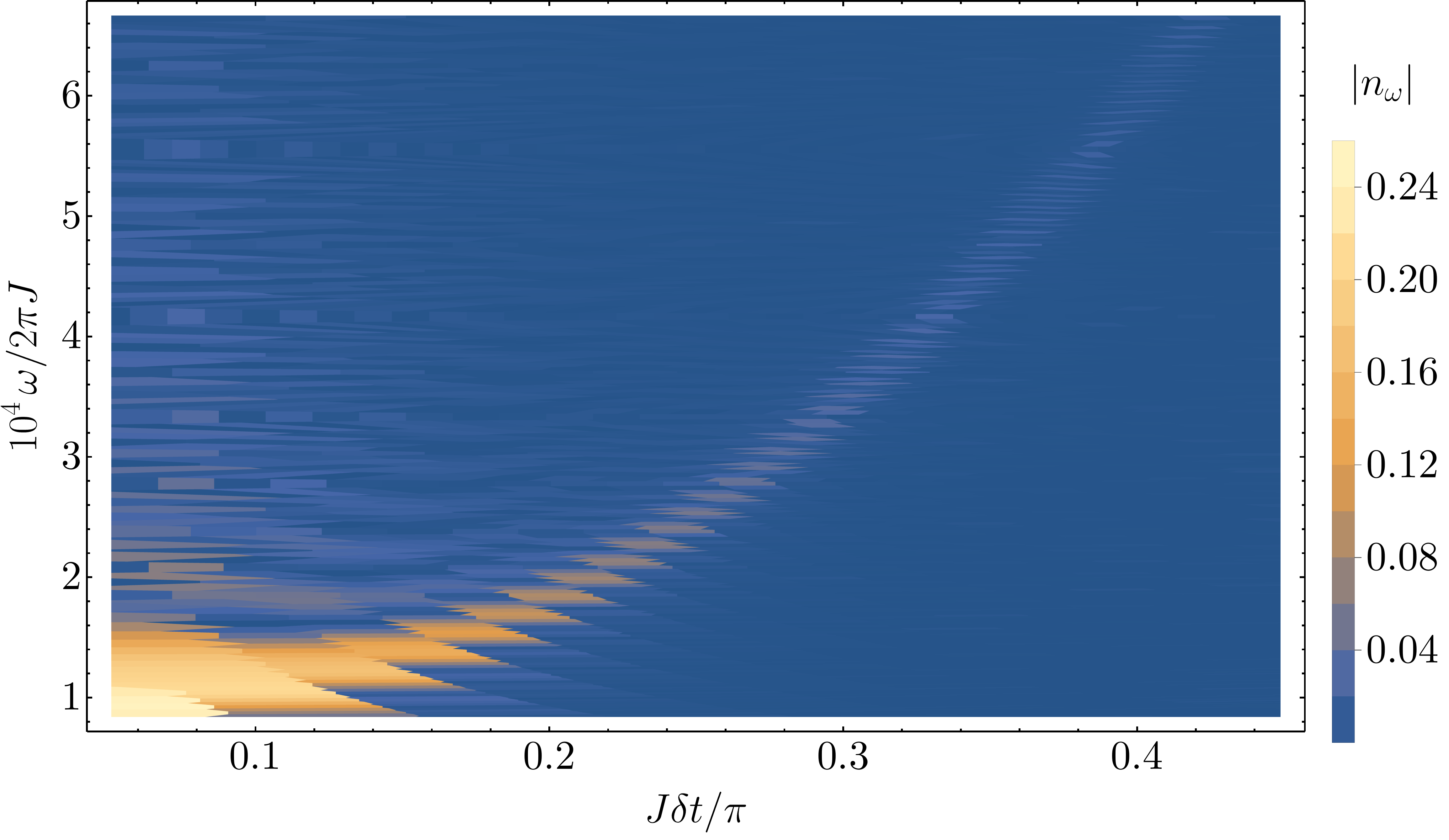}\hfill{}\includegraphics[width=0.46\textwidth]{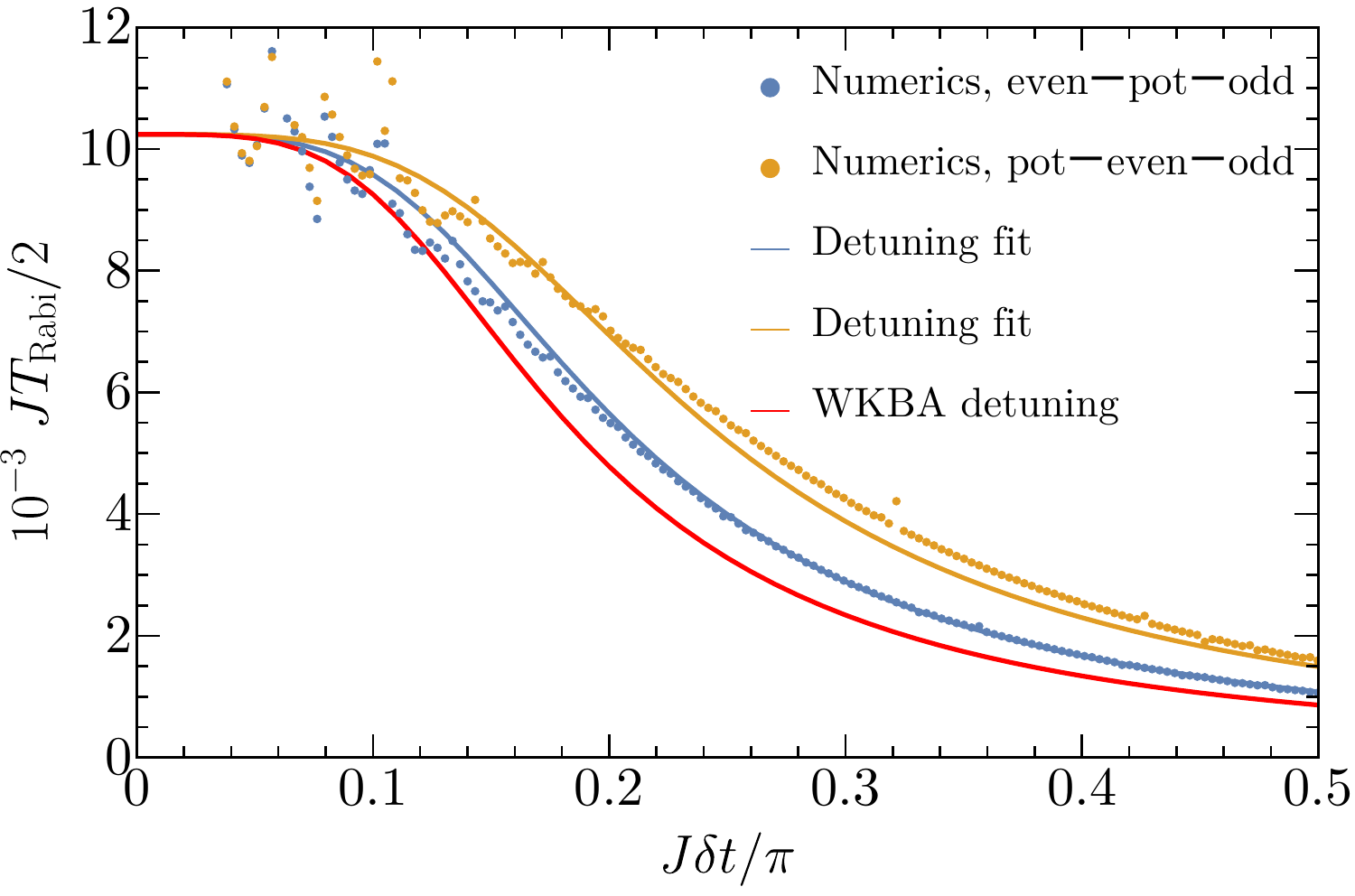}
\par\end{centering}
\caption{\emph{Left}: The density map of the absolute value of the Fourier
harmonic $\left|n_{\omega}\right|$ of the occupation number in the
left well, Eq.~(\ref{eq:left-well_occupation-number}), as a function
of angular frequency~$\omega$ and Trotter step~$\delta t$. The
evolution is performed for the parameters of \figref{experimental-eigensystem_asymmetric}.
In contrast to \figref{particle-number_density-plot}, the resonance
visibility sharply decreases with $\delta t$ due to detuning, according
to Eq.~(\ref{eq:Detuned-Rabi-oscillations_visibility}). \emph{Right:}
The dependence of the Rabi oscillation period $T_{\text{Rabi}}$ on
$\delta t$ for the same parameters. The dots are extracted from the
numerical simulations for operator orderings $\left\{ K_{\text{even}},\,P,\,K_{\text{odd}}\right\} $
(blue; corresponding to the left panel) and $\left\{ P,\,K_{\text{even}},\,K_{\text{odd}}\right\} $
(orange). The lines correspond to Eq.~(\ref{eq:Detuned-Rabi-oscillations_period}),
with $T(0)$ extracted from exact diagonalization at $\delta t=0$,
and $\beta\approx\left(2.45\pm0.15\right)\times10^{-3}$ (blue curve)
and $\beta=\left(1.61\pm0.07\right)\times10^{-3}$ (orange curve)
obtained from fitting. Jumps in the numerical data correspond to poor
extraction of the period due to decreased visibility (cf. \emph{left}
panel). The red curve is Eq.~(\ref{eq:Detuned-Rabi-oscillations_period})
with $\beta\approx2.943\times10^{-3}$ computed from WKBA Eqs.~(\ref{eq:correction-to-action})--(\ref{eq:semiclassical-energy-shift}).
The discrepancy with the numerical data is due to the steep potential
profile (cf.~\figref{experimental-eigensystem_asymmetric}), which
is also the reason why the two operator orderings yield quantitatively
different results. \label{fig:particle-number_density-plot_with-detuning}}
\end{figure}

The density map of $\left|n_{\omega}\right|$ is shown in \figref{particle-number_density-plot_with-detuning}.
Notably, the Rabi oscillation period decreases considerably faster
with $\delta t$ than in the case of a symmetric potential due to
finite detuning (cf. \figref{rabi-oscillation-period_vs_trotter-step}):
\begin{equation}
T(\delta t)=\frac{T(0)}{\sqrt{1+\left(\delta E\,T(0)/2\pi\right)^{2}}},\,\,\,\delta E=\beta J\,\left(J\delta t\right)^{2},
\label{eq:Detuned-Rabi-oscillations_period}
\end{equation}
where $T(0)$ is the period of oscillations in the target system (i.e.,
for $J\delta t=0$), and $\delta E$ is the perturbation-induced energy
shift, quadratic in $\delta t$ in accordance with \subsecref{Correction-to-semiclassics}.
Moreover, according to Eq.~(\ref{eq:detuned-Rabi-oscillations}),
the magnitude of the Rabi oscillations also decreases due to detuning,
which is reflected in the reduced visibility of the maximum Fourier
harmonic $\left|n_{\omega,\max}\right|$ of the occupation number:
\begin{equation}
\left|n_{\omega,\max}\right|\approx n_{0}\left[\frac{T(\delta t)}{T(0)}\right]^{2},
\label{eq:Detuned-Rabi-oscillations_visibility}
\end{equation}
where $n_{0}$ is the magnitude at $J\delta t=0$, and $T(\delta t)$
is found from Eq.~(\ref{eq:Detuned-Rabi-oscillations_period}).

\subsection{The Effect of Finite Gate Precision}

Lastly, we estimate the required gate precision, by which we understand
the tolerance to random shifts in the phases of one- and two-qubit
gates that are constant during each realization of the experiment,
as opposed to time-dependent errors that induce decoherence and thus
require more careful study. As follows from \subsecref{Rabi-oscillations_detuning},
the main effect of the gate error is that it causes uncontrolled energy
shifts of the bound states in each well, thus detuning the resonance.
Below, we consider uncorrelated disorder in the single-qubit gates
as an example. The latter translates into potential disorder in the
model and thus produces a random energy shift in energy level $E_{N}$
of the bound state. The dispersion of this quantity can be estimated
perturbatively:
\begin{equation}
\left\langle \delta E^{2}\right\rangle =\delta V^{2}\sum_{i}\left|\left\langle \psi_{N}|i\right\rangle \right|^{4}\sim\delta V^{2}\frac{1}{n_{\text{cl}}},
\label{eq:energy-level-dispersion_example}
\end{equation}
where we estimated the inverse participation ratio $\sum_{i}\left|\left\langle \psi_{N}|i\right\rangle \right|^{4}$
of the bound state inside the well via the inverse number of sites
in the allowed region. The expectation value of the total occupation
number in the left well is given by the average of the quantum expectation
value~(\ref{eq:detuned-Rabi-oscillations}) over the distribution
of the detuning $\epsilon$, whose dispersion~$\sigma$ is given
by Eq.~(\ref{eq:energy-level-dispersion_example}):
\begin{align}
\overline{n}(t) & =\overline{\left(\frac{\epsilon^{2}/4}{\eta^{2}+\epsilon^{2}/4}\right)+\left\{ 1-\left(\frac{\epsilon^{2}/4}{\eta^{2}+\epsilon^{2}/4}\right)\right\} \cos^{2}\sqrt{\eta^{2}+\epsilon^{2}/4}t}\nonumber \\
 & =\begin{cases}
1+\tau^{2}+\frac{1}{3}\tau^{4}\left(1+\frac{\sigma^{2}}{4}\right), & \tau\ll1,\\
F\left(\sigma(\sigma)\right)+\frac{1}{2}\text{Re}\left\{ \sqrt{\frac{4-2\sigma^{2}i\tau}{\pi\sigma^{2}}}\exp\left\{ 2i\tau+\frac{\left(4-2\sigma^{2}i\tau\right)^{2}}{8\sigma^{2}}\right\} \,K_{\frac{1}{4}}\left(\frac{\left(4-2\sigma^{2}i\tau\right)^{2}}{8\sigma^{2}}\right)\,\left[1+O\left(\sigma^{2}\right)\right]\right\} , & \sigma^{2}\tau\ll1,\\
F(\sigma)-\frac{\sin2\tau}{\sqrt{2\tau}}\left[1+O\left(\frac{1}{\tau}\right)\right], & \sigma^{2}\tau\gg1,
\end{cases}
\label{eq:averaged-occupation-number}
\end{align}
\begin{equation}
F(\sigma)=1-\sqrt{\frac{\pi}{2\sigma^{2}}}\,e^{\frac{2}{\sigma^{2}}}\text{erfc}\left(\frac{\sqrt{2}}{\sigma}\right)=\frac{1}{2}+\frac{\sigma^{2}}{8}+O\left(\sigma^{4}\right)\,\,\,(\sigma\ll1),
\end{equation}
where $\sigma^{2}=\left\langle \delta E^{2}\right\rangle /\eta^{2}$,
$\tau=\eta t$, $K_{n}(z)$ is the modified Bessel function of the
second kind of order $n$, and we have assumed Gaussian distribution
of $\epsilon$ to compute the asymptotic expressions.

As is clear both from Eq.~(\ref{eq:averaged-occupation-number})
and from the uncertainty principle, if the standard deviation $\sqrt{\left\langle \delta E^{2}\right\rangle }$
is smaller than the target tunnel splitting $\eta=2\pi/T$, tunneling
will remain visible. This yields the following condition on the typical
error in the single-qubit gate phase:
\begin{equation}
\sqrt{\left\langle \delta\phi^{2}\right\rangle }=\sqrt{\left\langle \delta V^{2}\right\rangle }\delta t\apprle\eta\delta t\,\sqrt{n_{\text{cl}}}\sim\sqrt{20}\times10^{-3}J\delta t.
\label{eq:phase-error-limitation}
\end{equation}
where the last estimation uses the parameters proposed in this Appendix.
Similar perturbative estimates can be made for the two-qubit gates,
in which case we can also express the result~(\ref{eq:phase-error-limitation})
as a bound on the relative gate precision: $\sqrt{\left\langle \delta\phi^{2}\right\rangle }/J\delta t\sim\sqrt{20}\times10^{-3}\sim5\times10^{-3}$.
For existing devices, condition~(\ref{eq:phase-error-limitation})
is satisfied even for the smallest value of $J\delta t=0.2$ used
in the simulations presented above, as the existing calibration procedure
\citep{neill-2021_fermionic-ring-simulation} allows achieving precision
of order $1\times10^{-4}$.

\section{Trotterization as a Periodic High-Frequency Perturbation and Prethermalization\label{app:Trotterization_as-periodic-perturbation_and_thermalization}}

The STA of \subsecref{Model_Suzuki-Trotter-alg} can be described
by the following piecewise-constant time-dependent Hamiltonian:
\begin{equation}
H_{\text{appr.}}(t)=n\times\begin{cases}
A_{n}, & \frac{1}{2n}<t/\delta t\le\frac{1}{n},\\
...,\\
A_{k}, & \frac{n-k}{2n}<t/\delta t\le\frac{n-k+1}{2n},\\
...,\\
A_{1}, & \frac{n-1}{2n}<t/\delta t\le\frac{n+1}{2n},\\
...,\\
A_{n}, & \frac{2n-1}{2n}<t/\delta t\le1,
\end{cases}
\label{eq:Suzuki-Trotter_exact-Hamiltonian}
\end{equation}
which is periodically repeated every Trotter step, with the overall
factor of $n$ introduced to identify the ``system time'' $t$ describing
the evolution of the target model~(\ref{eq:Hamiltonian}) with the
actual physical time on the device. Averaging Eq.~(\ref{eq:Suzuki-Trotter_exact-Hamiltonian})
over one Trotter step produces the target Hamiltonian:
\begin{equation}
\intop^{\delta t}_{0}dt\,H_{\text{appr.}}(t)=\delta t\,H.
\end{equation}
The difference $H_{\text{appr.}}(t)-H$ can be regarded as a high-frequency
perturbation in the form of a cosine series with the main frequency
$\Omega=2\pi/\delta t$ (due to the symmetry $t\leftrightarrow\delta t-t$
in $H_{\text{appr.}}(t)$, sine components are absent).

Because $H_{\text{appr.}}(t)$ is periodic with period $\delta t$,
the Floquet theorem guarantees the existence of the effective time-independent
Hamiltonian $H_{\text{eff}}$ that describes the evolution of the
system at stroboscopic times $t_{n}=n\delta t,\,\,n\in\mathbb{N}$:
$\psi\left(t_{n}\right)=\exp\left\{ -it_{n}H_{\text{eff}}\right\} \psi(0)$.
One can then derive a perturbative expansion~\citep[Sec. 3]{ref-for-high-frequency-expansion_bukov-2015}
to express the effective Hamiltonian $H_{\text{eff}}$ as a series
in powers of $1/\Omega$. In particular, Eq.~(\ref{eq:Suzuki-Trotter_leading-correction_recursion})
describes the leading term of this expansion of order $\Omega^{-2}$
(with the $\Omega^{-1}$ term vanishing due to the symmetry $t\leftrightarrow\delta t-t$).

The question then arises whether the high-frequency expansion is applicable
in the thermodynamic limit. Indeed, one expects a system under periodic
drive to absorb energy, unless the latter is in the many-body localized
phase~\citep[Ch. 8]{mahanManyParticlePhysics2000}. However, heating
is clearly inconsistent with the conservation of any quasi-local Hamiltonian
$H_{\text{eff}}$ with finite energy density at stroboscopic times.
This implies that the power series for $H_{\text{eff}}$ must be asymptotic,
i.e., diverge beyond a certain order. However, as shown in Ref.~\citep{prethermalization_abanin-2017},
truncating the power series at the optimally chosen order $n_{*}$
allows one to correctly describe stroboscopic evolution up to a time~$\tau_{*}$
that is exponentially large in the frequency, viz., $\ln\tau_{*}\sim\Omega/\Lambda$,
where $\Lambda$ is the typical local energy scale of the Hamiltonian,
e.g., $\Lambda=\max\left\{ J,h_{i}\right\} $ for Hamiltonian~(\ref{eq:Hamiltonian}).
We therefore expect our analysis to be qualitatively valid for sufficiently
small Trotter step~$\delta t$ if the simulation is carried out for
times $t$ shorter than the prethermalization time $\tau_{*}$. In
particular, the latter must be longer than the characteristic tunneling
times of interest.

Crucially, for the model considered in this work, the issue of heating
is absent. The Hamiltonian~(\ref{eq:Hamiltonian}) is mapped to non-interacting
fermions, and the exact dynamics of the many-body wave function is
fully determined by the single-particle evolution operator $\mathcal{U}$
{[}see Eq.~(\ref{eq:fermionic-evolution-operator}){]}. Since $\mathcal{U}$
now describes a single particle with bounded spectrum, no heating
is present, and the associated high-frequency expansion of the effective
single-particle Hamiltonian $\mathcal{H}_{\text{eff}}$ is thus convergent,
allowing one to infer its properties from perturbation theory in powers
of $\Omega^{-1}$. As a result, our analysis is quantitatively correct
for arbitrarily long times for the system~(\ref{eq:Hamiltonian})
in question. However, coupling to a bath will violate this conclusion
by introducing finite level width, which will inevitably cause heating
at sufficiently high frequency~\citep[Ch. 8]{mahanManyParticlePhysics2000}.

\end{document}